\documentclass[twocolumn]{aastex62}

\usepackage{amsmath}
\usepackage{longtable}
\usepackage{natbib}
 \usepackage{appendix}

\shorttitle{Atomic Spectral Survey in WASP-76b}
\shortauthors{Kesseli et al.}
\pdfoutput=1

\begin{document}

\title{An Atomic Spectral Survey of WASP-76b: Resolving Chemical Gradients and Asymmetries }

\correspondingauthor{Aurora Y. Kesseli}
\email{kesseli@strw.leidenuniv.nl}
\author[0000-0002-3239-5989]{Aurora Y. Kesseli}
\affiliation{Leiden Observatory, Leiden University, Postbus 9513, 2300 RA, Leiden, The Netherlands}

\author[0000-0003-1624-3667]{I.A.G. Snellen}
\affiliation{Leiden Observatory, Leiden University, Postbus 9513, 2300 RA, Leiden, The Netherlands}

\author[0000-0002-2891-8222]{N. Casasayas-Barris}
\affiliation{Leiden Observatory, Leiden University, Postbus 9513, 2300 RA, Leiden, The Netherlands}

\author[0000-0003-4096-7067]{P. Molli\`ere}
\affiliation{Max-Planck-Institut f\"ur Astronomie, K\"onigstuhl 17, D-69117 Heidelberg, Germany}

\author[0000-0002-0516-7956]{A. S\'{a}nchez-L\'{o}pez}
\affiliation{Leiden Observatory, Leiden University, Postbus 9513, 2300 RA, Leiden, The Netherlands}

\begin{abstract}

Ultra-hot Jupiters are gas giants that orbit so close to their host star that they are tidally locked, causing a permanent hot dayside and a cooler nightside. 
Signatures of their nonuniform atmospheres can be observed with high-resolution transit transmission spectroscopy by resolving time-dependent velocity shifts as the planet rotates and varying areas of the evening and morning terminator are probed. These asymmetric shifts were seen for the first time in iron absorption in WASP-76b.
Here, we search for other atoms/ions in the planet’s transmission spectrum and study the asymmetries in their signals. We detect Li I, Na I, Mg I, Ca II, V I, Cr I, Mn I, Fe I, Ni I, and Sr II, and tentatively detect H I, K I, and Co I, of which V, Cr, Ni, Sr II, and Co have not been reported before. We notably do not detect Ti or Al, even though these species should be readily observable, and hypothesize this could be due to condensation or cold trapping. We find that the observed signal asymmetries in the detected species can be explained in different ways. We find a relation between the expected condensation or ionization temperatures and the strength of the observed asymmetry, which could indicate rain-out or recombination on the nightside. However, we also find a dependence on the signal broadening, which could imply a two-zoned atmospheric model, in which the lower atmosphere is dominated by a day-to-night wind, while the upper atmosphere is dominated by a vertical wind or outflow. These observations provide a new level of modeling constraint and will aid our understanding of atmospheric dynamics in highly irradiated planets.

\vspace{25pt}

\end{abstract}

\section{Introduction}

Ultra-hot Jupiters (UHJs) are an exotic class of giant, highly irradiated exoplanet and have dayside temperatures greater than 2200 K \citep{Parmentier2018}. These planets are likely tidally locked and are characterized by clear atmospheres dominated by atomic and ionic features from dissociated molecules on the daysides, and cooler cloudy atmospheres on the nightsides. The large scale heights of UHJs make them ideal laboratories to study how tidally locked planets absorb and redistribute the intense radiation from their host stars.

High-resolution transit transmission spectroscopy has proven a powerful tool to probe the atmospheres of UHJs due to its ability to resolve many narrow-line absorbers (Fe, Ti, V, Mg, etc.) that are invisible at low-spectral resolution. This technique has been used to confirm the presence of a vast assortment of atoms and ions thus showing that most molecules and condensates are dissociated on the limbs of the UHJs. \citet{Hoeijmakers2018, Hoeijmakers2019} first explored the atomic/ionic composition, up to atomic number 78, of the hottest known exoplanet, KELT-9b ($T_{eq}$ = 4050 K; \citealt{Gaudi2017}), and uncovered signals from neutral Fe I, Mg I, and Na I, as well as ionized Sc II, Cr II, Y II, Fe II, and Ti II. Subsequent studies of WASP-121b ($T_{eq}$ = 2358 K; \citealt{Delrez2016}), MASCARA-2b ($T_{eq}$ = 2260 K; \citealt{Talens2018}), and WASP-76b ($T_{eq}$ = 2160 K; \citealt{West2016}) have extended this work to cooler UHJs. Recent papers have built upon each other to detect more new species and confirm previous detections of atoms and ions in the atmosphere of WASP-121b, and the list of detected species now includes Mg I, Na I, Ca I, Ca II, Cr I, Fe I, Fe II, Ni I, V I, H I, Li I, K I and Sc II \citep{Ben-Yami2020, Hoeijmakers2020, Borsa2021, Merritt2021}. In MASCARA-2b, neutral Fe I, Na I, and Mg I, as well as ionized Ca II, Fe II, and Cr II have been identified \citep{Casasayas2019, Stangret2020, Nugroho2020, Hoeijmakers2020M2}. \citet{Tabernero2021} detected absorption from neutral Li I, Na I, Mg I, Fe I, K I and Mn I, and ionized Ca II in the atmosphere of WASP-76b. Planets cooler than UHJs exhibit absorption from far fewer metals \citep{Allart2020, Casasayas2021}, seemingly because many elements have condensed into molecules, clouds, and hazes \citep[e.g.,][]{Kataria2016, Gao2021}. 

Recently, \citet{Ehrenreich2020} explored the transition region between the hot dissociated dayside and cool cloudy nightside of WASP-76b using the Echelle Spectrograph for Rocky Exoplanets and Stable Spectroscopic Observations (ESPRESSO) on the Very Large Telescope (VLT). They observed asymmetric absorption in the cross correlation signal from Fe I, where, at the end of the transit, the signal was blueshifted by $\sim$10 km s$^{-1}$ from its expected position, while at the beginning of the transit, the signal was at rest.
This same asymmetry was confirmed in \citet{Kesseli2021} using four HARPS transits and a different analysis method. As WASP-76b transits in front of its host star, its viewing angle changes by $\phi=30\arcdeg$ \citep{Ehrenreich2020} and assuming tidal locking, more of the illuminated morningside is visible at the start of the transit, while more of its illuminated eveningside is visible at the end of the transit. Due to the interplay between the planet's rotation and the measured global wind blowing from the hot dayside to the cool nightside \citep{Seidel2021}, the eveningside of the planet is expected to have a larger blueshift than the morningside, but this velocity change would only be able to be resolved if the atmospheric temperature profile or abundance of Fe I were not uniform over the surface of the planet. Recent in-depth global circulation modeling (GCM) by \citet{Wardenier2021} and \citet{Savel2021} confirmed that a hotter evening terminator, a cooler morning terminator, and either condensation of Fe I or clouds obscuring the cooler morningside were necessary to explain the asymmetry.

By detecting more of these asymmetric transit signals, we can directly measure abundances of individual elements across the surface of exoplanets, and hence indirectly infer which cloud species are forming. \citet{Lothringer2020} suggested a similar method to probe condensate formation and rainout by using low-resolution near-UV spectral indices, as most metals absorb strongly at these wavelengths. Furthermore, the high-spectral resolution technique discussed here is complementary to proposed techniques for resolving the different temperature structures and low-resolution spectra of the morning and evening terminators with \textit{JWST} \citep{Espinoza2021}. By combining information from both low-resolution and high-resolution techniques, we can extract information on the dynamics and velocity structure, temperatures, and abundances of a range of species in the morning and evening terminators separately.

In this paper, we aim to search for asymmetric transit signals within the full inventory of atoms and ions that are accessible at the visible wavelengths covered by ESPRESSO, using the same dataset that was used for the original Fe condensation result \citep{Ehrenreich2020}. In Section \ref{s:atoms}, we will assess which atoms and ions contribute significant opacity at visible wavelengths and the temperatures seen in UHJs. We will then use these findings to search for all observable species in the ESPRESSO transit observations of WASP-76b. We discuss the data and our pre-processing steps in Section \ref{s:data}. We then explain how we made the models of each species in Section \ref{s:model} and the subsequent cross correlation analysis in Section \ref{s:cc}. We present the results of the cross correlation search in Section \ref{s:results}, and discuss which of the detected species exhibit this asymmetric signal. We discuss the implication and what physical information about the exoplanet atmosphere can be extrapolated in Section \ref{s:discussion}. Finally, we conclude in Section~\ref{s:conclusions}.

\section{Observability of Atomic Absorption} 
\label{s:atoms}

\begin{figure*}
\begin{center}
\includegraphics[width= 0.8\linewidth]{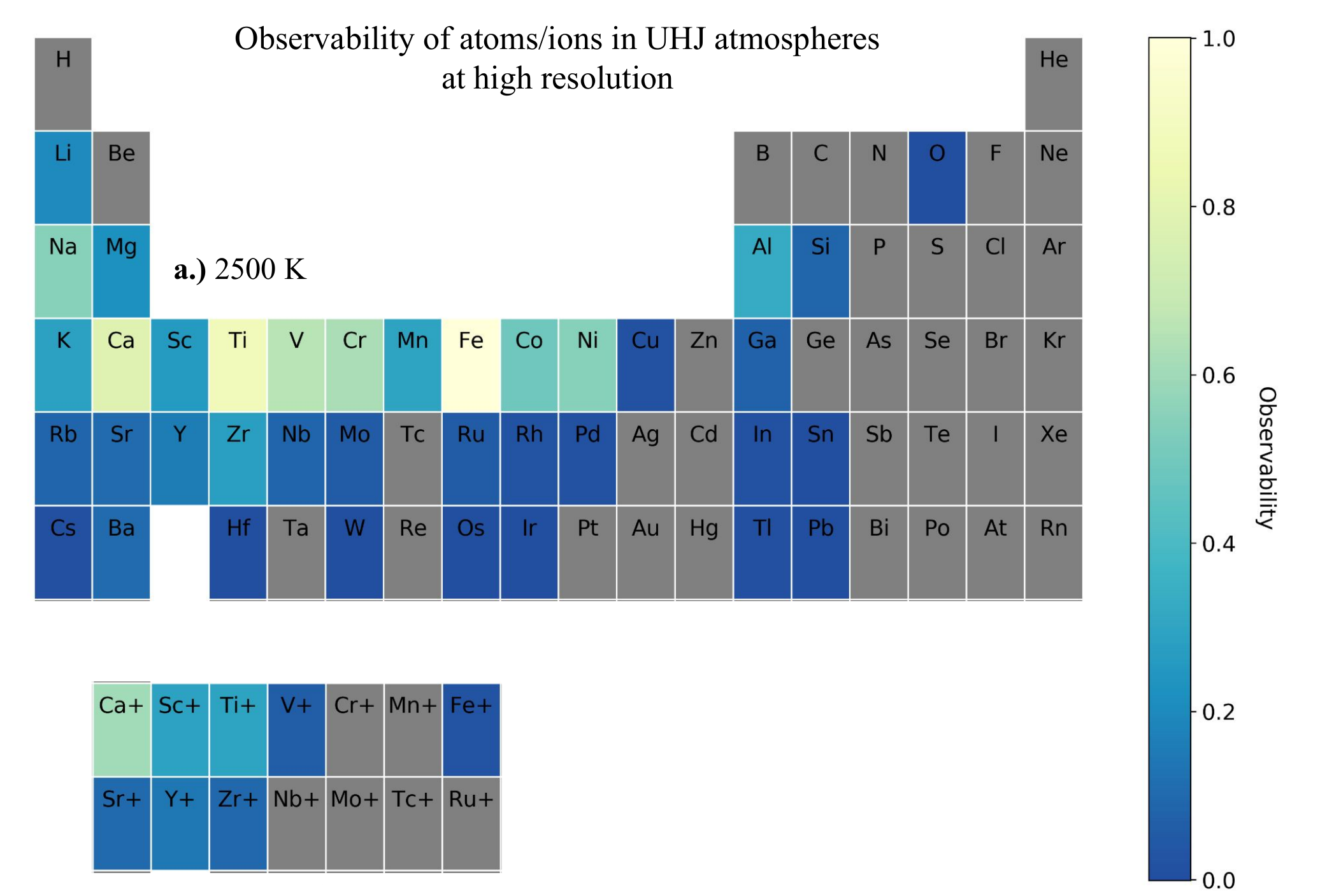}
\includegraphics[width= 0.8\linewidth]{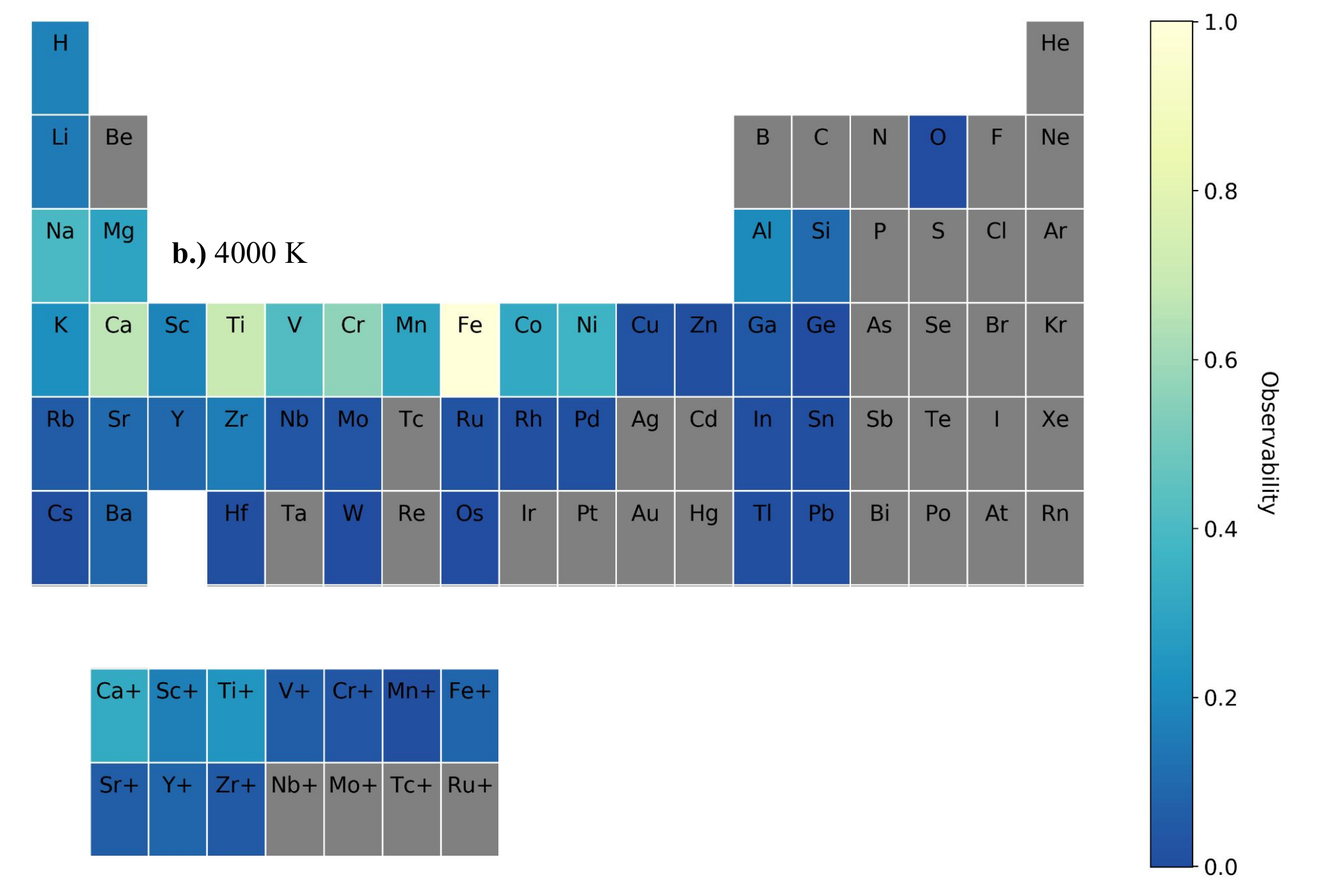}
\caption{\small 
Results of our test on the observability for each element and select ions that have absorption lines at optical wavelengths for \textbf{a.)} a temperature of 2500 K and \textbf{b.)} 4000 K. We only show a portion of the periodic table for the ions (bottom of both panels \textbf{a.)} and \textbf{b.)}) as these ions are the only ones that have significant absorption lines in the optical. Neutral atoms and ions that do not have any absorption lines above the expected continuum, set by the opacity due to collision-induced absorption and H$^{-}$, are shown in gray. Elements and ions with higher observability have deeper and more numerous absorption lines and are abundant enough that these lines significantly contribute to the overall opacity. For both cases, transition metals, alkali metals, and alkaline metals are particularly observable.  }
\label{f:periodic}
\end{center}
\end{figure*}

Since our goal is to search for asymmetric transit signals in the cross correlation maps of each element, we began by exploring which elements would produce detectable signals so as to avoid the unnecessary task of performing the full cross correlation analysis for elements that would not be detectable. Elements with a low observability score should either have no absorption lines in the visible portion of the spectrum or are expected to be so under-abundant that even if they do present features, the signal would be smaller than that of the base noise level or continuum opacity sources. A similar analysis that also aimed to determine which elements and ions would be detectable at high spectral resolution was performed in \citet{Hoeijmakers2019}.

To determine an observability score for each element, we used atomic opacities from the DACE Opacity database\footnote{\url{https://dace.unige.ch/opacityDatabase/?}}. These opacities were generated with the open-source HELIOS-K opacity calculator \citep{Grimm2015, Grimm2021} using input from the Kurucz \citep{Kurucz2018}, NIST \citep{Kramida2019}, and VALD3 \citep{vald3} spectroscopic databases. Each database contains opacities for atoms and ions at temperatures ranging from 2500 to 6100 K and a single pressure of 10$^{-8}$ bars. We chose to use the opacities generated using NIST because it was the only database that contained opacities for all of the atoms and ions (both the Kurucz and VALD3 databases did not contain opacity data for Br, Kr, and I). We downloaded data for two temperatures, 2500 and 4000 K, which spans the expected temperature range of UHJs. \citet{Hoeijmakers2019} found that the average altitude where most of the detected absorption lines formed in the UHJ Kelt-9b was between $\sim10^{-4}$ and $10^{-6}$ bars, and that the cores of the strongest absorption lines could be produced at even higher altitudes. While 10$^{-8}$ bars is higher in the atmosphere than the expected formation location, it should give a reasonable estimate for each element as pressure broadening is not expected to be too important at these high altitudes and low pressures. 

We choose not to implement chemical modeling and instead simply scaled each element by its solar abundance from \citet{Asplund2009}, which will give us a useful upper limit on detectability of each species. Chemistry in hot Jupiters is complicated, and correctly modeling these processes would require a much more in-depth atmospheric model including full temperature pressure profiles at a range of longitudes, extensive chemical reactions and condensation information, and assumptions about equilibrium versus nonequilibrium processes, which is beyond the scope of this paper. This type of in-depth chemical study has been performed previously on other UHJs \citep[e.g.,][]{Kataria2016, Gao2021}, and in later sections, we will refer to these results to help interpret any deviations that arise between the data and the predictions presented here. To estimate the abundances, we scaled the opacities by each element's expected mass density given solar abundances from \citet{Asplund2009}. Tc, Po, At, and Rn are not included in \citet{Asplund2009} and so we did not include them in our calculations (shown in gray in Figure \ref{f:periodic}). We used the values derived from the solar photosphere when available, except for Li. We used the Li abundance derived from solar system meteorites, which is also given in \citet{Asplund2009}, since photospheric abundance of Li drastically changes with age and the current photospheric abundance does not represent the initial value that would have been available in the disk during planet formation \citep{Thevenin2017}. 

We also added a continuum opacity, below which each element's opacity would no longer contribute. We used petitRADTRANS \citep{Molliere2019} to find the expected levels of continuum opacity from H$_2-$H$_2$ and H$_2 - $ He collisions, as well as H$^{-}$ at 3 mbar, which is where the continuum opacity is expected to begin to become optically thick in hot Jupiter atmospheres \citep{Hoeijmakers2019}. This resulted in a base opacity near 10$^{-3}$ cm$^2$ g$^{-1}$. We did not include Rayleigh scattering as a source of continuum opacity, as \citet{Fu2020} found that Rayleigh scattering was not consistent with the observed slope of the Hubble Space Telescope (HST) transmission spectra of WASP-76b, which may be too hot to host significant aerosols. We experimented with slightly different base opacity levels, but found that this did not significantly change our results as most of the signal comes from the strong absorption lines that are well above the continuum.

We calculated an observability score for each species, $x$, following similar methods as \citet{Molliere2019isotope}, and given by: 

\begin{equation}
    Observability_x =  \int_{3500 \AA}^{8000 \AA} ln(\tau _{all}) - \int_{3500 \AA}^{8000 \AA} ln(\tau _{all - x})
\end{equation}

\noindent where $\tau _{all}$ is the combined scaled opacities of all of the species and the continuum opacity, and $\tau _{all-x}$ is the same combined opacities except without the contribution from the relevant species, $x$. These scores were scaled so the most observable species was normalized to 1. 
The wavelength region of 3500 to 8000 \AA\ corresponds to the ESPRESSO wavelength coverage (as well as similar wavelength coverage of other high-resolution spectrographs like HARPS and EXPRES, barring the reddest 1000 \AA). Figure \ref{f:periodic} shows the results of this test. As expected, we find that many of the elements that are the most observable are those that are readily detected in hot-Jupiter atmospheres, such as Fe, Na, and Ca. 

We note that atoms with few, but strong, absorption lines (e.g., H, Li, Na, Mg, Al and K) are shown to have a lower observability than atoms with many lines (e.g., Fe, Ti, V) because our method sums line strengths over a wide wavelength region. However, by focusing only on the region around these single strong lines they can be readily observed.

\section{The Data}
\label{s:data}

\begin{center}
\begin{table}[ht]
\centering
\caption{WASP-76 system parameters from \citet{Ehrenreich2020}} 
\begin{tabular}{l l}
\hline
Parameter & Value \\ 
\hline
\textbf{Star} & \\
\hline
$M_*$ ($M_{Sun}$) & $1.458\pm0.021$ \\
$R_*$ ($R_{Sun}$) & $1.756\pm0.071$ \\
$T_\mathrm{eff}$ (star; K) & $6329\pm65$ \\
$v_{sys}$ (km s$^{-1}$) & -1.16 \\
$v \sin i$ (km s$^{-1}$) & $1.48\pm0.28$ \\
$K_*$ (m s$^{-1}$) & $116.02 \substack{+1.29 \\-1.35}$ \\
\hline
\textbf{Planet} & \\
\hline
$M_{p}$ ($M_{J}$) & $0.894\substack{+0.014 \\-0.013}$ \\
$R_{p}$ ($R_{J}$) & $1.854\substack{+0.077\\-0.076}$  \\
$T_{eq}$ (K) & 2228$\pm$122 \\
$P_{orb}$ (d) & $1.80988198\substack{+0.00000064\\-0.00000056}$ \\
$T_0$ (d) & $2458080.626165\substack{+0.000418\\-0.000367}$ \\
$a$ (au) & 0.0330$\pm$0.0002 \\
$K_p$ (km s$^{-1}$) & $196.52\pm0.94$ \\
$i$ (deg) & 89.623$\substack{+0.005\\-0.034}$ \\
\hline
\end{tabular}
\label{t:WASP76}
\end{table}
\end{center}

We now turn to the ESPRESSO dataset of WASP-76b to search for all of the elements that we have deemed observable and to search for asymmetries in the transits of each. This is the ideal dataset for our study as this same dataset led to the first detection of an asymmetry in the absorption signal of Fe \citep{Ehrenreich2020}. The combination of ESPRESSO's stability and high spectral resolution (R$\sim$138,000; \citealt{Pepe2010}) with the collecting area of the European Southern Observatory's VLT (8m) is unprecedented, and this type of study is only possible because of the unique combination.

WASP-76b is a well-studied UHJ that orbits a bright (V=9.5) F7 star (see Table \ref{t:WASP76} for the system parameters). Two transits of WASP-76b were observed on 2018 September 2 (night 1) and 2018 October 30 (night 2). 35 spectra were recorded with an exposure time of 600s on the night 1, while 70 spectra were recorded with an exposure time of 300s on night 2. These spectra cover about 30 minutes before the transit, the full transit duration, and about 30 minutes after the transit. The data were reduced with version 1.3.2 of the ESPRESSO pipeline to produce 1D blaze-corrected, stitched spectra. The given wavelength solutions have already been corrected for the barycentric velocity. More information about the reduction procedure and the data can be found in the original publication where the data were presented \citep{Ehrenreich2020}. This dataset has also been used in two subsequent papers, \citet{Tabernero2021} and \citet{Seidel2021}. \citet{Tabernero2021} identified individual absorption lines from Li I, Na I, Mg I, Ca II, Mn I, K I, and Fe I, and also searched for Fe I, Ti I, Cr I, Ni I, TiO, VO, and ZrO using the cross correlation technique, but only detected the previously seen Fe I signal. \citet{Seidel2021} performed a concentrated analysis of the Na I doublet and found evidence for a uniform day-to-night wind, as well as a vertical wind in the upper atmosphere, which acts to significantly broaden the Na I doublet.

Using the 1D spectra from the pipeline, we performed a series of steps to ready the spectra for cross correlation.

\subsection{Telluric Correction}
Before any other steps were performed, we corrected the spectra for any telluric contamination due to H$_2$O and O$_2$ in the Earth's atmosphere using \texttt{molecfit} \citep{Smette2015}. This method was first used by \citet{Allart2017}, and has subsequently been used in many studies including those on ESPRESSO data \citep[e.g.,][]{Tabernero2021, Casasayas2021}. This method is successful at removing weak H$_2$O lines, but fails at removing the deepest absorption lines, such as those seen in the line cores of the O$_2$ bands, and so even after the \texttt{molecfit} correction, we masked any pixels where the transmission through the atmosphere was less than 40\%. A threshold of 40\% is standard in other cross correlation analyses \citep[e.g.,][]{SanchezLopez2019} and nicely masks the deep line cores that are not corrected well. 

\subsection{Spectral cleaning}
After telluric correction, we performed a series of steps to clean the stellar spectra so that they would be uniform in time and minimal stellar residuals would remain after their removal at a later point in the analysis. We started by correcting for cosmic rays and performed a sigma clipping, interpolating over any pixels that were more than 5$\sigma$ different than the other pixels in that same pixel channel.
To correct for the variations in the velocity of the star over time, the telluric-corrected spectra were shifted into the host star's rest frame by correcting each spectrum for the reflex motion of the star due to the exoplanet ($K_*$) and the system velocity ($v_{sys}$). After this step, all of the spectra were interpolated onto a uniform wavelength grid, masking the few missing values at the very ends of the spectra caused by the shifts in velocity, to make a uniform 440350 (wavelength pixels in each spectrum) by N (number of spectra taken over the course of the night) grid in time for each night.  

We then corrected for any changes in flux over time due to changing airmass or signal-to-noise by simply normalizing all of the spectra. While this corrects for large-scale changes in the flux over time, small differences in the shape of the spectrum due to variations in the blaze function are still present and will dominate the residuals if they are not corrected. To remove these differences while keeping the overall shape of the spectrum, we followed \citet{Merritt2020} 
to place all of the spectra on a ``common" blaze function. This process entailed dividing each spectrum by the average stellar spectrum in time, and then applying a Gaussian filter with a width of 200 pixels ($\sim$100 km s$^{-1}$). Each original spectrum is then divided by the resulting filtered and continuum subtracted spectrum. Any low-frequency noise is removed, while the higher-frequency signal from the exoplanet is preserved. This step also removes the sinusoidal noise pattern that has been seen in ESPRESSO data \citep{Tabernero2021, Borsa2021}, which was found to have a period of 30$-$40 \AA ($> 1000$ km s$^{-1}$). 

Finally, we completely removed any wavelength columns that that had large standard deviations due to hot pixels, imperfect telluric removal, or low signal-to-noise ratios. We found that by removing about 2\% of all the columns we could decrease the noise while still preserving the signal. At this point the host star's spectrum should be completely uniform over the course of the observations, and the only differences that arise should be due to the transit of the exoplanet. 

\section{Atmospheric Transmission Models}
\label{s:model}

\begin{figure*}
\begin{center}
\includegraphics[width=0.75\linewidth]{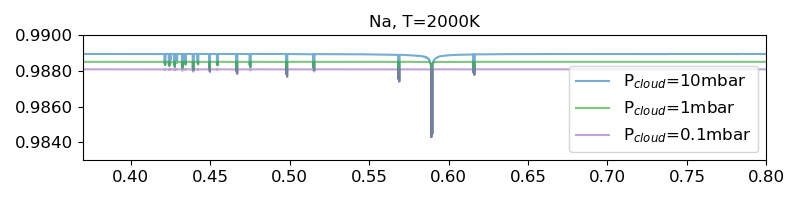}
\includegraphics[width=0.75\linewidth]{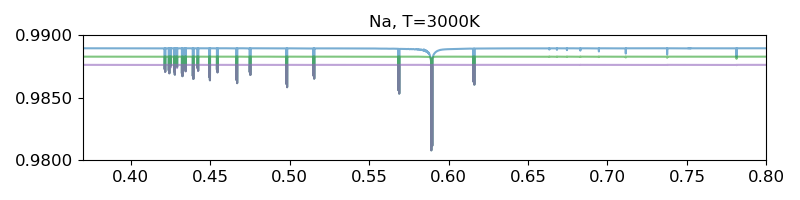}
\includegraphics[width=0.75\linewidth]{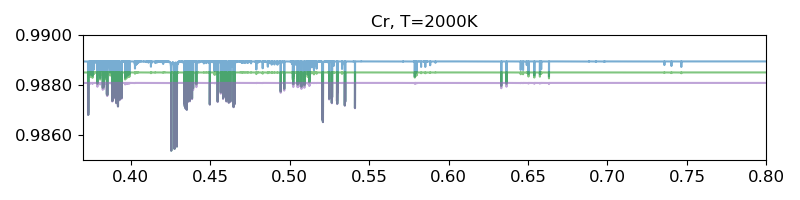}
\includegraphics[width=0.75\linewidth]{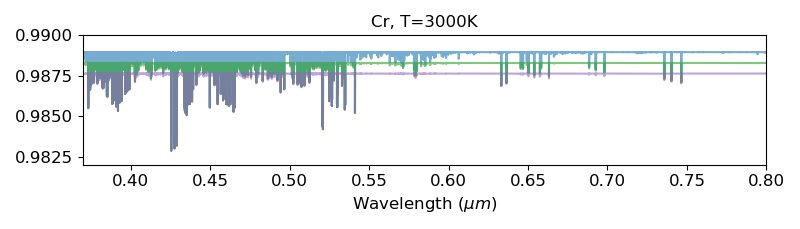}
\caption{\small Example model atmospheres showing the relative transit depth vs. wavelength for Na and Cr in WASP-76b. We show how changing the two free parameters, temperature and pressure at the opacity base due to continuum opacity sources (P$_{cloud}$), change the relative absorption of each species. At higher temperatures, more absorption lines are present, and the scale height of the atmosphere is increased. When the base opacity is moved higher in the atmosphere, the weaker lines and line wings are hidden. Moving the base opacity higher in the atmosphere has the same effect as decreasing the abundance. The results we present for each species use a temperature of 3000 K and a base opacity at 1 mbar (unless stated otherwise), but in Section \ref{s:otherparams} we discuss how different models might affect our results.   }
\label{f:models}
\end{center}
\end{figure*}

We created a separate transmission model for each observable atom and ion using petitRADTRANS \citep{Molliere2019}. petitRADTRANS calculates transmission spectra of exoplanet atmospheres at either low- or high-spectral resolution, given a pressure-temperature (PT) profile, the exoplanet's radius and surface gravity, the mean molecular weight of the atmosphere, and the abundances of the requested atoms or molecules. The code has successfully been used on many previous high-resolution transit transmission observations \citep[e.g.,][]{Kesseli2020, Kesseli2021, Landman2021}. 

We input the planetary radius and surface gravity for WASP-76b calculated from the parameters in Table \ref{t:WASP76}. We chose to use an isothermal PT profile with a temperature of 3000 K. An isothermal PT profile is often used for transmission spectroscopy as this type of spectrum is not very dependent on the underlying PT profile. Indeed, \citet{Seidel2021} found that an isothermal profile fit this ESPRESSO data well and was preferred over the more complex profiles that they tested. A temperature of 3000 K was motivated as an intermediate choice between \citet{Landman2021}, who retrieved a best-fit temperature of 2701 $\substack{+448\\-279}$, and \citet{Seidel2021}, who found a best-fit temperature of 3389 $\pm$ 227 K. Figure \ref{f:models} shows how changing the temperature changes the resulting model spectrum, and in Section \ref{s:otherparams} we test how our results change if we use isothermal PT profiles of 2000 and 4000 K.

As has been discussed in many recent papers, the abundance and the pressure below which the atmosphere cannot be probed are degenerate because we are only sensitive to the line strength above the continuum \citep{Welbanks2019, Seidel2021, Merritt2021}. We therefore chose to use solar abundance as our input for each species, and nominally set a uniform gray cloud deck (P$_{cloud}$) at 1 mbar. This gray cloud deck serves to mimic continuum opacity from a variety of sources, including collisionally induced opacity, H$^-$ opacity, and clouds. \citet{Hoeijmakers2019} found that at the temperatures seen in UHJs, the atmosphere generally becomes opaque around millibar pressures due to H$^-$, which motivated our selection of this gray atmospheric base.  For the majority of the species, the neutral atomic abundance is expected to be similar to the solar abundance of that species because a temperature of 3000 K is too cool to ionize most atoms, but too warm for significant condensation. In Section \ref{s:otherparams} we explore how using different pressure cloud decks can alter our results.

petitRADTRANS has pre-computed opacity files for a range of neutral and ionized atoms, as well as many molecules. Not every atom that we wanted to test was available. For those, we utilized the same opacity data from DACE that we used in Section \ref{s:atoms}. These files can be used to create the necessary input files to run petitRADTRANS. While the database contains opacities calculated for a large range of temperatures, unfortunately, it only has atomic data at a single pressure (10$^{-8}$ bar). Therefore, the pressure broadening profiles will not be exactly correct for those species that we added ourselves into petitRADTRANS. We tested the extent to which this simplification affected our results by creating our own input files using DACE opacities for Cr I and Na I, and compared them to the models made with the pre-computed petitRADTRANS files. We found that the vast majority of features had no change, but as expected, the wings of the strongest three absorption features of Cr I and the Na I D doublet had slight deviations of a few percent the height of the feature, but no noticeable changes to the resulting cross correlation grids. Therefore we do not expect any significant difference in signals between the petitRADTRANS pre-computed opacities with those we computed from the DACE opacities. 

Before we used the models in the cross correlation analysis, we reduced the spectral resolution to match that of the ESPRESSO instrument by convolving each model with a Gaussian kernel with a FWHM of 2.2 km s$^{-1}$. Since the asymmetric transit signature of Fe I is best modeled as iron condensing out of the atmosphere on the morningside, we do not rotationally broaden the model spectra because we do not expect the atoms to be uniformly distributed across the limbs of the planet and thus create a net broadening effect. Examples of Cr I and Na I models that have been convolved to ESPRESSO resolution and are ready for cross correlation are shown in Figure \ref{f:models}.

\section{Cross Correlation Analysis} 
\label{s:cc}

\begin{figure*}
\begin{center}
\includegraphics[width= 0.48\linewidth]{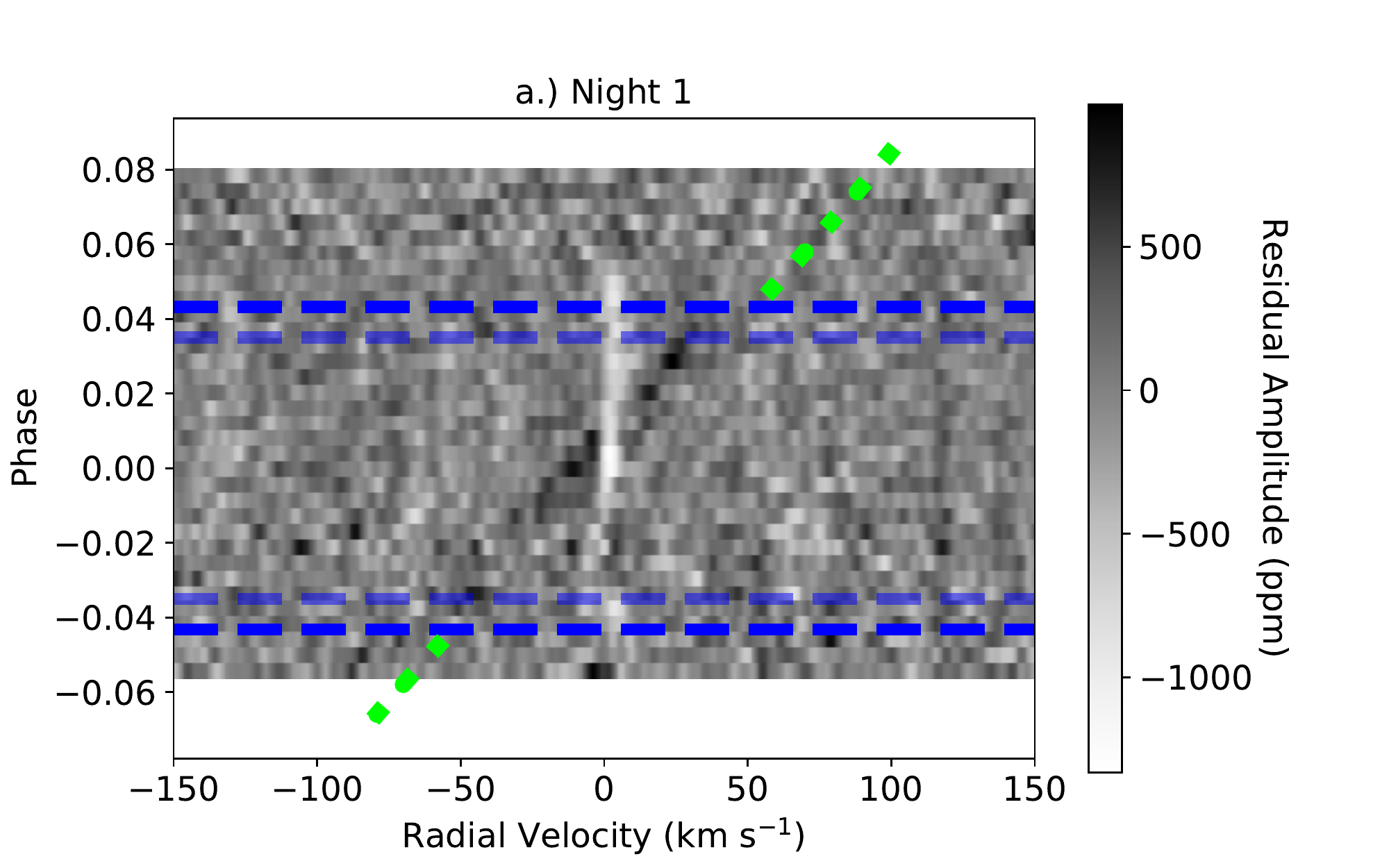}
\includegraphics[width= 0.48\linewidth]{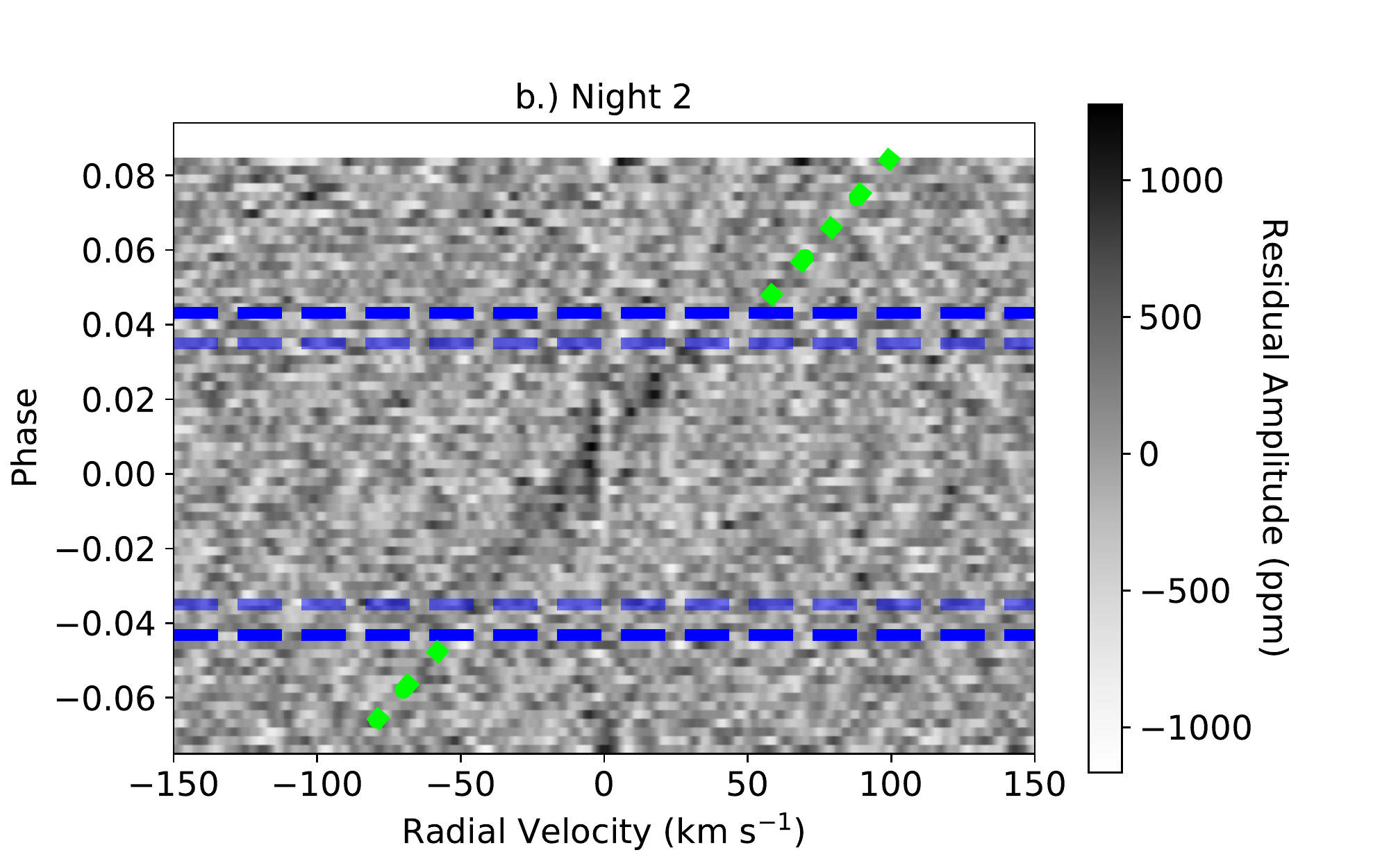}
\includegraphics[width= 0.48\linewidth]{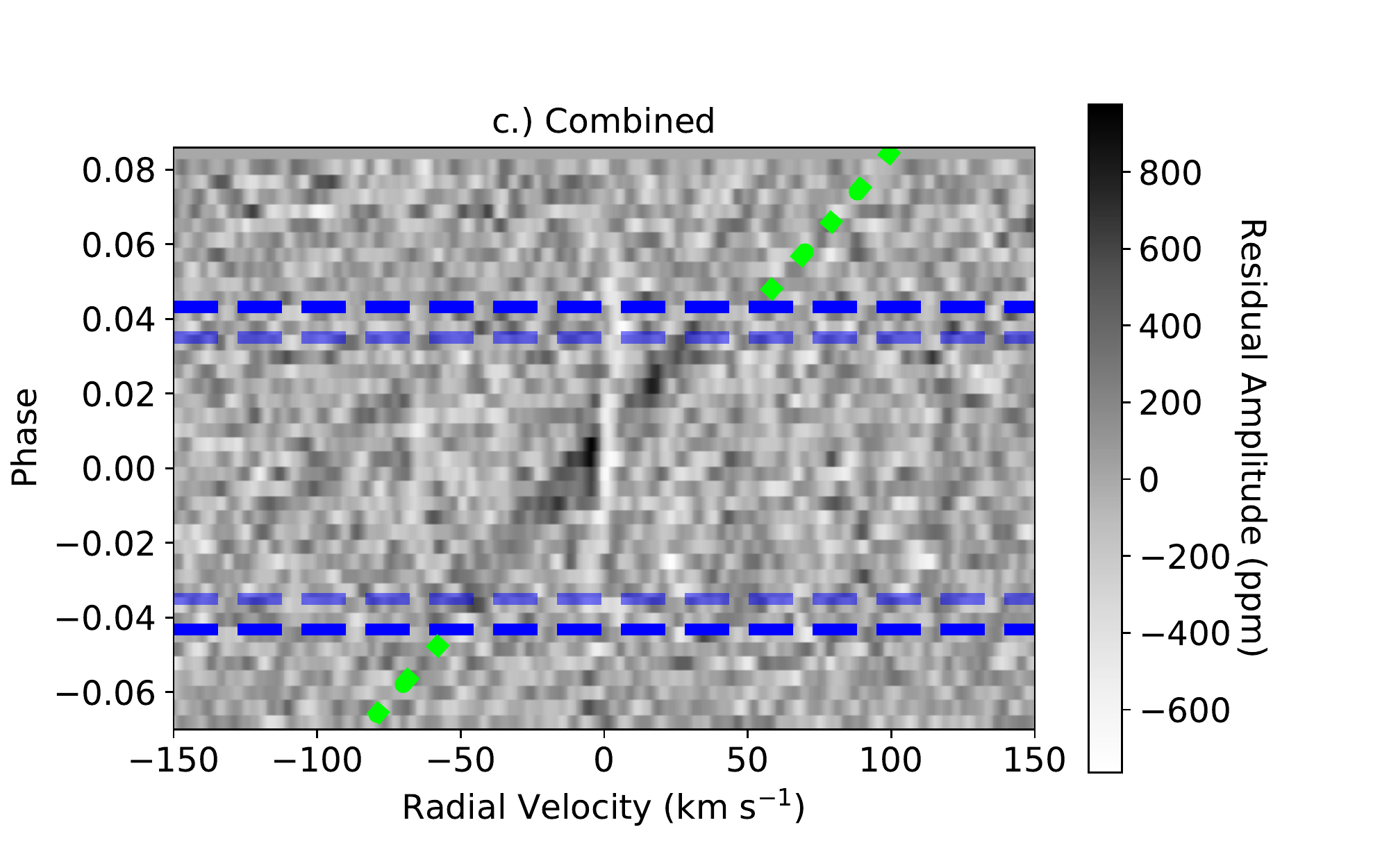}
\includegraphics[width= 0.48\linewidth]{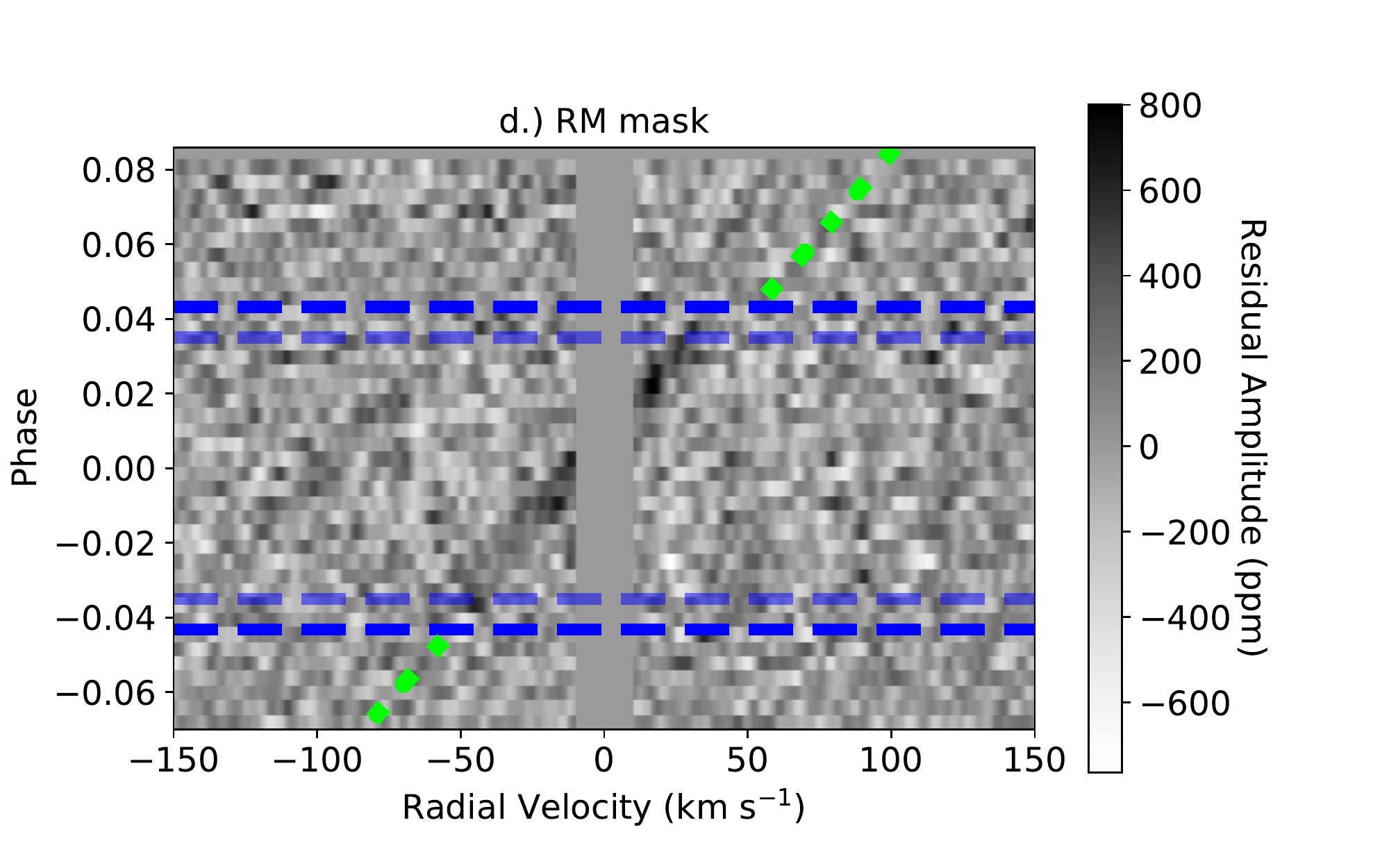}
\caption{\small 
Example showing how we combined the two nights and the resulting cross correlation grids for Cr I. \textbf{(a.)} The resulting cross correlation grid from night 1. Each row in the 2D cross correlation grid is a separate cross correlation function in time between an ESPRESSO spectrum and the atmospheric transition model for Cr, divided by the average out-of-transit cross correlation function. The four blue horizontal lines from bottom to top show the start of ingress, the end of ingress, the start of egress and the end of egress, respectively. The green dashed line shows the expected exoplanet's velocity. Black residuals near the exoplanet's expected velocity are due to excess absorption from Cr in the exoplanet's atmosphere. Any velocity offset from the planet's expected position is due to wind and dynamics in the planet's atmosphere and will be quantified in Section \ref{s:results}. White residuals near 0 km s$^{-1}$ are stellar residuals due to a combination of the Rossiter-McLaughlin effect, center-to-limb variations, and low flux within the stellar Cr line cores which causes increased noise near 0 km s$^{-1}$. \textbf{(b.)} Same as (a.), but for night 2. \textbf{(c.)} The two nights have been combined by interpolating each onto a uniform grid in phase space and averaging them together, weighting them by their average signal-to-noise ratio and the number of spectra taken during the transit. \textbf{d.)} The central region between $-10$ and $+10$ km s$^{-1}$ has been masked as a simple method to remove all of the pixels affected by the stellar residuals. }
\label{f:cc_example}
\end{center}
\end{figure*}

To determine which species are present in the atmosphere of WASP-76b, we cross correlated each cleaned spectrum from night 1 and night 2 with the $>$ 40 individual models of each neutral atom or ion. Cross correlation has proven extremely successful at uncovering the very small signals present in exoplanet atmospheres (on the level of $\sim$10$-$1000 parts per million) because it effectively combines the weak signals from tens to hundreds of individual absorption lines \citep[e.g.,][]{Snellen2010}. For each individual neutral atom or ion, we followed similar steps as those laid out in \citet{Hoeijmakers2020} and \citet{Kesseli2021} to perform the cross correlation analysis, using the following equation: 

\begin{equation} 
\label{e:1}
c(\nu, t) = \sum_{i=0}^{M} - x_i(t) T_i(\nu). 
\end{equation}

\noindent Here, $c(\nu, t)$ is the 2D cross correlation grid at each radial velocity ($\nu$) and observation time ($t$). $x_i(t)$ is the observed spectrum at each time, and $T_i(\nu)$ is the model of each element, shifted to every radial velocity. The radial velocities range from $-300$ to $+300$ km s$^{-1}$ in 0.5 km s$^{-1}$ steps. The models are normalized such that $\sum_{i=0}^{M} T_i(\nu) = 1$. The negative sign is included so that excess absorption from the exoplanet will appear as a positive residual, as is standard for cross correlation. In the end, the cross correlation function acts as a weighted sum for each spectrum, and pixels where the absorption is strong in the model ($T_i(\nu)$) are weighted much higher than pixels where there is little or no absorption from the relevant atom, creating an excess signal when the model is aligned with absorption lines in the planet.

The spectra used in the cross correlation still contained the contribution from the host star, so the cross correlation grids are dominated by any signal from it. To remove this signal, we averaged together all of the cross correlation functions that did not occur as the planet was transiting, and then divided each cross correlation function by this average out-of-transit function. The final cross correlation grid for each night is multiplied by $10^6$ to convert the residual cross correlations into parts per million (ppm). The resulting residual cross correlation grid for each night of observations using the Cr I model is shown in panels (a.) and (b.) of Figure \ref{f:cc_example}. 

We chose to report the cross correlation values in ppm, as in \citet{Tabernero2021}. The residual amplitude values from the cross correlation functions reported here are simply a weighted average of the transmission depths from the residual spectrum within all of the lines of the relevant atom (Equation \ref{e:1}). Larger residual amplitudes correspond to deeper lines in the planet's spectrum and larger transit depths. Therefore, the reported cross correlation amplitudes give a general idea of where in the atmosphere the majority of the signal arises, and larger residual amplitudes mean the signal was produced on average higher in the atmosphere than smaller residual amplitudes. We do note, however, that unlike the residual amplitudes reported for single lines, these residual amplitudes are model dependent as the model ($T_i(\nu)$) acts as the weights and does not directly correlate to the formation location of any individual line or lines. To ensure that all of the trends we measured are model independent, we tested models with different temperatures, abundances, and continuum opacities, which is discussed in Section \ref{s:discussion}.  

We then combined the two nights together into a single 2D cross correlation grid to maximize the signal-to-noise ratio (SNR). We interpolated each cross correlation grid onto a new grid that was uniformly spaced in phases with a step size of 0.004. We chose this step size as it approximately corresponded with the largest phase step in both nights and so none of the data will be extrapolated at a finer spacing than was observed. After both nights were interpolated onto the same grid, they were averaged together. Night 2 had finer temporal spacing, and so to account for this, we weighted the nights by the number of spectra that were taken during the transit on each night. We also weighted the nights by the average SNR of the stellar spectra over the full course of observations. The combined cross correlation grid for the same example with Cr is shown in panel (c.) of Figure \ref{f:cc_example}.

\subsection{Removal of the Rossiter-McLaughlin Effect and Other Stellar Residuals}

As the planet transits in front of the host star it blocks part of the disk of the star. Because the stellar disk is not uniform in brightness (center-to-limb effect; CLV) and is composed of different radial velocity components due to the stellar rotation (Rossiter-McLaughlin effect; RM), as the planet blocks different parts of the star, the overall stellar line shape will change throughout the transit. By removing the out-of-transit average that does not contain these effects, stellar residuals are left in the cross correlation grid. These two effects will be present in all of the cross correlation grids where the atom we are searching for is also present in the host star.  
Out of all of the elements we tested, we detected each of them in the spectrum of the host star except for Li and some of the more rare species that are expected to be mostly ionized at the stellar effective temperature including Sc, Rb, Sr, Y, Cs, and Zr. Furthermore, when the atom is present in the host star, the flux in the cores of the spectral lines are reduced and the SNR of the residual spectrum in these locations is decreased leading to excess noise in the residual cross correlation grid around 0 km s$^{-1}$ (see panels (a.) and (b.) of Figure \ref{f:cc_example}). 

Fortunately in the case of WASP-76, all of these effects are confined to velocities around 0 km s$^{-1}$ due to the orientation of the system and the slow rotation of the host star \citep[see][]{Ehrenreich2020}. Even so, panel (c.) of Figure \ref{f:cc_example} clearly shows stellar residuals (white pixels around 0 km s$^{-1}$) in the expected pattern given by the combination of the RM and CLV variations. 

Instead of modeling this effect and attempting to completely remove it as in other studies \citep{Yan2018, Casasayas2019, Kesseli2021}, we chose to simply mask out the pixels in the center 20 km s$^{-1}$. As discussed in \citet{Seidel2021} and \citet{Casasayas2021}, different stellar models and methods for removing the RM and CLV effects introduce systematic errors and uncertainties, and so \citet{Seidel2021} also chooses to mask out the affected pixels. Because we are mostly interested in resolving asymmetries between the beginning and the end of the transit for each species, masking the signal around 0 km s$^{-1}$ will not affect our results. This masking can be seen in panel (d.) of Figure \ref{f:cc_example}, which successfully removed all of the stellar residuals.

\section{Results} 
\label{s:results}

\begin{figure*}
\begin{center}
\includegraphics[width=\linewidth]{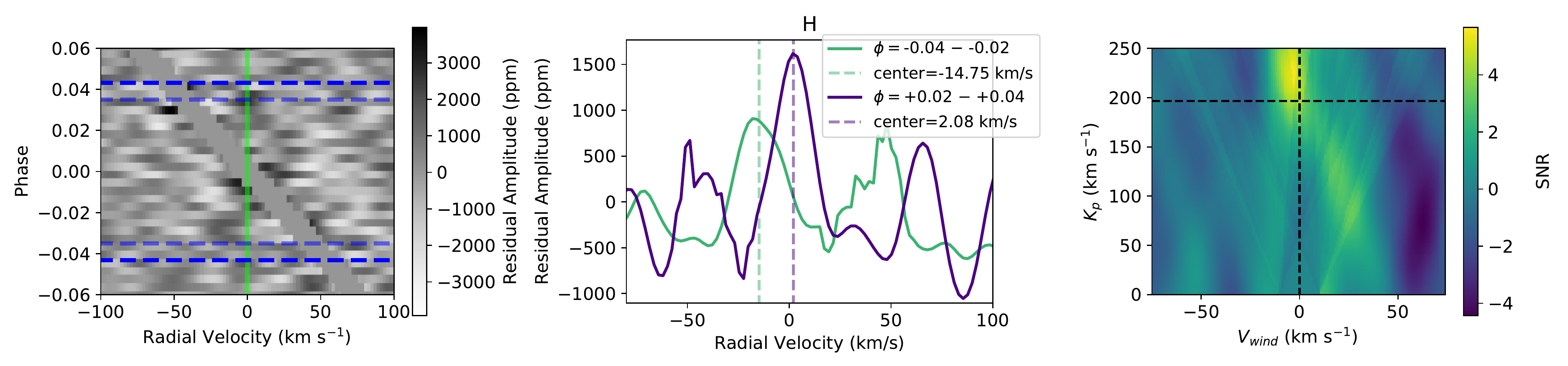}
\includegraphics[width=\linewidth]{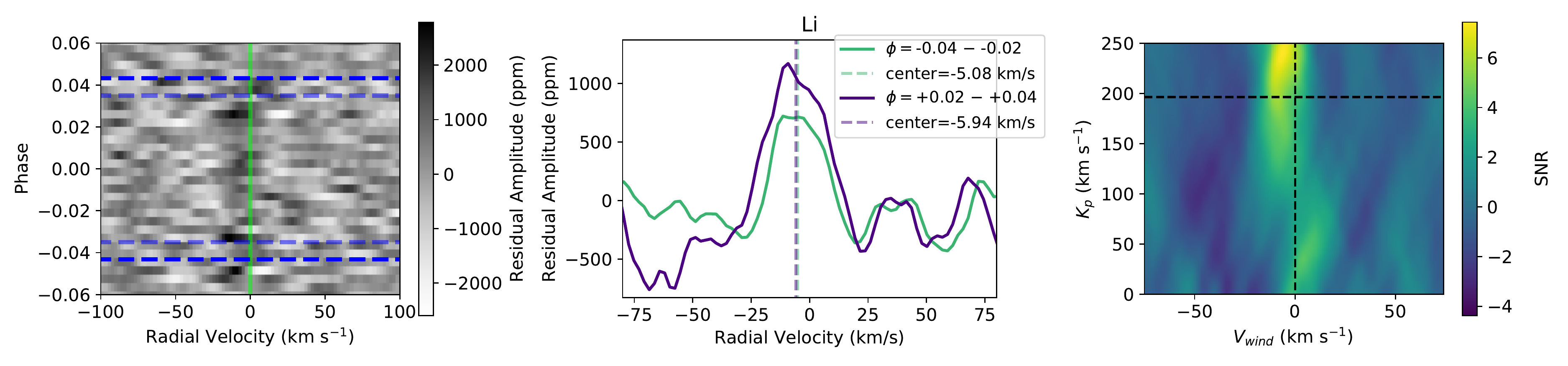}
\includegraphics[width=\linewidth]{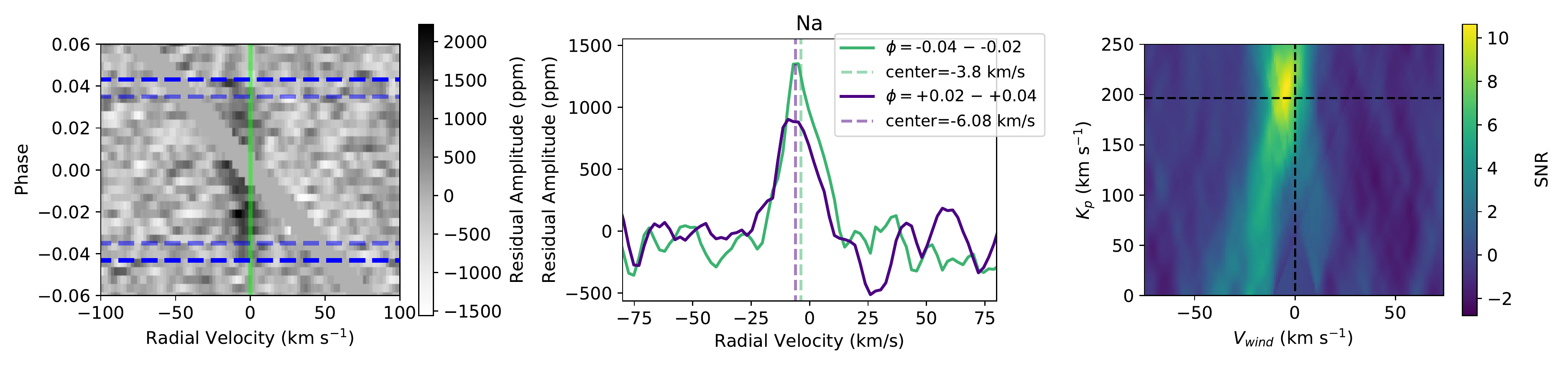}
\includegraphics[width=\linewidth]{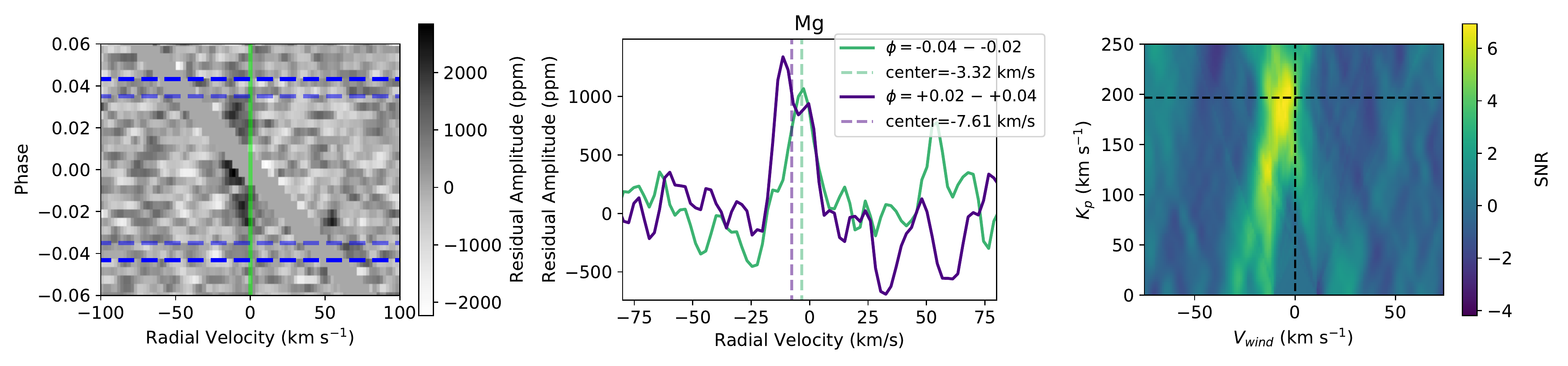}
\includegraphics[width=\linewidth]{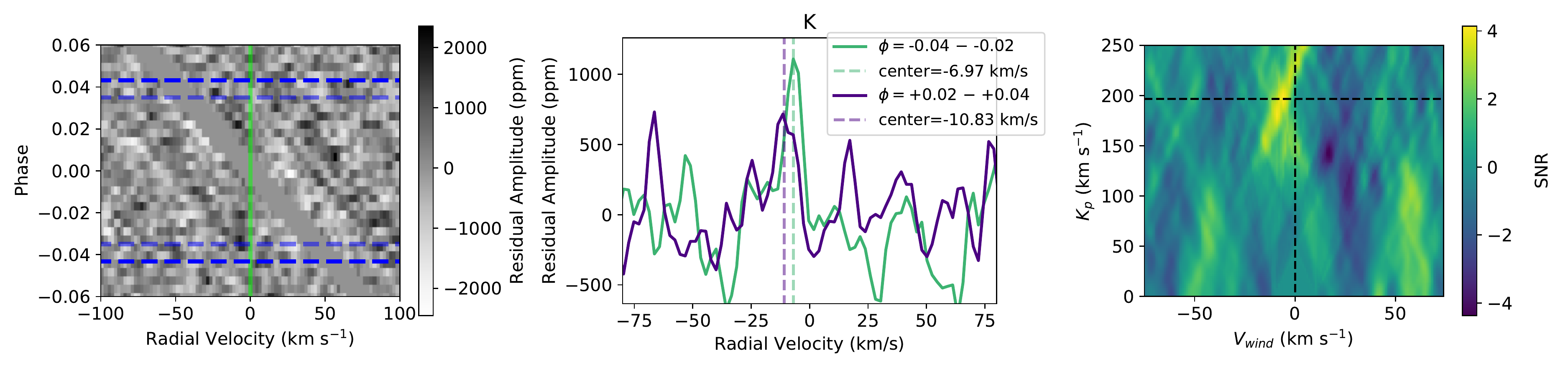}
\label{f:cc_1}
\end{center}
\end{figure*}

\begin{figure*}
\begin{center}
\includegraphics[width=\linewidth]{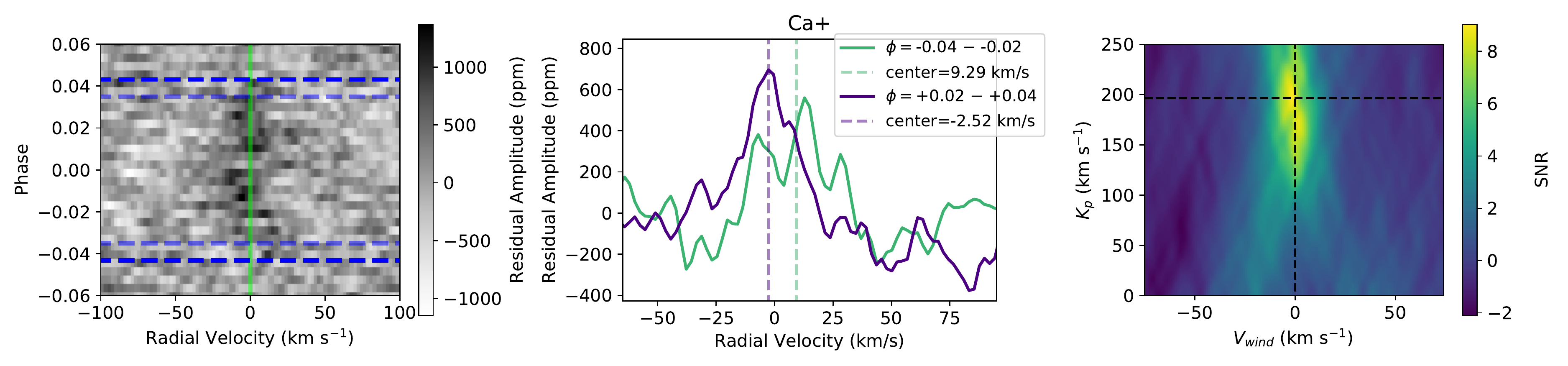}
\includegraphics[width=\linewidth]{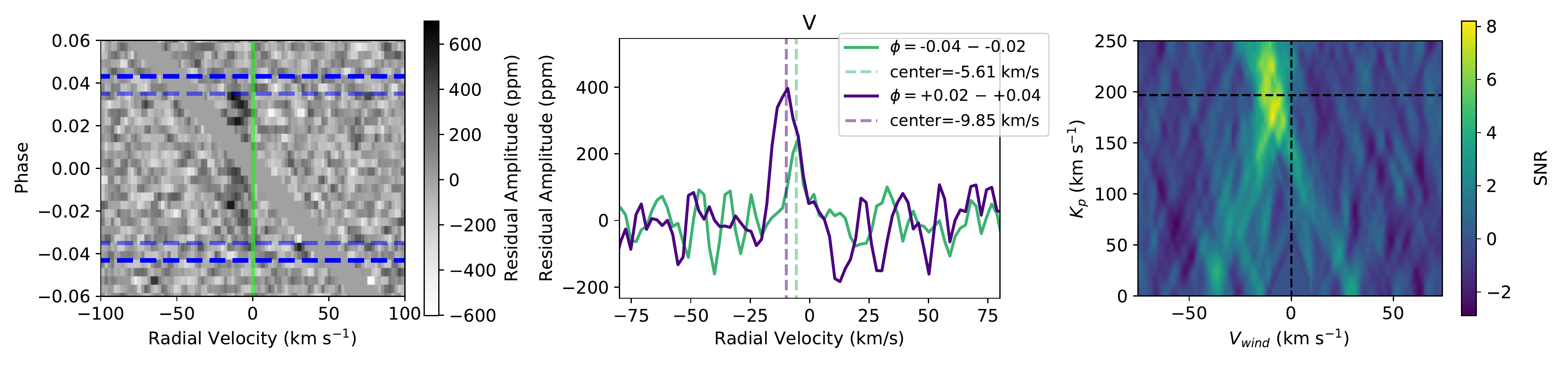}
\includegraphics[width=\linewidth]{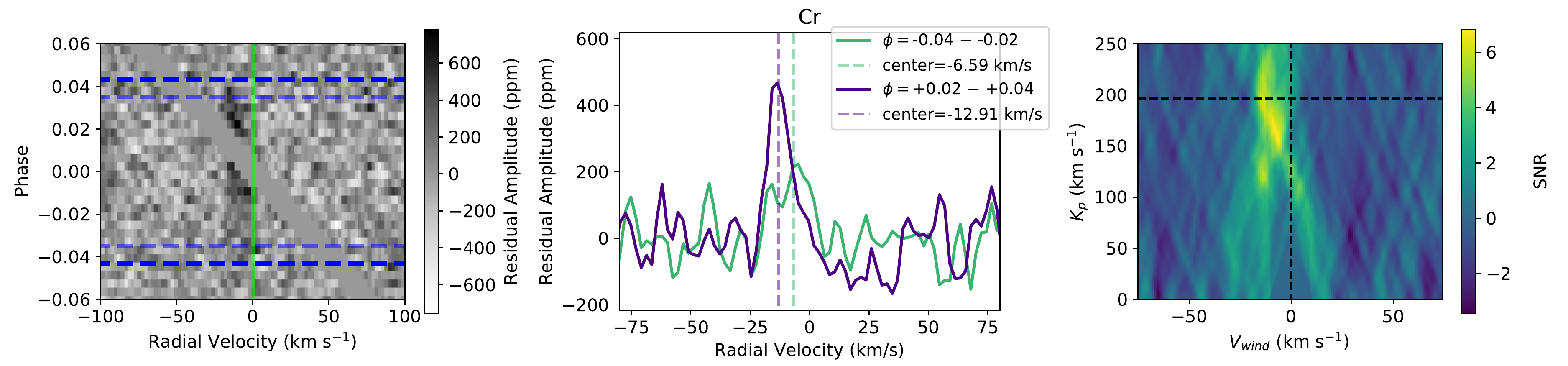}
\includegraphics[width=\linewidth]{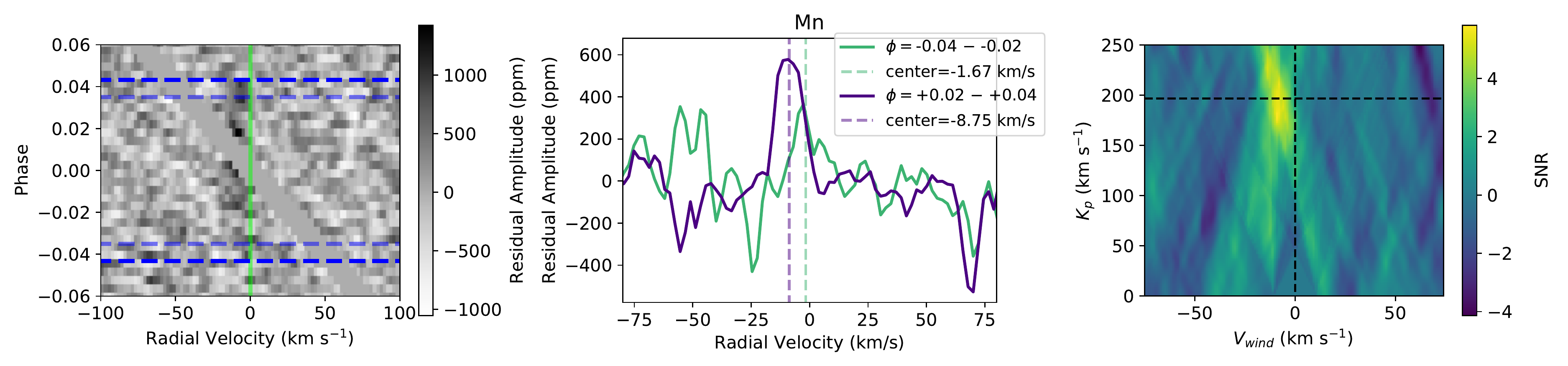}
\includegraphics[width=\linewidth]{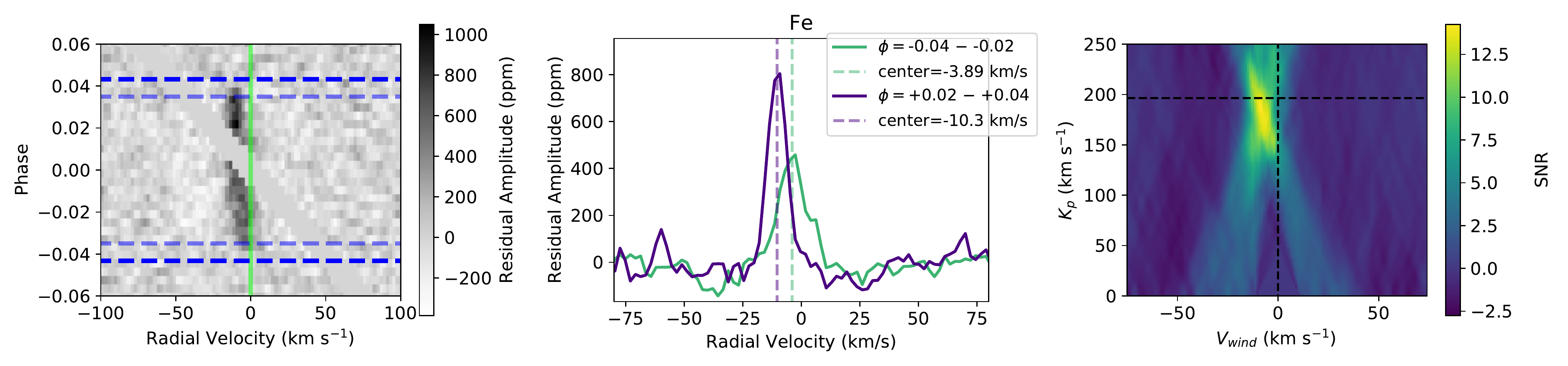}
\label{f:cc_metals1}
\end{center}
\end{figure*}

\begin{figure*}
\begin{center}
\includegraphics[width=\linewidth]{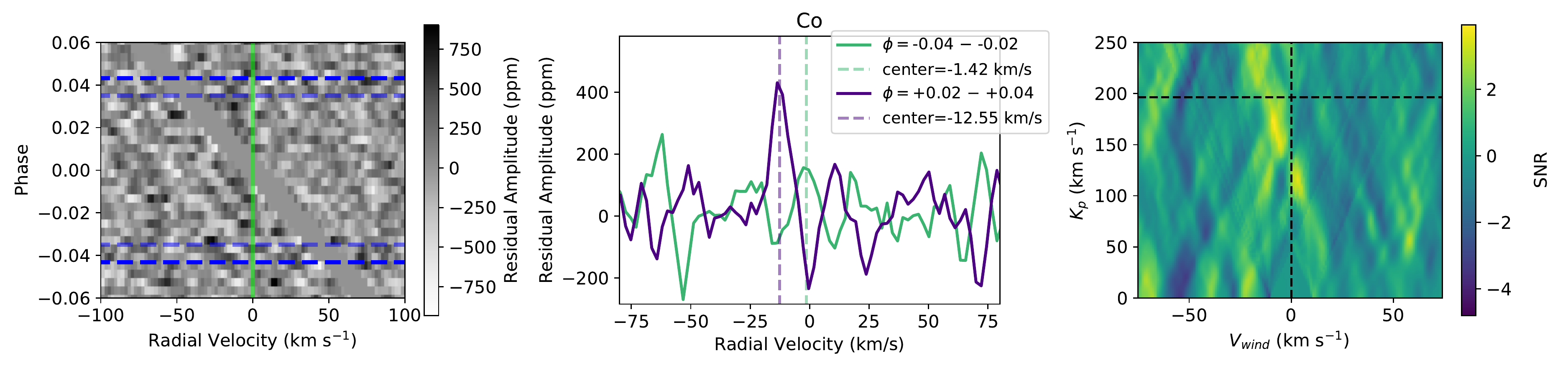}
\includegraphics[width=\linewidth]{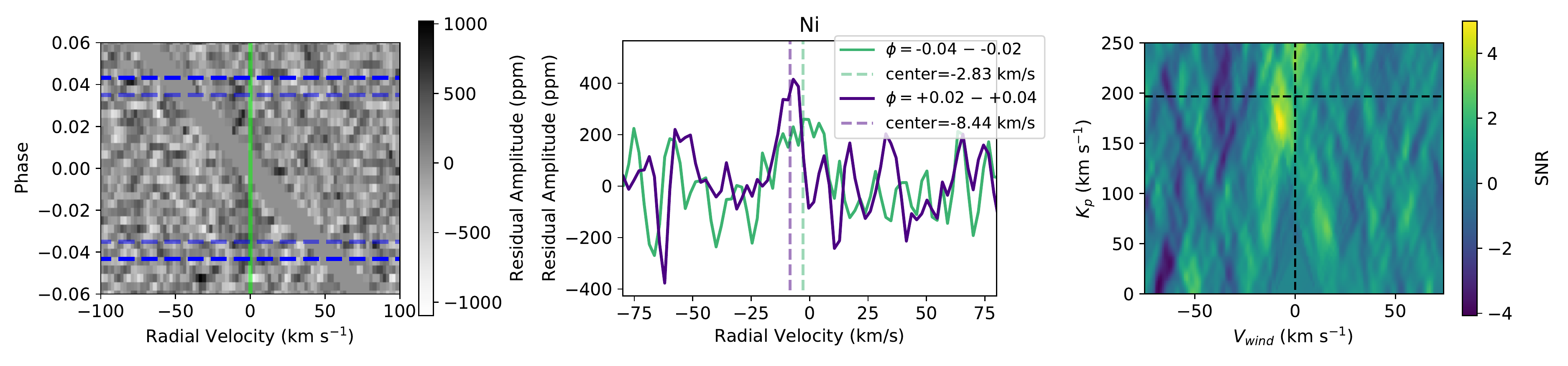}
\includegraphics[width=\linewidth]{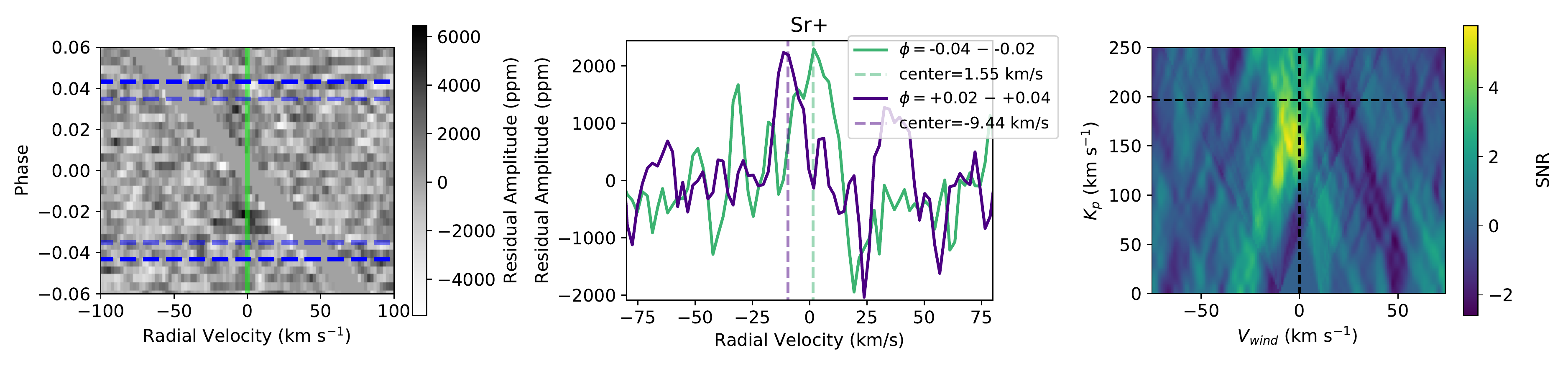}
\caption{\small 
Results of our cross correlation analysis for all of the detected and tentatively detected atoms and ions. \textbf{Left}: for each atom, we show the 2D cross correlation functions, as shown in panel (d.) of Figure \ref{f:cc_example}, except they have now been shifted to the rest frame of the planet using the expected $K_p$. Therefore, any excess absorption from the planet should appear as black residuals at 0 km s$^{-1}$ (shown by a green line). \textbf{Center}: the many cross correlation functions in the left panel have been averaged together at the start of the transit ($\phi=-$0.04 to $ -$0.02, turquoise) and end of the transit ($\phi=+0.04$ to $+0.02$, dark purple) separately. The dashed vertical lines show the radial velocity of the center of a Gaussian that was fit to each average 1D cross correlation function. \textbf{Right}: we also searched in $K_p$ and $v_{wind}$ parameter space to assess the significance of the detection and to determine the $K_p$ and $v_{wind}$ values that gave the highest significance detection. The black dotted lines show the expected parameters of the system neglecting atmospheric dynamics. In many cases the most significant values deviate from the expected values. \citet{Wardenier2021} found that this deviation can be explained by including global winds and condensation of the species on the morningside of the planet.}
\label{f:cc_metals2}
\end{center}
\end{figure*}

\begin{center}
\begin{table*}
\centering
\caption{Parameters of Detected Species} 
\begin{tabular}{c c c c c c c c}
\hline
Atom & Peak & RV$_1$ & RV$_2$ & Amplitude 1 & Amplitude 2 & K$_p$ & FWHM \\ 
 & SNR & (km s$^{-1}$) & (km s$^{-1}$) & (ppm) & (ppm) & (km s$^{-1}$) & (km s$^{-1}$) \\ 
\hline
H I & 5.31 & $-14\pm3.51$ & $2.08\pm1.78$ & $989\pm141$ & $1709\pm229$ & $219\substack{+40\\-20}$ & $15.25\pm2.26$\\
Li I& 6.03 & $-5.06\pm2.07$ & $-5.28\pm2.89$& $822\pm94$ & $1195\pm156$ & $235\substack{+16\\-24}$  & $21.23\pm3.39$\\
Na I& 10.63 & $-3.80\pm0.79$& $-6.08\pm1.49$& $1285\pm135$ & $941\pm116$ & $209\substack{+14\\-20}$  & $17.37\pm2.01$\\
Mg I& 6.94 & $-3.32\pm1.29$ & $-7.61\pm1.25$ & $1049\pm128$ & $1240\pm163$ & $189\substack{+26\\-30}$  & $15.4\pm2.02$\\
K I& 4.22 & $-6.97\pm0.83$ & $-10.83\pm2.04$ & $1157\pm166$ & $743\pm138$ & $197\substack{+14\\-12}$ & $9.35\pm2.0$\\
Ca II& 7.8 & $8.05\pm2.3$ & $-2.52\pm1.53$ & $387\pm53$ & $653\pm78$ & $197\substack{+16\\-38}$  & $22.5\pm2.28$\\
V I& 8.19 & $-5.61\pm0.68$ & $-9.85\pm0.85$ & $228\pm29$ & $411\pm52$ & $175\substack{+14\\-10}$  & $10.89\pm1.41$ \\
Cr I& 6.82 & $-6.59\pm2.17$ & $-12.91\pm0.68$ & $207\pm31$ & $482\pm60$ & $173\substack{+12\\-14}$  & $10.03\pm1.21$ \\
Mn I& 5.82 & $-1.67\pm2.1$ & $-8.75\pm1.21$ & $305\pm68$ & $639\pm83$ & $179\substack{+40\\-10}$  & $11.98\pm1.62$\\
Fe I& 14.24 & $-3.89\pm0.48$ & $-10.26\pm0.24$ & $440\pm47$ & $833\pm83$ & $188\substack{+10\\-16}$  & $9.05\pm0.5$\\
Co I& 4.03 & $-1.42\pm2.16$ & $-12.55\pm0.8$ & $173\pm53$ & $434\pm65$ & $173\substack{+14\\-16}$ & $8.13\pm1.5$ \\
Ni I& 5.01 & $-2.83\pm3.02$ & $-8.44\pm1.31$ & $259\pm43$ & $429\pm69$ & $173\substack{+16\\-12}$ & $9.21\pm2.18$ \\
Sr II& 5.77 & $1.55\pm1.6$ & $-9.44\pm1.3$ & $2268\pm327$ & $2319\pm376$ & $152\substack{+20\\-8}$ & $11.32\pm2.2$ \\
\hline
\end{tabular}
\label{t:results}
\end{table*}
\end{center}

We performed the same cross correlation analysis for each neutral atom or ion. We detected Li I, Na I, Mg I, Ca II, V I, Cr I, Mn I, Fe I, Ni I, and Sr II with a SNR $>$ 5, and we tentatively detected Co I and K I with a peak SNR $>$ 4. We detected H I with a SNR of 5.77, but there is significant noise around the stellar rest frame due to the low SNR of the spectra near the center of the H $\alpha$ feature, and so we only call this a tentative detection. \citet{Tabernero2021} also found tentative evidence for H $\alpha$ absorption in their single-line analysis, which agrees with our tentative detection. Fe I, Mn I, Na I, Ca II, Li I, K I and Mg I were previously detected in \citet{Tabernero2021}, but V I, Cr I, Ni I and Sr II are detected, and Co I is tentatively detected here, for the first time in WASP-76b. 

All of the detections and tentative detections are shown in Figure \ref{f:cc_metals2}. Each row contains three plots with results of the cross correlation analysis of a different neutral atom or ion. The left panel shows the 2D cross correlation grid in the rest frame of the planet, the middle panel shows the averaged 1D cross correlation function for the beginning of the transit (turquoise) and end of the transit (dark purple) separately, and the right panel shows our systematic search for all prominent signals in the 2D cross correlation grid  for a range of $K_p$ and $v_{wind}$ values.

To determine which of these detections and tentative detections exhibit an asymmetry, we fit Gaussians to the 1D cross correlation functions from the beginning and end of the transit separately (middle panel of Figure \ref{f:cc_metals2}), using the python module \texttt{lmfit}. We record the central radial velocities in Table \ref{t:results}, where RV$_1$ is the central radial velocity of the Gaussian fit to the average cross correlation function between phases $\phi =-0.04$ and $-0.02$, and RV$_2$ is the central radial velocity for the phases $\phi =+0.02$ to $+0.04$. The uncertainties that we record use the same method as \citet{Kesseli2021}, and are computed by converting the FWHM of the fitted Gaussian to the standard deviation and then dividing by the SNR of the peak. We tested that these values gave a true measurement of uncertainty by recording the central radial velocity measurements for each individual Fe I cross correlation function in the same phase ranges for the nights individually before they had been interpolated onto the uniform phase grids. We then took the standard deviation of all of these individual measurements and divided by the square root of the number of measurements, as expected for the standard deviation of a sample. We found uncertainties that were consistent within 20\% for these two methods. We also tested using the uncertainties returned by \texttt{lmfit}, but found that these were about half the size of the uncertainties we calculated and the ones given by our test using the many measurements of Fe I, and so we did not use these. For all of the species that we detected confidently (SNR$>5$), the measured RV offsets between the first and second half of the transit are discrepant by more than $2\sigma$ for Ca II, V I, Cr I, Mn I, Fe I, and Sr II. The two RVs are discrepant by $1\sigma$ for Mg I and Ni I, while the RVs are consistent for Li I and Na I. 

Table \ref{t:results} also contains the peak amplitude of the Gaussians fit to the 1D cross correlation functions for the beginning and end of the transit separately. The uncertainties for these measurements are given by taking the standard deviation of the noise outside of the peak and dividing by the square root of the number of pixels within the peak. We have also listed the FWHM values of the Gaussians and the associated errors produced by \texttt{lmfit}. Finally, we list the values of $K_p$ (planet semi-amplitude velocity) that give the largest SNR, as seen in the right panels of Figure \ref{f:cc_metals2} for each detection or tentative detection. The uncertainties here are simply the $K_p$ values corresponding to a decrease in SNR of 1 from the peak. For any of the plots that were multi-peaked (i.e. Ca II), we made sure that the uncertainties extended to include both peaks. 

\subsection{Dependence on Model Parameters}
\label{s:otherparams}

WASP-76b has been thoroughly studied, and there are many different temperature estimates. Phase curve observations presented in \citet{May2021} infer a dayside disk-integrated brightness temperature of $\sim2400 - 2700$ K. \citet{vonEssen2020} analyzed two HST transits and retrieved a terminator temperature of $2300\substack{+412\\-392}$ K. \citet{Edwards2020} used a different HST transit of WASP-76b and retrieved a temperature of $2231\substack{+265\\-283}$ for the terminator. Finally, \citet{Fu2020} analyzed both of these datasets in combination with two other HST transits and two Spitzer transits and found that the residual spectrum was best fit by a forward model with a temperature of 2000 K. As mentioned before, with high-resolution, \citet{Seidel2019} determined a best-fit temperature of 3389 $\pm$ 227 K, but warned that this is unusually high, while \citet{Landman2021} recovered a temperature between 2700 and 3700 K. Using low-resolution HST emission spectra, \citet{Fu2020} found that temperatures in the upper regions of the atmosphere where high-resolution observations probe ranged from 2700 and 3000 K. With all of these differing results, we tested whether using models with temperatures of 2000 and 4000 K would alter any of our results. 

In addition to changing the temperature, we also tested moving the gray cloud opacity below which we do not probe to higher and lower points in the atmosphere. Again, this parameter does not necessarily mean that there are clouds in the atmosphere, but acts as an approximation of a range of continuum opacity sources. As this parameter is degenerate with the abundance, changing the cloud opacity in this way acts to test how relaxing our assumption of solar abundance changes our detections.

We found that the RVs measured when using the different models are within the uncertainties given in Table \ref{t:results}. Alternatively, the amplitudes are model dependent, and by using different models, we recover different amplitudes that are not always within the stated uncertainties, but we find that the amplitudes of both parts increase or decrease at the same rate and so the relative amplitude difference is model independent within our uncertainties. Because of the model-dependence of the amplitudes, we refrain from plotting the individual amplitudes or average amplitude and only draw conclusions about the relative change between the amplitude of the first and second halves of the transit. The average amplitude of the cross correlation of each species can only give a general idea of where in the atmosphere the signal originates, but does not correspond to an exact physical location and so the individual values of amplitude 1 and amplitude 2 in Table \ref{t:results} should not be interpreted separately or directly compared to the single-line values from \citet{Tabernero2021} and \citet{Seidel2021}. After these tests, we can confidently conclude that any overall trends discussed next should not be dependent on any of our choices in parameters.

\section{Discussion} 
\label{s:discussion}

\begin{figure*}
\begin{center}
\includegraphics[width=0.42\linewidth]{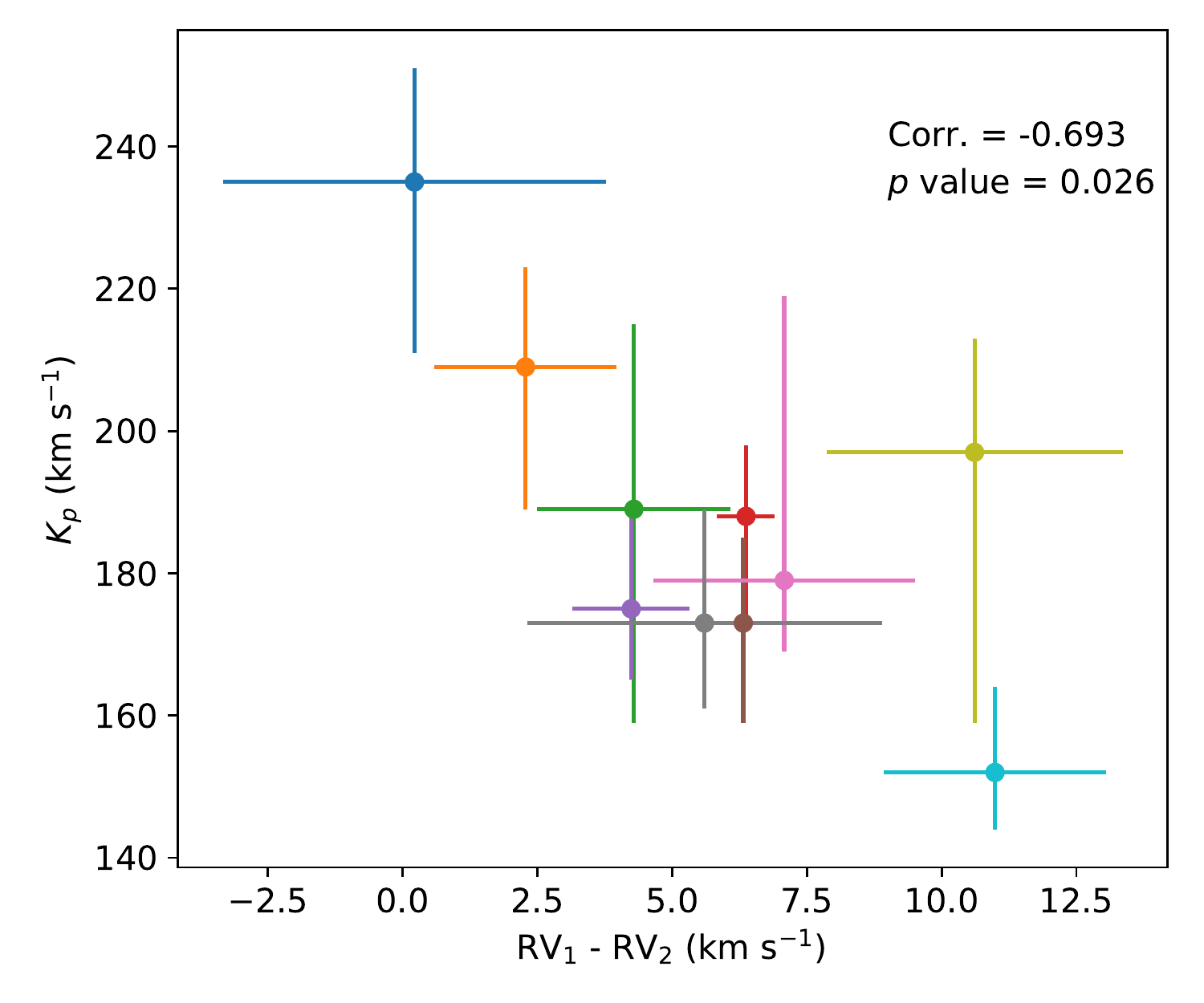}
\includegraphics[width=0.42\linewidth]{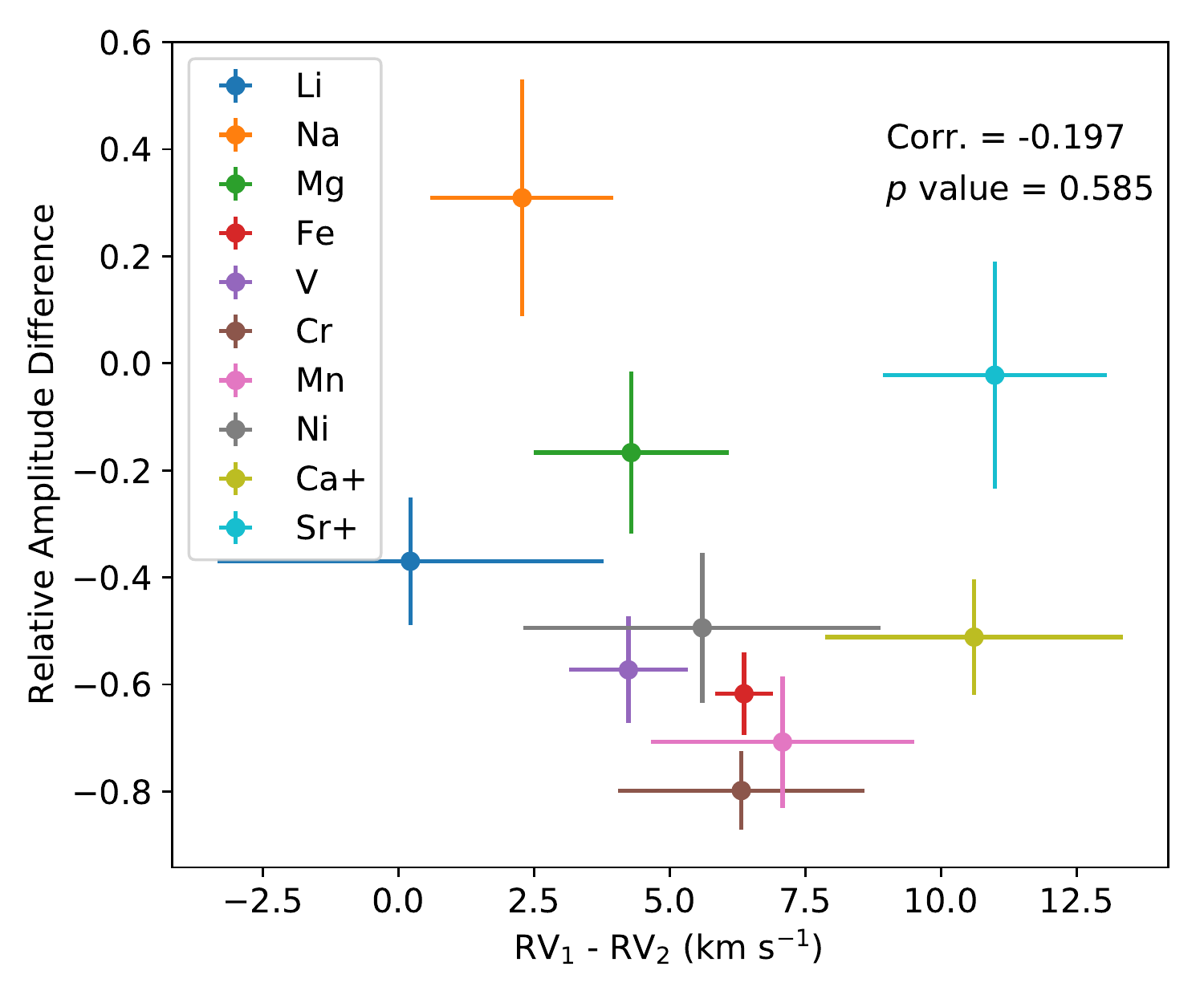}
\includegraphics[width=0.42\linewidth]{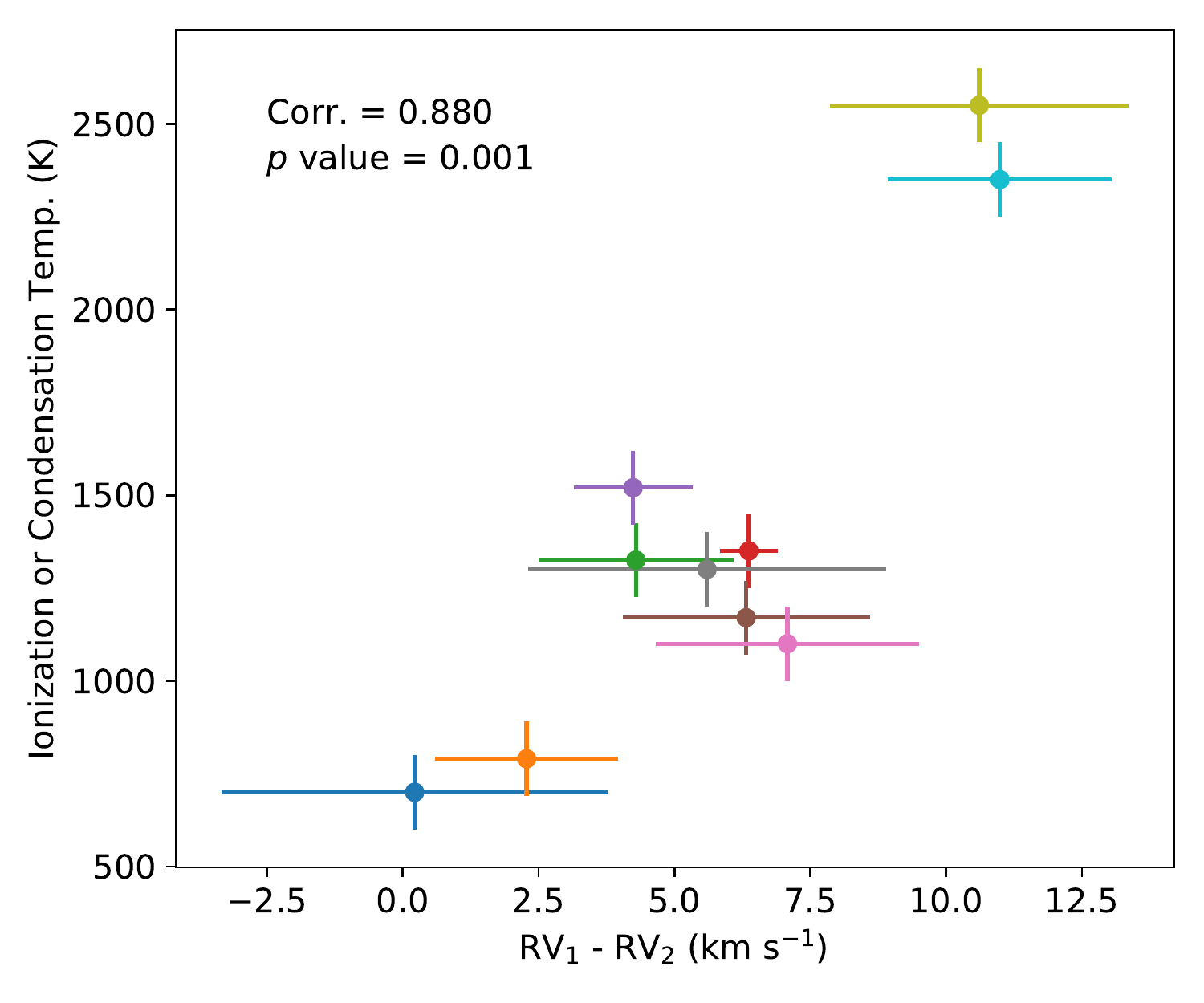}
\includegraphics[width=0.42\linewidth]{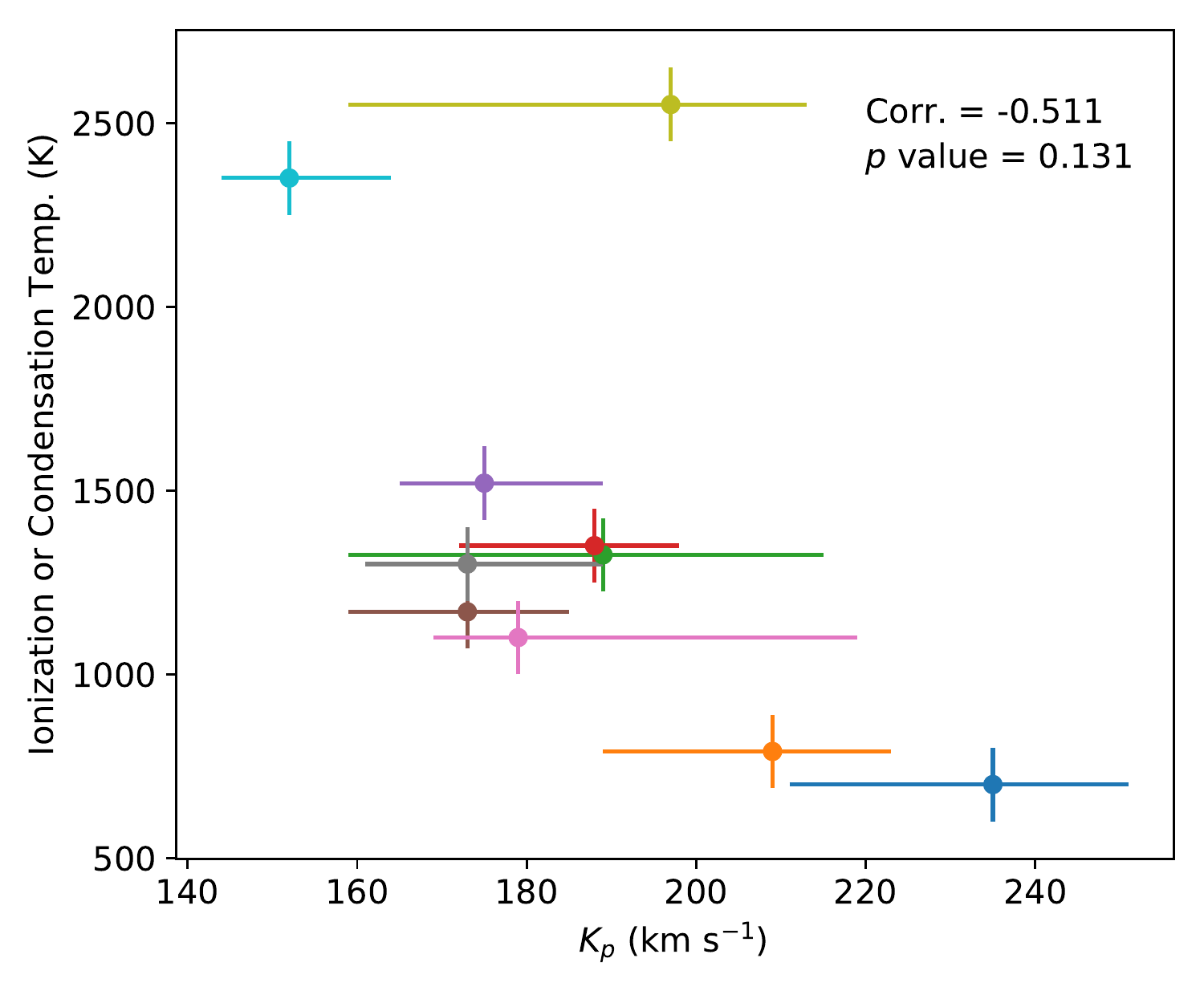}
\includegraphics[width=0.42\linewidth]{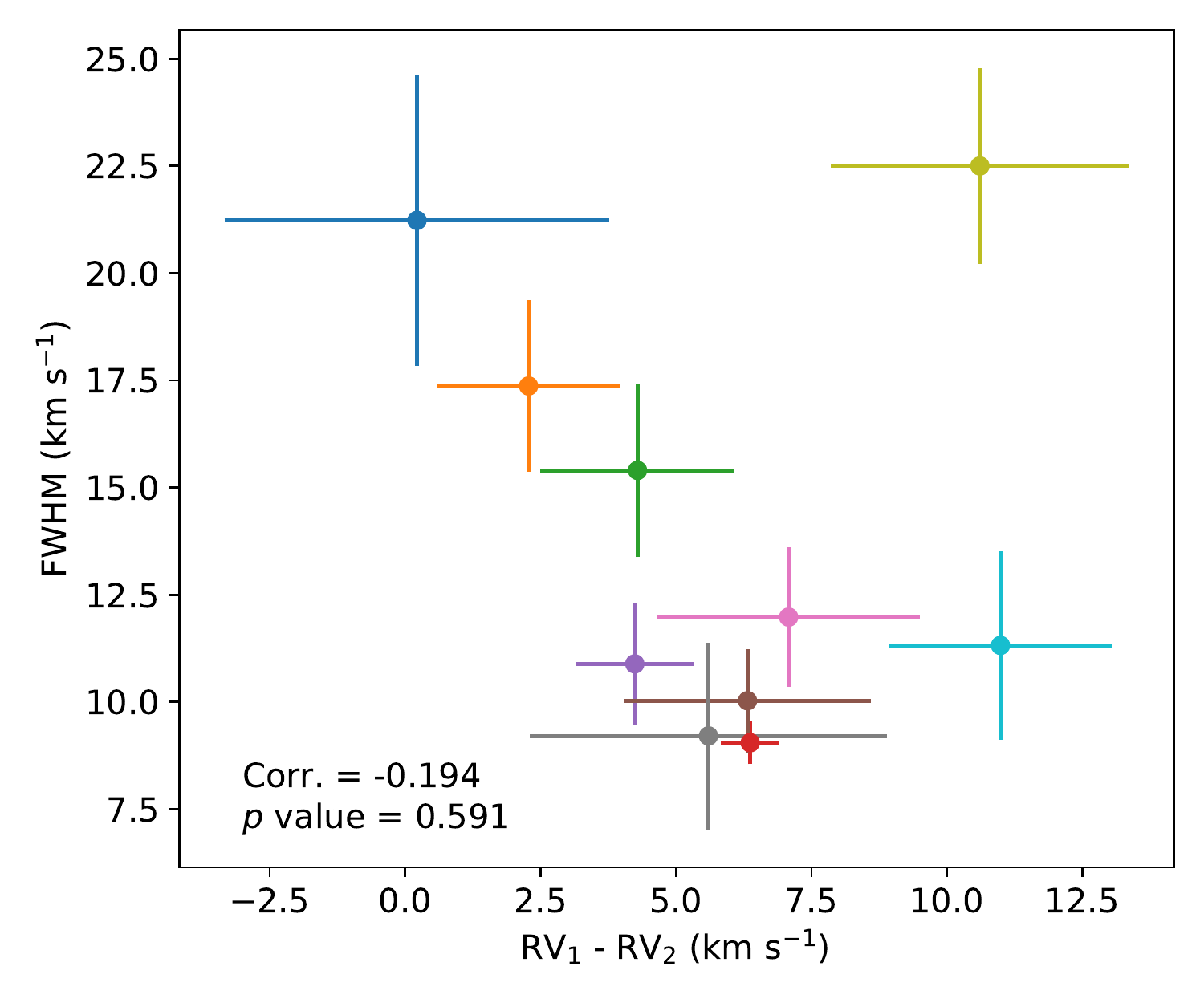}
\includegraphics[width=0.42\linewidth]{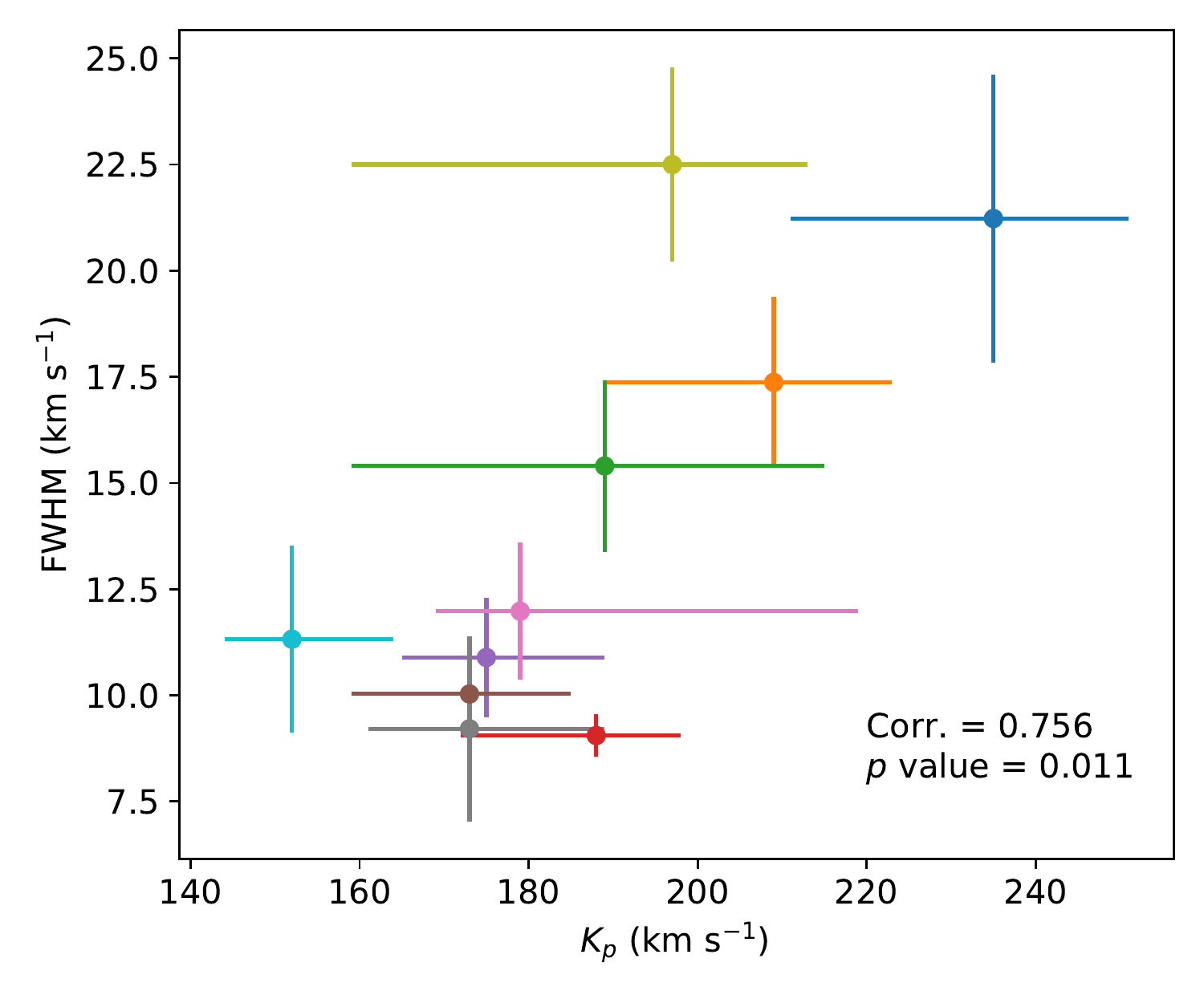}
\caption{\small 
Scatter plots showing the relations between the derived parameters from Table \ref{t:results} for all of the species detected with a SNR$>$5. \textbf{Top right:} the $K_p$ that exhibits the peak SNR from the right-most plots in Figure \ref{f:cc_metals2} vs. the radial velocity difference. \textbf{Top left:} amplitude 1 divided by amplitude 2 (relative amplitude difference) vs. the radial velocity difference between the first and second half of the transit. \textbf{Middle left:} the temperature below which the atom or ion is expected to no longer exist vs. the radial velocity difference. The temperatures for the ions are where the species are expected to become mostly neutral (ionization temperature) and are estimated using an equilibrium chemistry code from \citet{Molliere2017} assuming a pressure of $10^{-4}$ bars. The temperatures for the neutral atoms are the temperatures where each is expected to condense into a solid and are taken from either \citet{Gao2020} (Mg, Cr, Mn, Fe) or \citet{Lodders1999} (Li, Na) at a pressure of $10^{-4}$ bars. \citet{Lodders2002} stated that V condenses about 200 K above Fe, and so we use this as our estimate for V condensation. We were unable to find a condensation temperature for Ni, and so we do not include it in this plot, but given its similar behavior to the other transition metals, we expect it to condense around similar temperatures (1300 to 1500 K). We assume uncertainties of 100 K for each of these temperatures. \textbf{Middle right:} same temperatures, but now plotted against the peak $K_p$. \textbf{Bottom left:} FWHM of the 1D cross correlation functions vs. the radial velocity difference; or (\textbf{bottom right}) the peak $K_p$.}
\label{f:trends}
\end{center}
\end{figure*}

We plotted the results from Table \ref{t:results} in Figure \ref{f:trends} to determine if there were any overarching trends that spanned all of the detected species and would help to explain the results. For each relation, we calculate a Pearson correlation coefficient to determine if the two variable are correlated and the strength of the relation. Pearson correlation coefficient values larger than 0.5 or less than $-0.5$ are considered to show correlation, while values close to 0.0 are uncorrelated. The top-left plot shows the planet semi-amplitude velocity ($K_p$) versus the radial velocity difference. It is intuitive that $K_p$ and the radial velocity difference are correlated as they are essentially measuring the same thing; if RV$_2$ is more blueshifted than RV$_1$ the highest SNR would result from a shallower slope, and hence smaller $K_p$. We measured a Pearson correlation coefficient of $-0.69$, which denotes moderately strong negative correlation. The only point that does not clearly fit the trend (although it is still within the uncertainty) is Ca II, which has a double-peaked structure in both the $K_p$ versus $v_{wind}$ diagram and in the 1D cross correlation function for the first half of the transit. 

The top right panel of Figure \ref{f:trends} shows the relative amplitude difference versus the radial velocity difference. While there does not seem to be a clear trend between these parameters (Pearson coeff. = $-0.2$), by examining the difference in amplitude, we can gain information about the temperature structure of the planet. All of the transition metals (V I, Cr I, Mn I, Fe I, Ni I, and a tentative detection of Co I) and Ca II show stronger absorption during the second half of the transit when the eveningside of the planet is more in view. Li I, Mg I, and Sr II show slightly stronger absorption during the second half of the transit, while Na I and K I actually show stronger absorption during the first half of the transit when the morningside of the planet is more in view. Chemical modeling from \citet{Kataria2016} predicted that Na I and K I would be more abundant on the nightsides than the daysides of UHJs, and so we hypothesize that they are also more abundant on the cooler morningside of the planet than on the hotter eveningside, where Na and K are expected to be mostly ionized. In contrast, most species are expected to show stronger absorption on the eveningside where the temperature is higher, and hence scale height is larger.

In the middle panels of Figure \ref{f:trends}, we plot the ionization temperature (for ions) or the condensation temperature (for the neutral atoms) versus the radial velocity difference and the measured $K_p$. The relation containing the radial velocity difference shows the highest degree of correlation, with a Pearson value of 0.88, which signals strong positive correlation.
As expected, we find a similar relation between the ionization or condensation temperature and $K_p$, and the Pearson coefficient still points toward moderate correlation with a value of $-0.511$. 

Li I and Na I both show very little radial velocity difference between the beginning and end of the transit, and with condensation temperatures lower than 800 K both atoms are expected to remain in the gas phase even on the cold nightside of the planet. All of the transition metals have mildly blueshifted ($\sim-3$ km s$^{-1}$) absorption at the beginning of the transit and absorption that is more blueshifted ($\sim-10$ km s$^{-1}$) at the end of the transit. The radial velocities of the first half and second half are offset by more than $2\sigma$ for all except for Ni I, which has consistent radial velocities with the other species, but larger uncertainties due to the lower SNR of the detection. The transition metals also all have comparable condensation temperatures that would result in their removal from the gas phase on the nightside of the planet. The ions exhibit the largest discrepancies between their two radial velocities and also are expected to be highly temperature sensitive. The correlation between the ionization or condensation temperature and the strength of the asymmetry could point toward condensation or chemical gradients causing the asymmetries, as species with more uniform distributions across the planet's surface are expected to show less of an asymmetry \citep{Wardenier2021}.

Finally, in the bottom panels of Figure \ref{f:trends}, we plot the FWHM versus the radial velocity difference (left) and $K_p$ (right). The absorption spectra of atoms such as H I, Li I, Na I, and Ca II, are dominated by a few, very strong lines. The line cores of species that have stronger lines (or are more abundant) become optically thick at higher altitudes. Our measurements of the residual amplitudes of these species are on average larger than the residual amplitudes of the transition metals (thousands instead of hundreds), confirming that the signals originate higher in altitude. The FWHMs of these signals are also on average larger than those of the transition metals, which tend to have FWHMs of $\sim10$ km s$^{-1}$. These broad absorption lines were seen in \citet{Seidel2021}, and explained as originating from a vertical wind in the upper atmosphere, which significantly broadens the lines. The trend showing $K_p$ versus FWHM has a Pearson coefficient of 0.756, indicating strong correlation. The correlation can be explained by a day-to-night wind in the lower atmosphere that causes an asymmetry but minimal broadening, while the upper atmosphere (or exosphere) is dominated by a vertical wind or outflow and is not highly rotation phase sensitive but significantly broadens the lines. 

To disentangle these two physical scenarios and to better interpret these signals, more GCM and chemical modeling is required. As GCMs are generally used to model the lower atmosphere ($<$1 microbar) and absorption such as that from Ca II extend significantly higher in the atmosphere \citep{Tabernero2021}, non-hydrostatic GCM models that include atmospheric escape \citep[e.g.,][]{Dobbs2013, Mayne2014, Mendonca2016} may be needed to fully explain the signals we detect. As both the temperature relation and the FWHM exhibit outliers, a combination of both condensation and a two-layered atmosphere could also best explain the measurements outlined here.

\begin{figure*}
\begin{center}
\includegraphics[width=\linewidth]{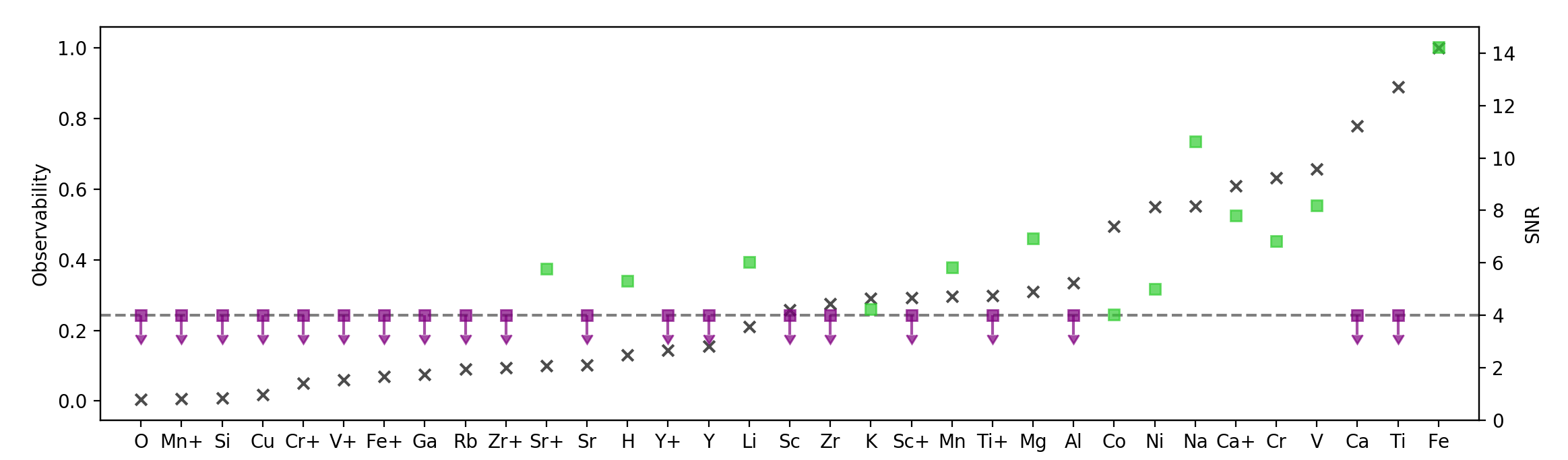}
\caption{\small 
Comparison between the expected observability for each atom or ion that was originally defined in Section \ref{s:atoms} (shown as black x marks) and the SNR for the detected species (green squares) or upper limits at an SNR of 4 (purple squares) for all the non-detections. The two y-axes have been normalized such that the Fe I signal's SNR is equal to an observability of 1. Most of the species that are expected to be observable have a detection, but a few notable exceptions are present and are discussed in detail in the text. }
\label{f:periodic_compare}
\end{center}
\end{figure*}

\subsection{Non-detections}

\begin{figure*}
\begin{center}
\includegraphics[width=0.32\linewidth]{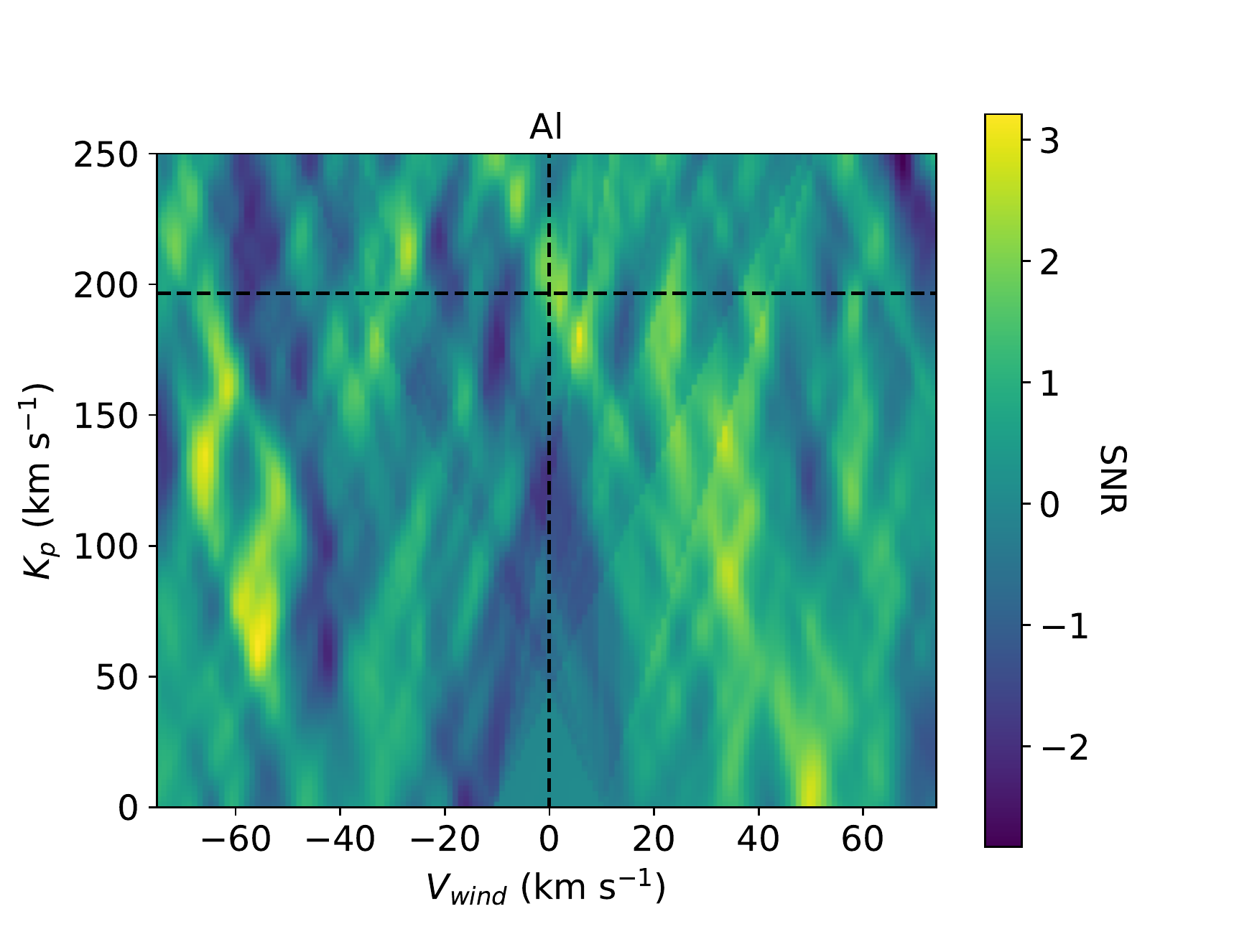}
\includegraphics[width=0.32\linewidth]{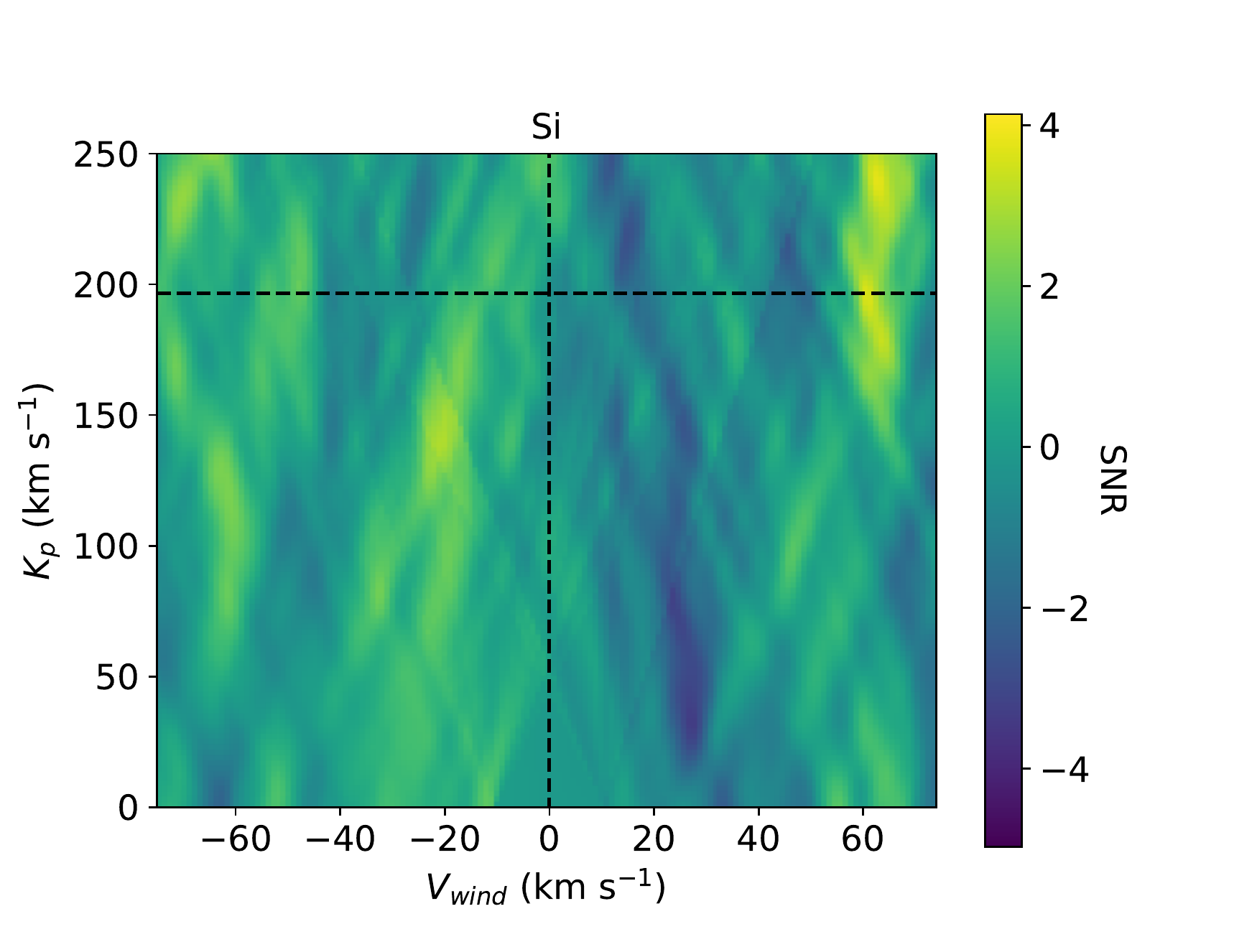}
\includegraphics[width=0.32\linewidth]{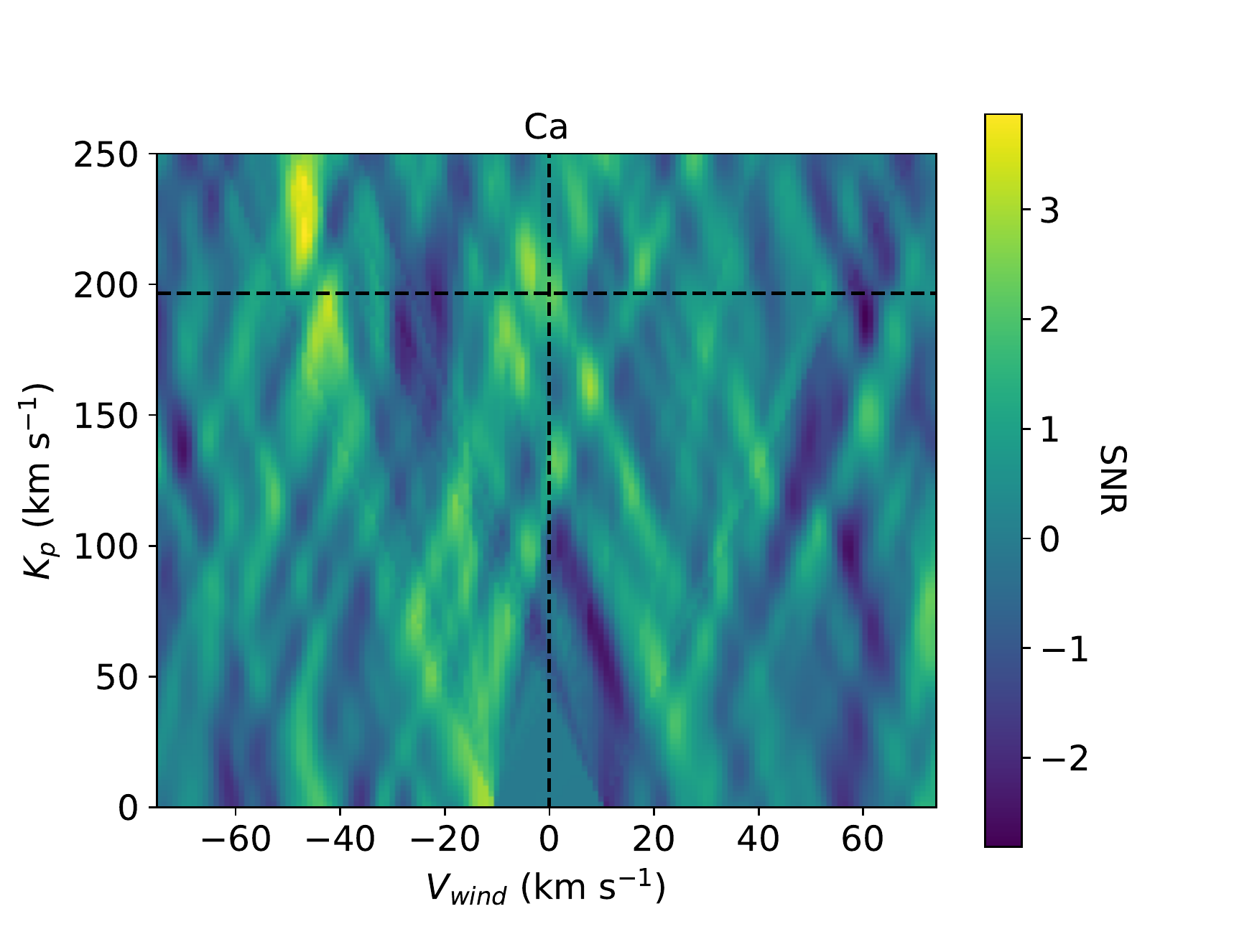}
\includegraphics[width=0.32\linewidth]{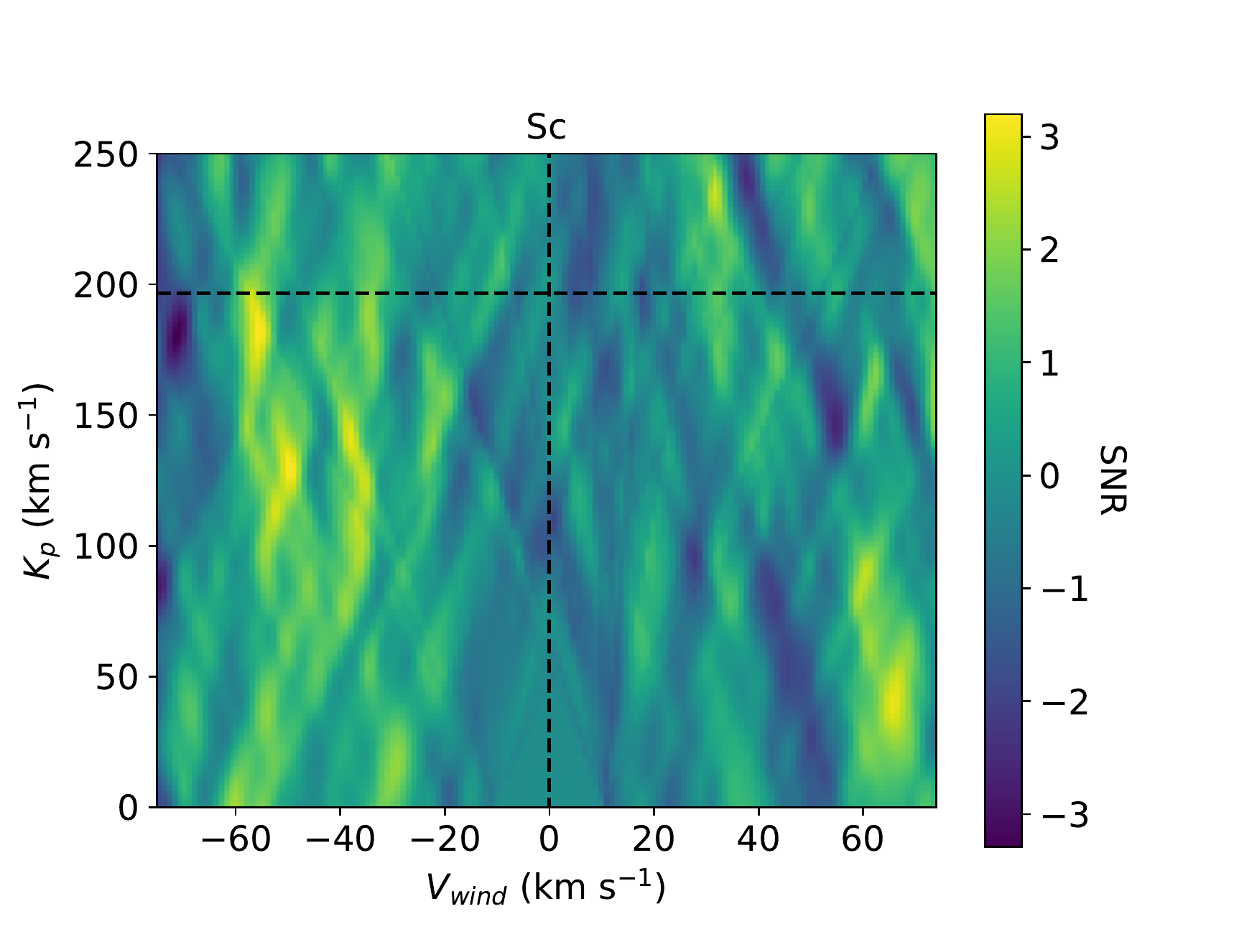}
\includegraphics[width=0.32\linewidth]{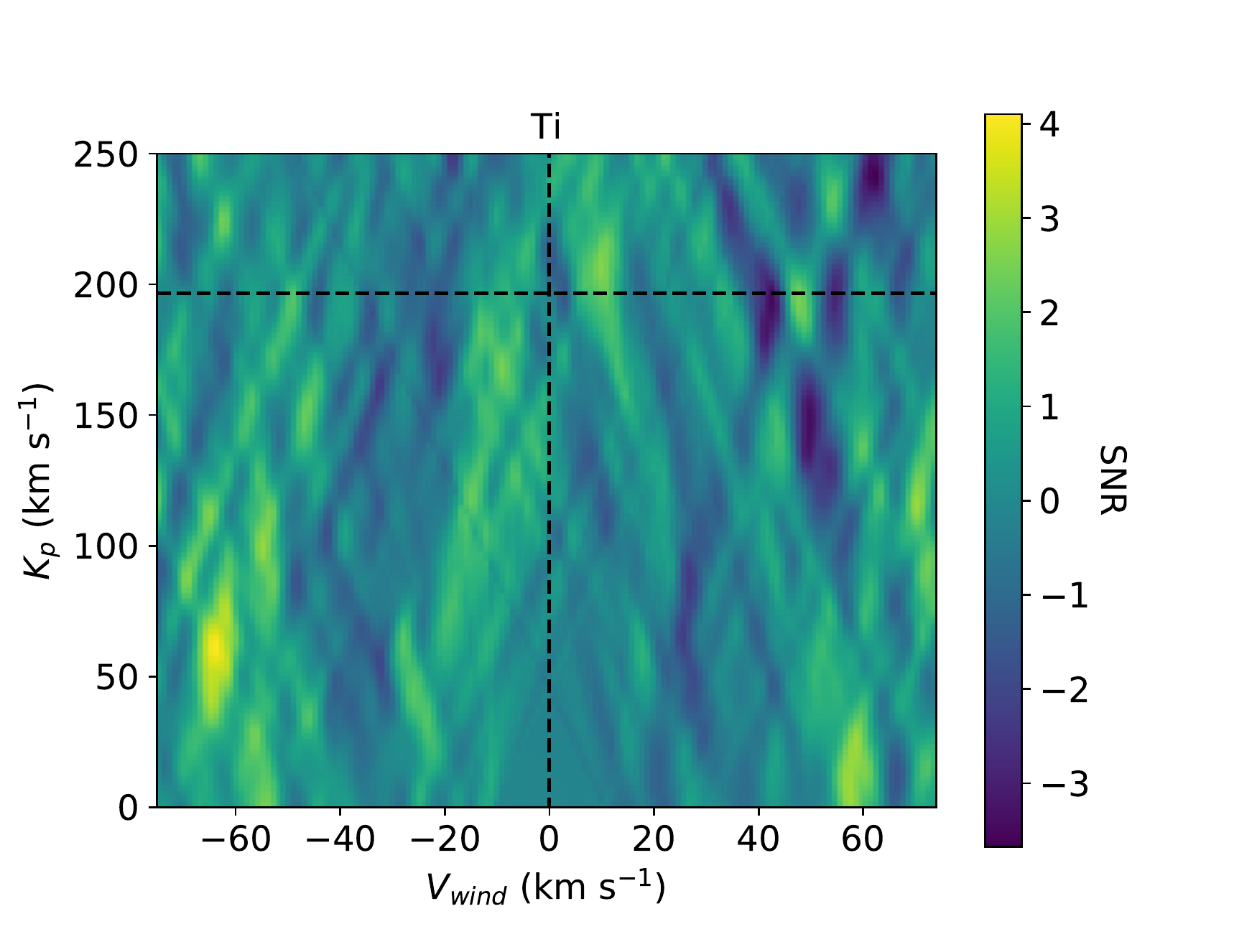}
\includegraphics[width=0.32\linewidth]{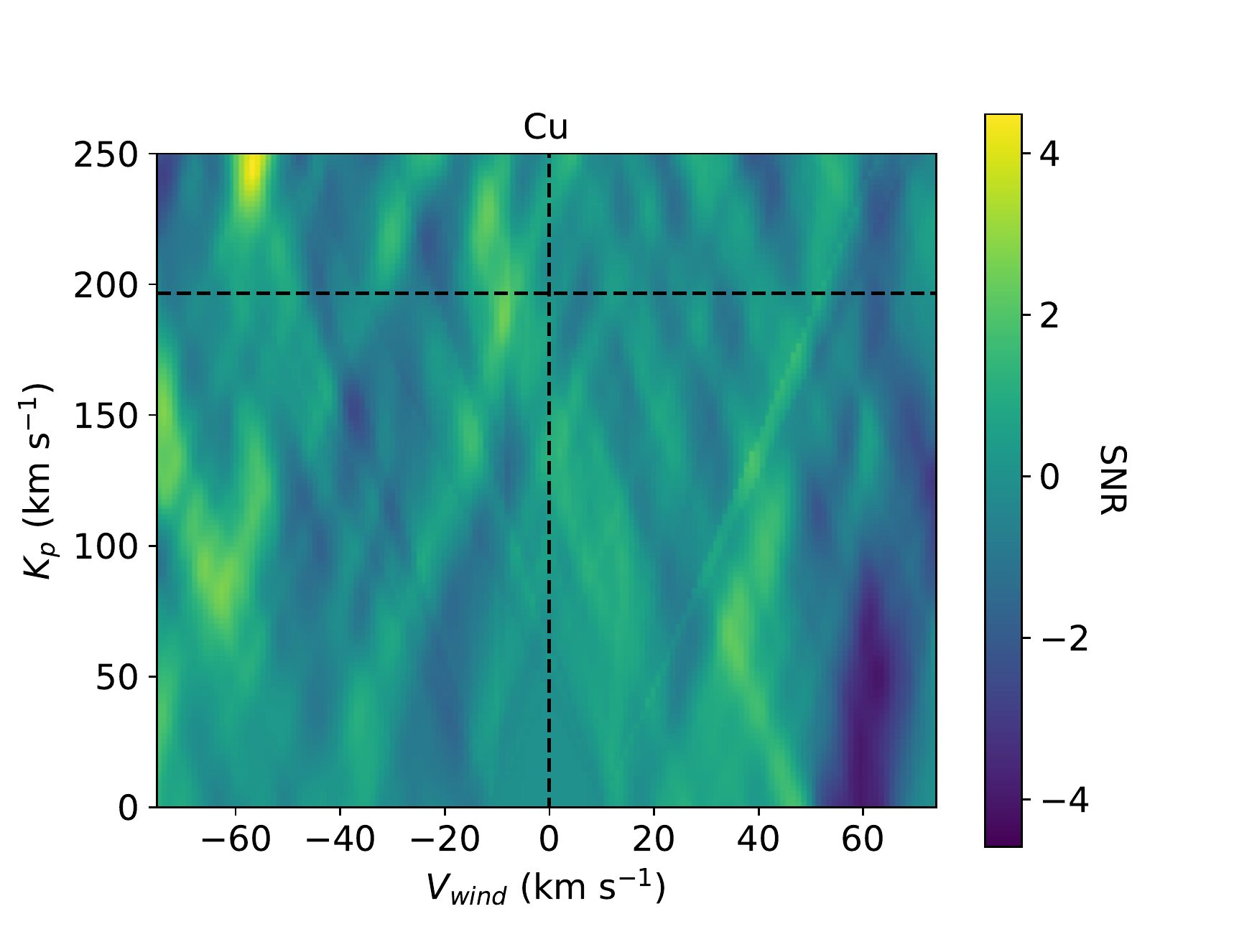}
\includegraphics[width=0.32\linewidth]{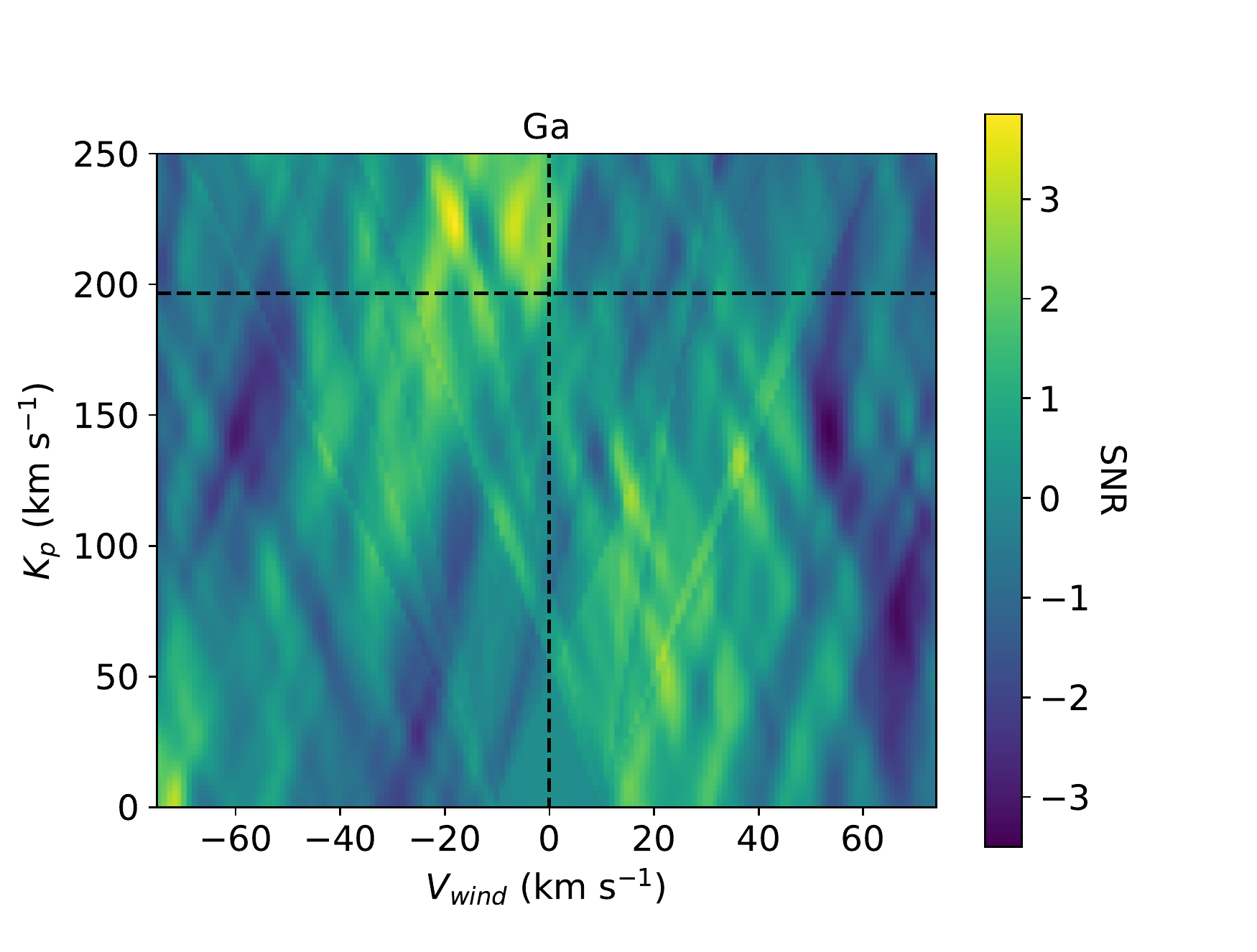}
\includegraphics[width=0.32\linewidth]{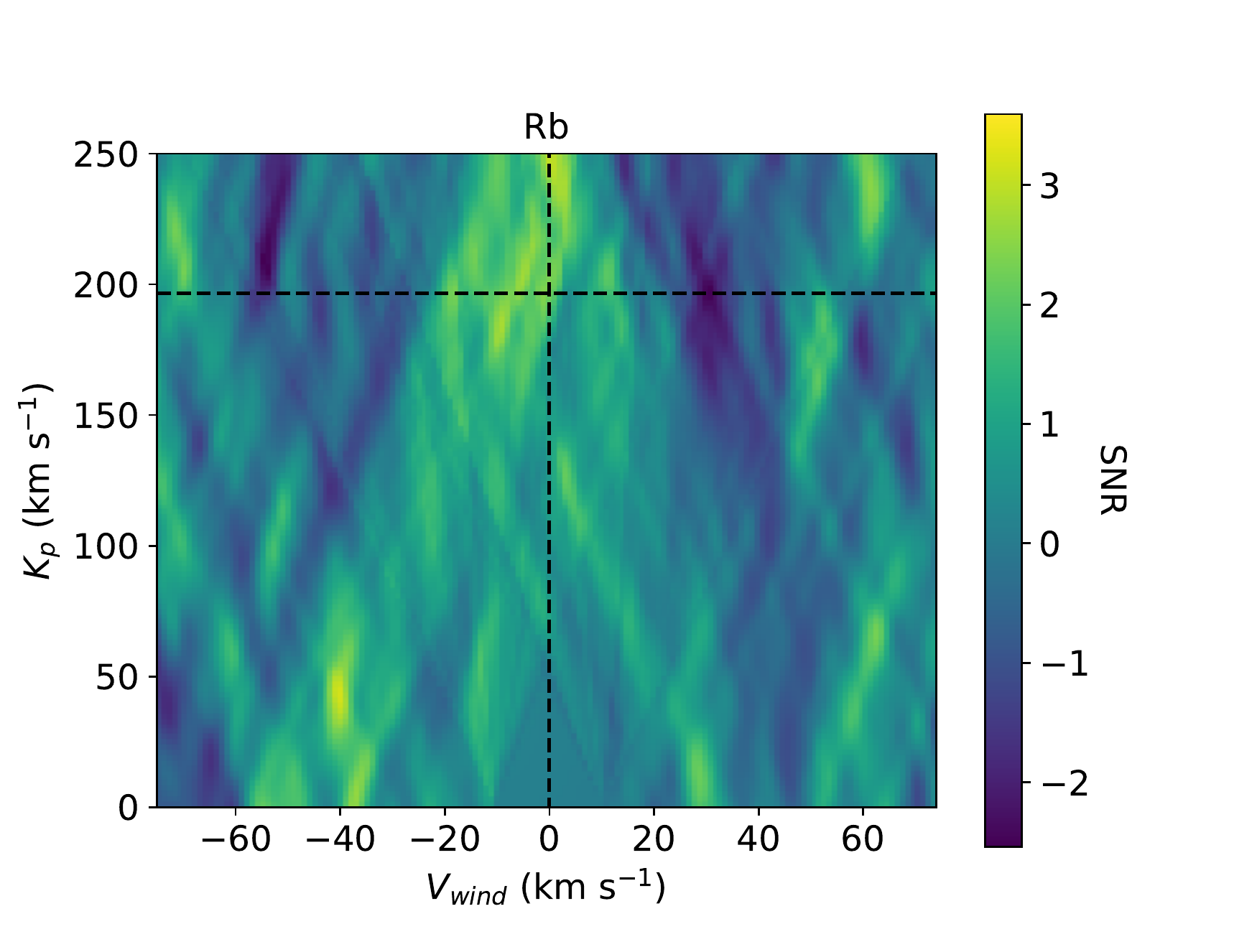}
\includegraphics[width=0.32\linewidth]{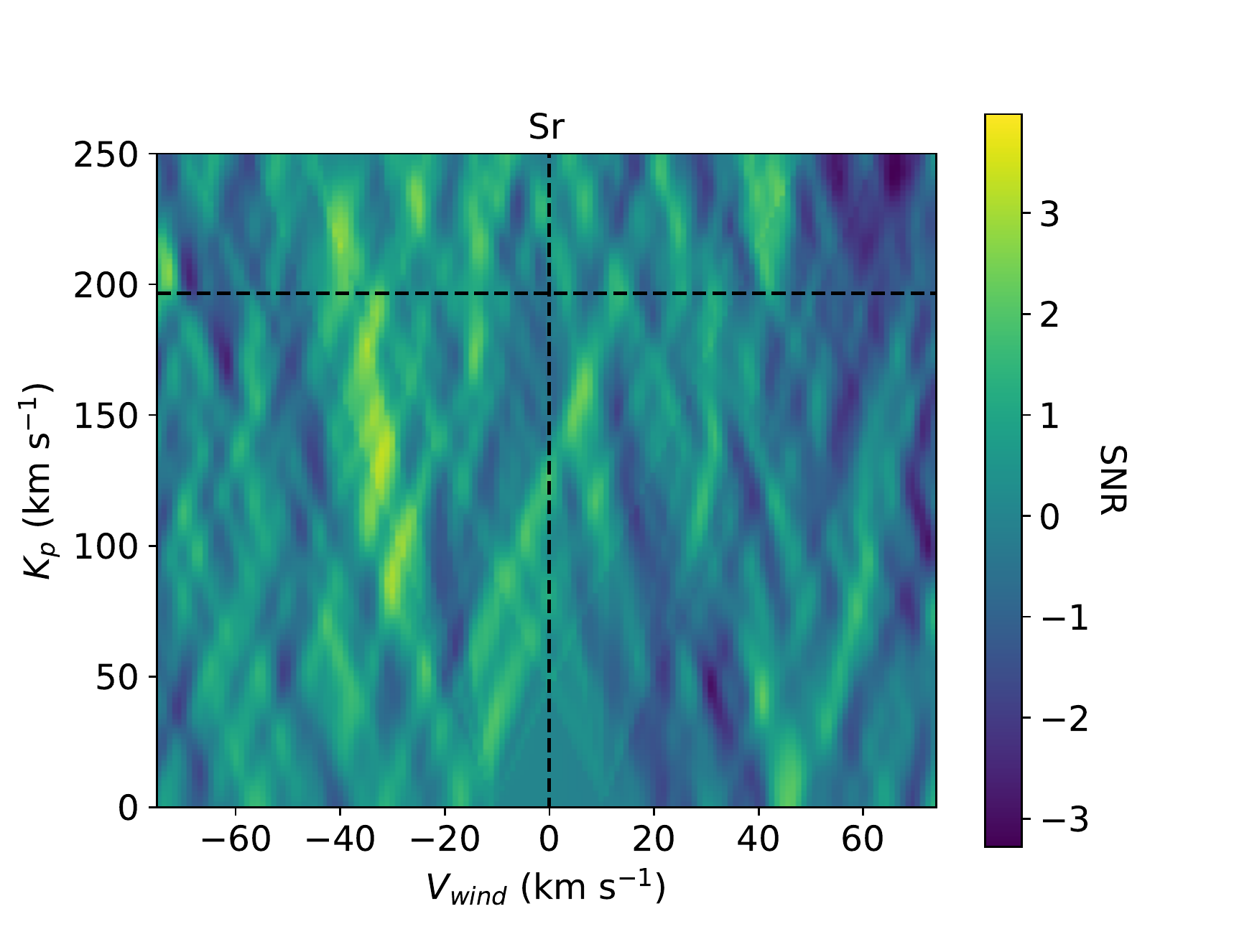}
\includegraphics[width=0.32\linewidth]{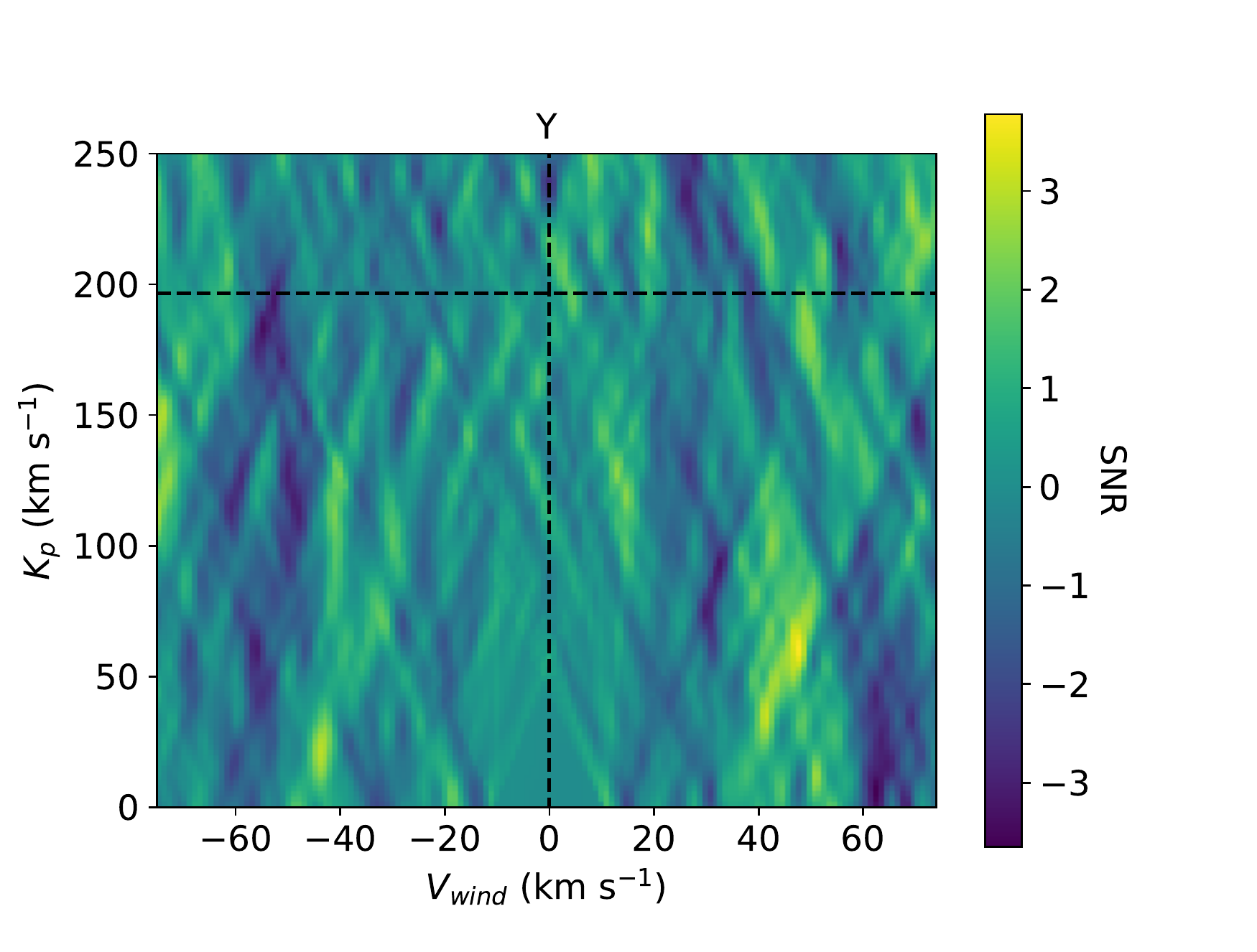}
\includegraphics[width=0.32\linewidth]{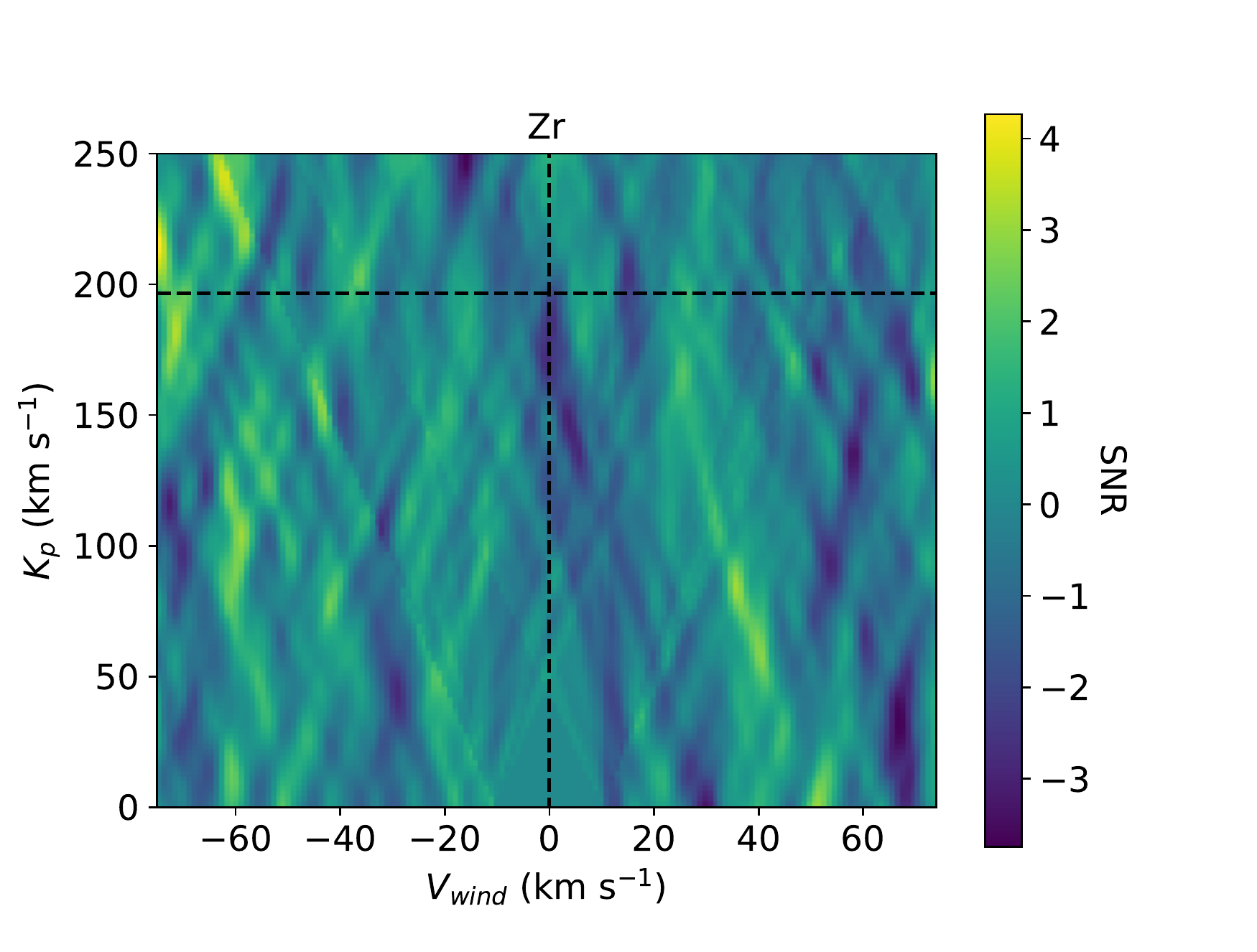}
\caption{\small 
$K_p$ versus $v_{wind}$ maps for each neutral atom that we searched for, but did not detect. These maps are the same as those in the right panels of Figure \ref{f:cc_metals2}. These are all considered non-detections, as none of them show a SNR$\geq$4 near the planet's expected $K_p$ and $v_{wind}$ value (shown by the dotted black lines). These plots show random noise peaks that often have SNRs of 3 to 4, but we do not see any SNR peaks of 5, which is why we chose the cutoff of 5 for our detections and only consider SNR$\geq$4 signals as tentative.  }
\label{f:cc_nodetect_atom}
\end{center}
\end{figure*}

\begin{figure*}
\begin{center}
\includegraphics[width=0.32\linewidth]{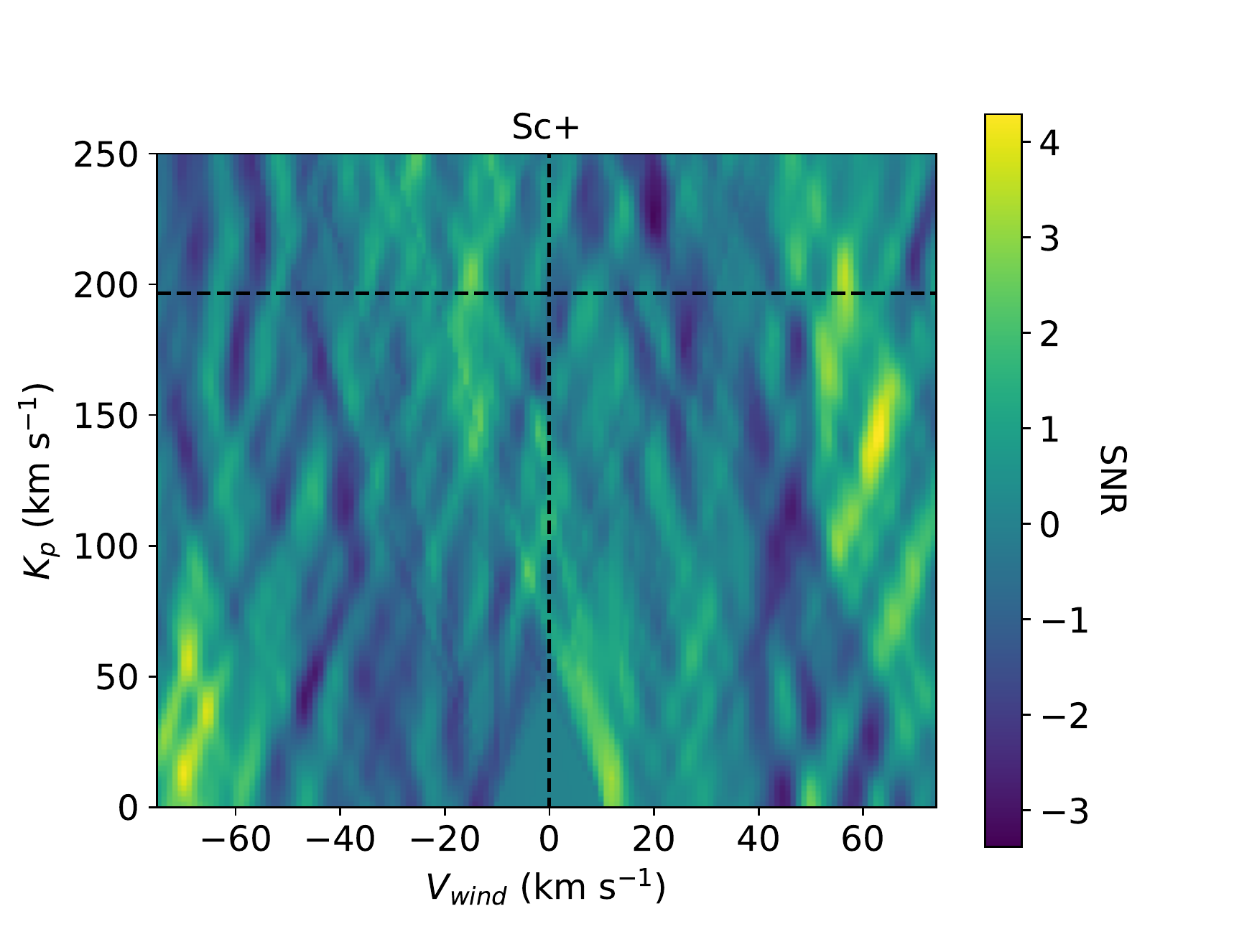}
\includegraphics[width=0.32\linewidth]{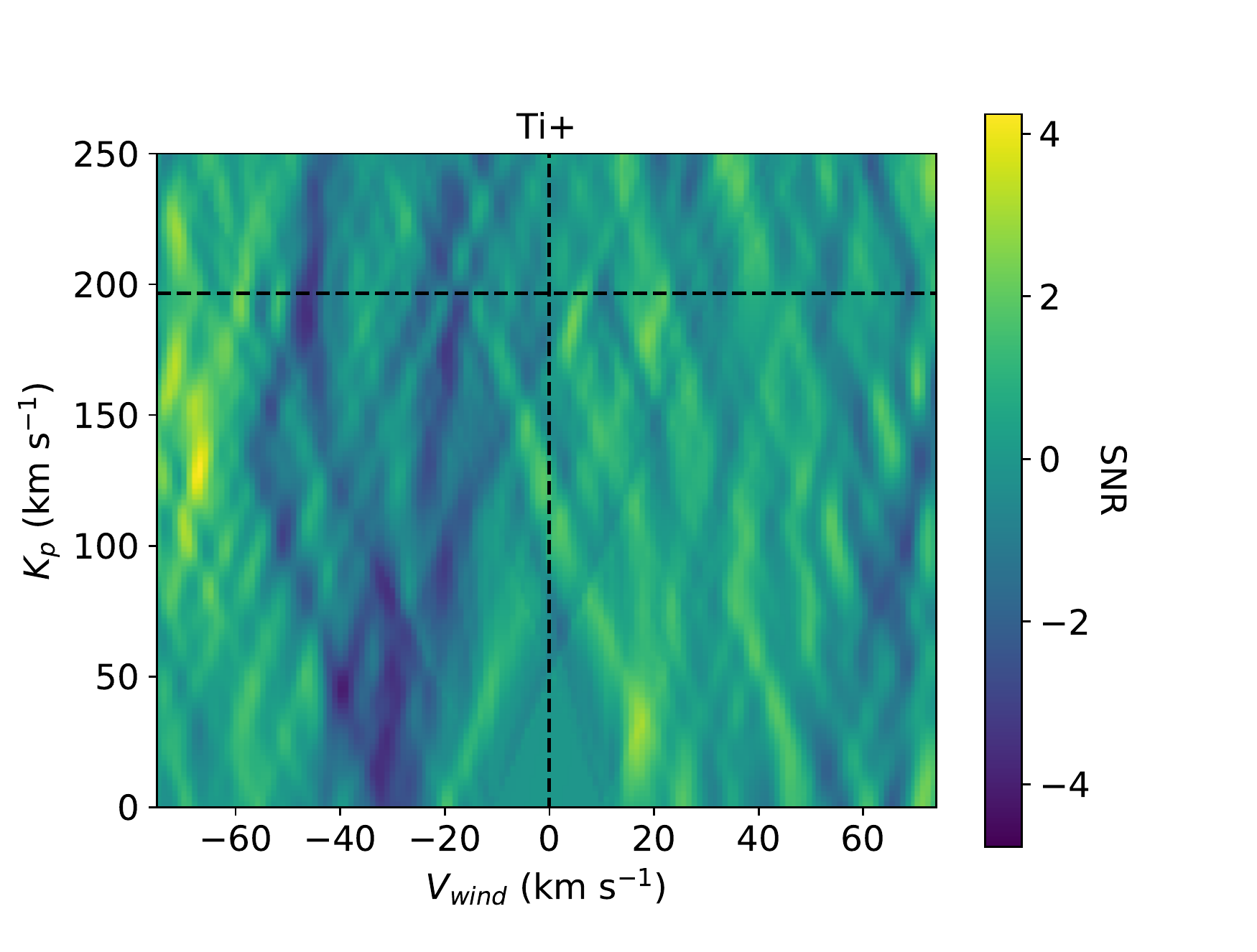}
\includegraphics[width=0.32\linewidth]{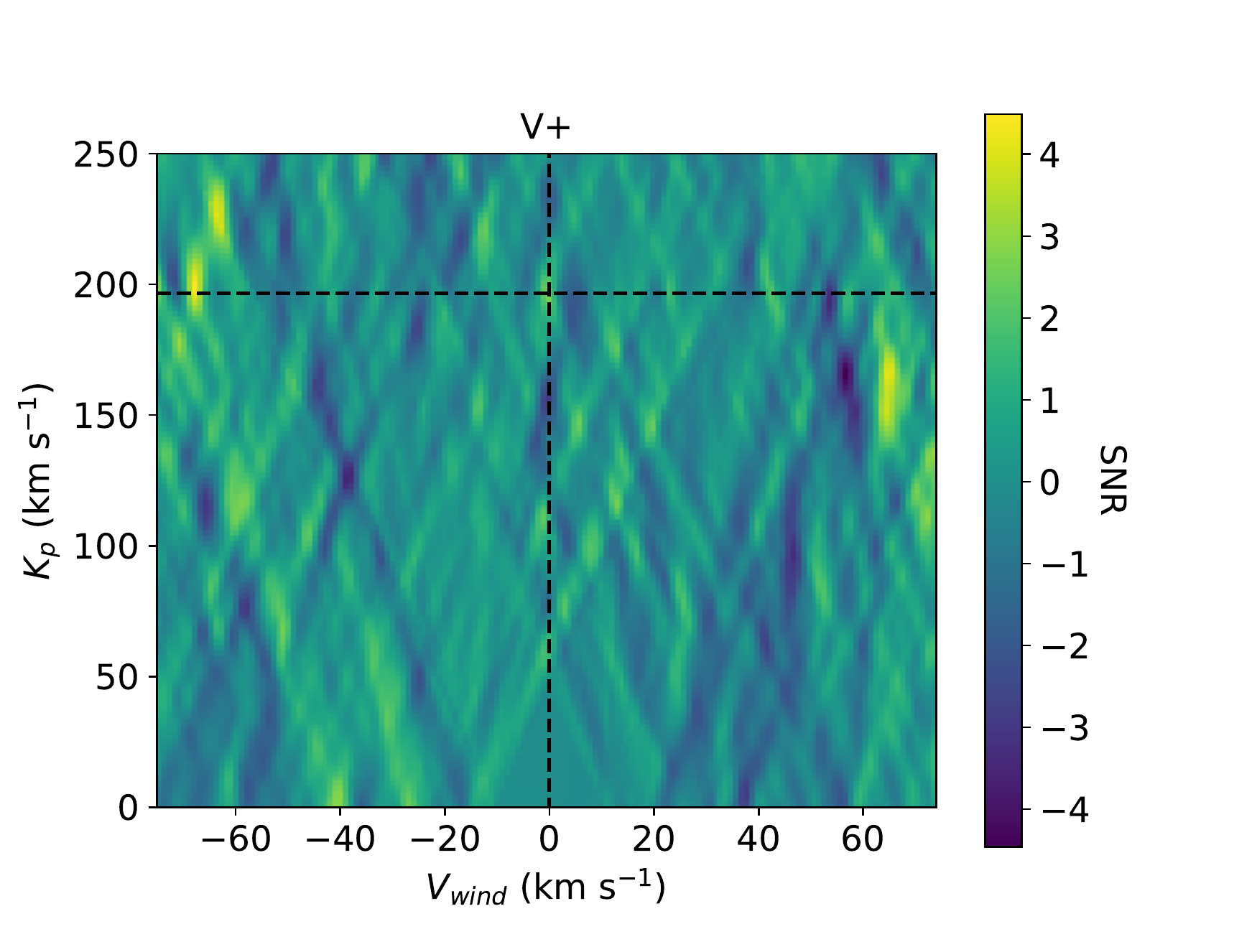}
\includegraphics[width=0.32\linewidth]{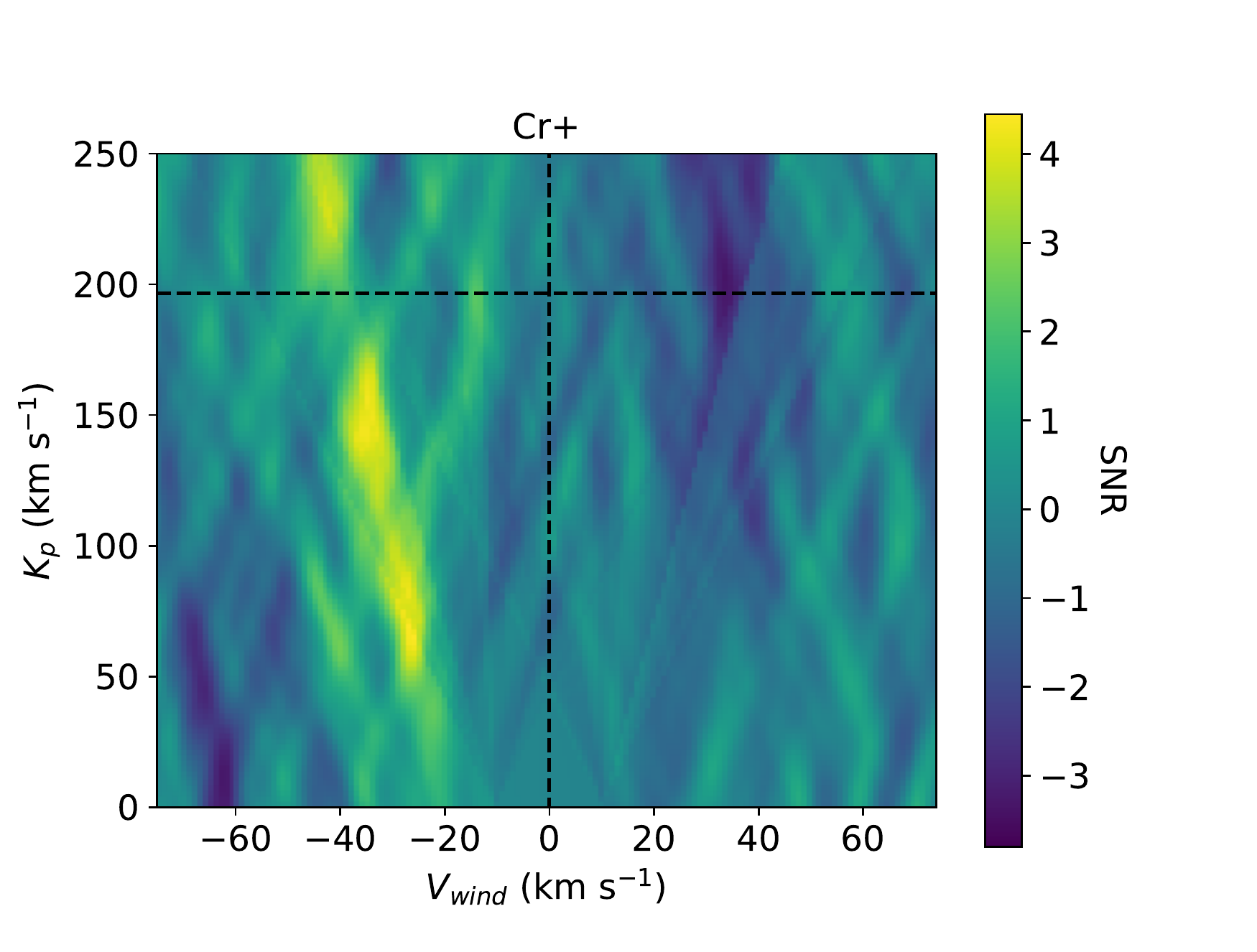}
\includegraphics[width=0.32\linewidth]{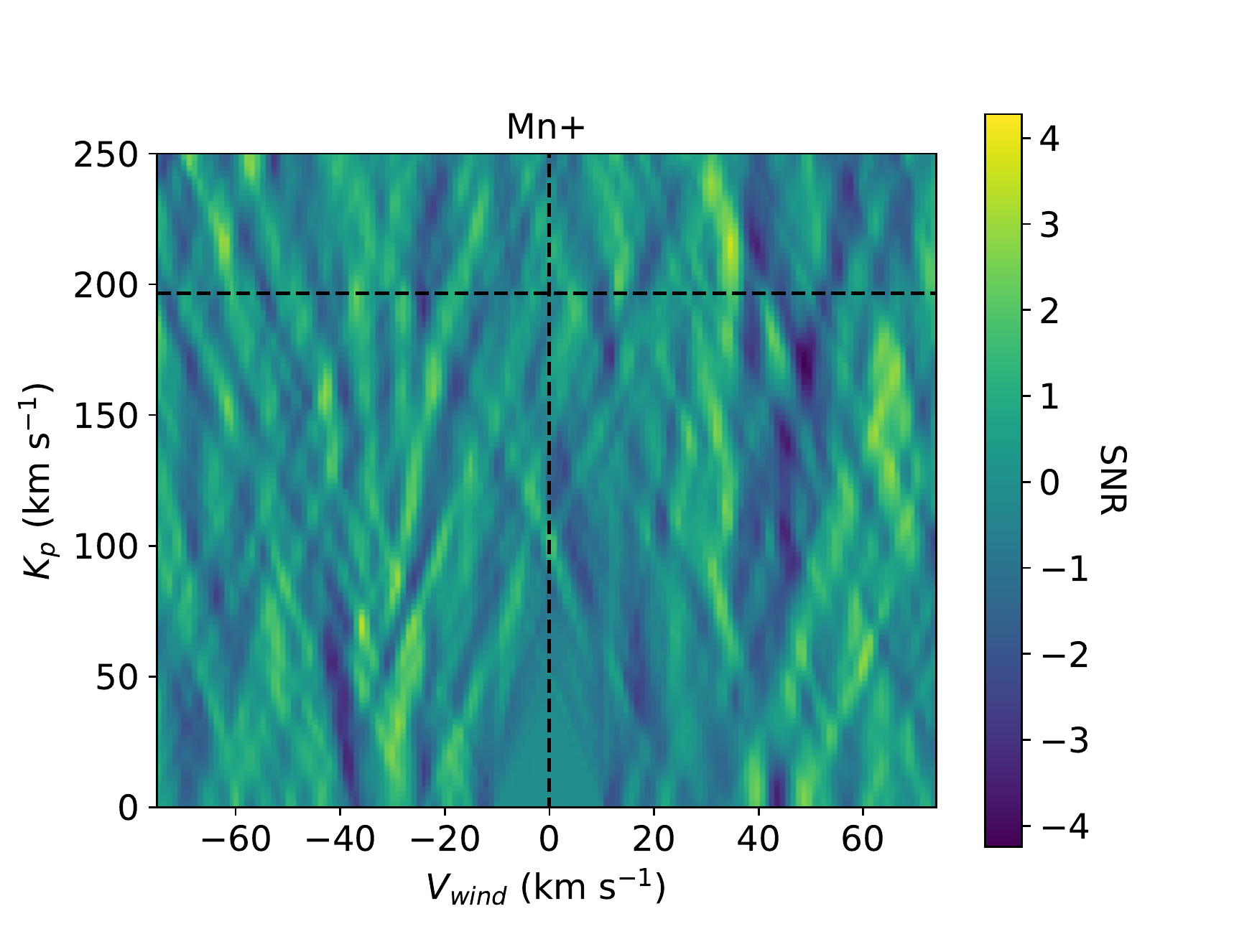}
\includegraphics[width=0.32\linewidth]{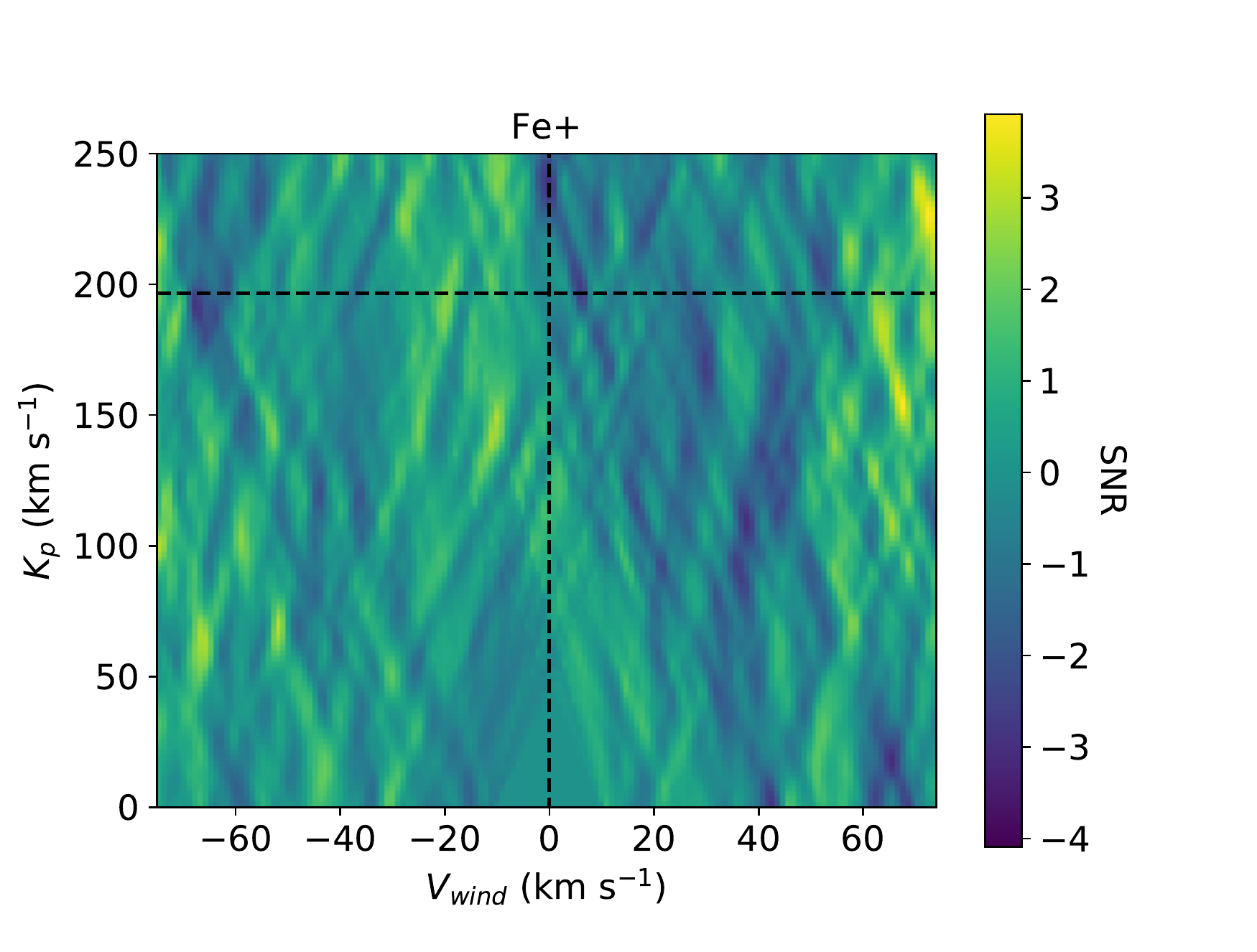}
\includegraphics[width=0.32\linewidth]{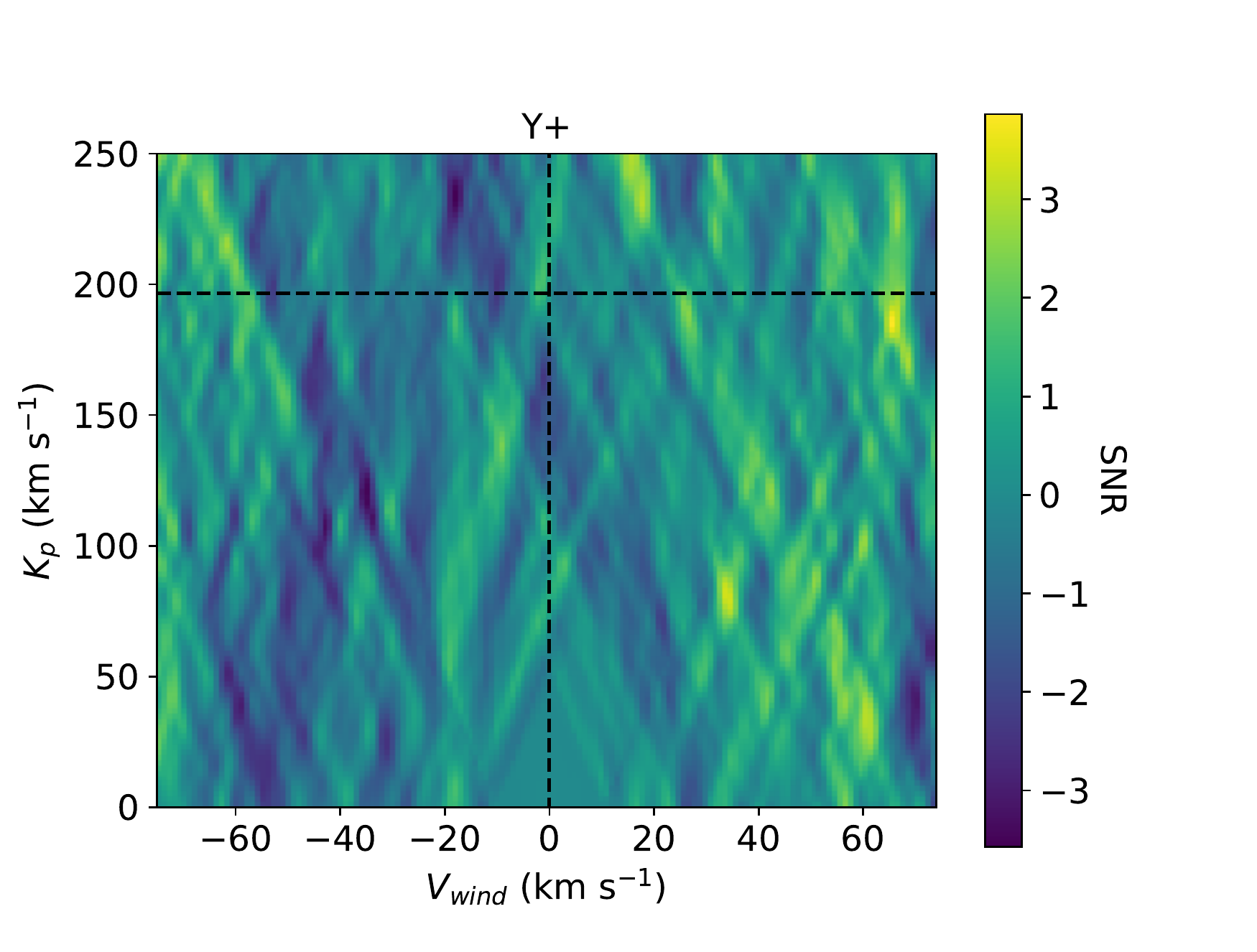}
\includegraphics[width=0.32\linewidth]{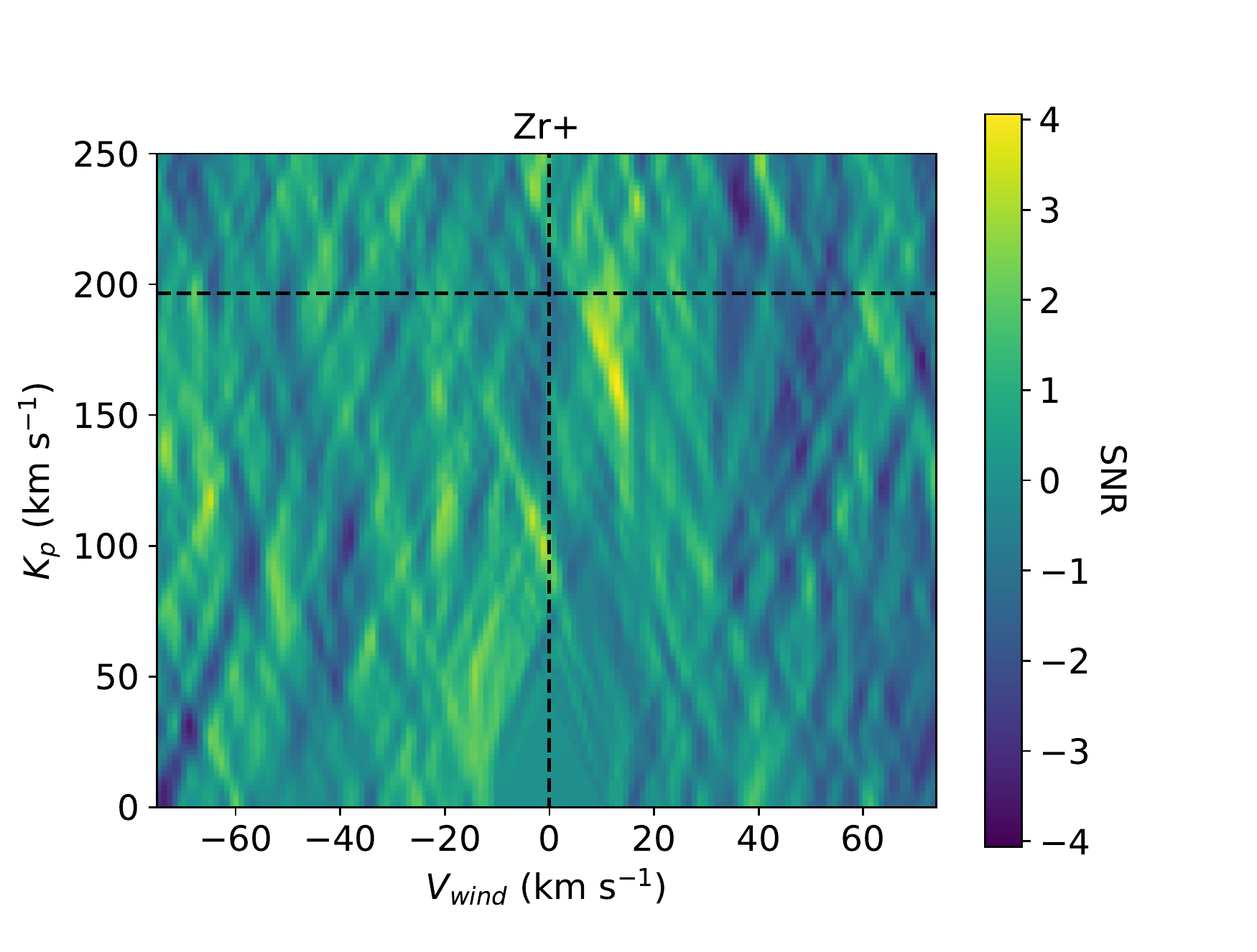}
\caption{\small 
Same as Figure \ref{f:cc_nodetect_atom}, but for the non-detections of each ion.  }
\label{f:cc_nodetect_ion}
\end{center}
\end{figure*}

We plot all of the observability scores, as defined in Section \ref{s:atoms} and shown in Figure \ref{f:periodic}, as well as our detections and non-detections in Figure \ref{f:periodic_compare}. For each non-detection, we did not find a peak with an SNR $>$ 4 near the exoplanet's expected velocity, and Figures \ref{f:cc_nodetect_atom} (neutral atoms) and \ref{f:cc_nodetect_ion} (ions) show the resulting $K_p$ versus $v_{wind}$ parameter space search. Most of the species that have high observability scores in Figure \ref{f:periodic} are detected. The only species that are not detected or tentatively detected that have an observability score above 0.25 that we do not find are: Ca I, Ti I, Ti II, Sc I, Sc II, Al I, and Zr I. Although we do not detect Ca I, we do detect Ca II, which leads us to conclude that Ca is mostly ionized at the temperatures and pressures probed in the terminator of WASP-76b. 

Since we expect to be able to detect both Ti I and II, as well as Sc I and II, these non-detections cannot be explained by ionization. Sc I and II are only on the edge of what we expect to be able to detect, and so it is not overly surprising that we do not find these species. In contrast, Ti I is expected to be the second most observable atom or ion after Fe I. Ti II has been observed in the hottest UHJ, Kelt-9b \citep{Hoeijmakers2019}, but neither Ti I nor II has been confidently detected in any of the other cooler UHJs, including WASP-121b \citep{Hoeijmakers2020, Merritt2021} and MASCARA-2b \citep{Hoeijmakers2020M2}. Furthermore, TiO was also not detected in WASP-121b \citep{Hoeijmakers2020, Merritt2020} or previously in WASP-76b \citep{Tabernero2021}. Ti I, Ti II, and TiO make up the main gas-phase Ti-bearing species for the range of temperatures (2000$-$4000 K) expected along the terminator of WASP-76b \citep{Lodders2002}, and as none of them are detected, this supports the hypothesis of Ti being trapped in condensates \citep[e.g.,][]{Spiegel2009, Parmentier2013}. 

Since we do not expect to measure ionized Al and Zr, a significant portion of both of these species may be ionized, and therefore the neutral atoms may be less observable than our estimates and explain the non-detections. Alternatively, Al could have condensed into clouds; Al$_2$O$_3$ condenses at the highest temperature ($\sim$2000 K) of all of the common condensate species shown in \citet{Gao2021}. As we will discuss later when we compare our results to recent GCM modeling results (Section \ref{s:modelcompare}), Al$_2$O$_3$ clouds are invoked in \citet{Savel2021} to explain the Fe I asymmetry. We recover strong detections of Mg and Li even though these species are expected to be of similar or even lower observability, and so our non-detection of the strong Al I doublet at 4000 \AA\ could be evidence supporting the cloud hypothesis put forth by \citet{Savel2021}. In addition to Al$_2$O$_3$ clouds, the cloud species with the second highest condensation temperature in \citet{Gao2021} is TiO$_2$, and so condensation could potentially explain the intriguing non-detections of both Ti and Al.

\subsection{Comparisons to previous observations of WASP-76b}

Three studies were recently published using HST data of WASP-76b. \citet{vonEssen2020} found evidence for Na in the optical, and a slope of increasing transit depth with decreasing wavelength. \citet{Fu2020} also detected this slope in the optical and argued that the slope is due to metal absorption and not Rayleigh scattering. Our detections of Fe I, V I, Cr I, Mn I, Co I, and Ni I support this hypothesis as all of these atoms strongly absorb between 0.2 and 0.4 $\mu$m, where the HST spectrum showed an increased slope. Both \citet{Fu2020} and \citet{Edwards2020} found evidence for TiO and/or VO. \citet{Hoeijmakers2020} argued that in WASP-121b their detection of V I suggests that VO would also be present, as the two are expected to coexist in chemical equilibrium. Our detection of V I here may also therefore suggest the presence of VO, but our non-detections of Ti I, Ti II, and the non-detection of TiO at high resolution by \citet{Tabernero2021} do not support the existence of TiO. We suggest that the increased opacity seen between 0.4 and 1.0 $\mu$m could be due to a combination of H$^{-}$ opacity \citep[e.g,][]{Lothringer2018, Rathcke2021}, other metal oxides or hydrides (VO, CaH, etc.) and absorption from metals such as Na I that absorb strongly in this region.

At high spectral resolution, \citet{Casasayas2021b} and \citet{Deibert2021} both published detections of the Ca II infrared triplet. \citet{Casasayas2021b} found a broad signal with an FWHM of $25\substack{+5\\-4}$ km s$^{-1}$ and tentative evidence for a double-peaked structure in the 1D cross correlation function. In their $K_p$ versus $v_{wind}$ diagram, they found a peak at 196 km s$^{-1}$, with an extended tail down to lower $K_p$ values. We also see a double-peaked structure in both the 1D cross correlation function and the $K_p$ versus $v_{wind}$ diagram, and all our measured values of the Ca II H and K lines are consistent within 1-$\sigma$. Furthermore, our separate analysis shows nearly identical results as the original Fe I signal seen in \citet{Ehrenreich2020}, and consistent results with the HARPS signal seen by \citet{Kesseli2021}. Our Na I detection shows a similar broad shape and velocity offset as what was measured in \citet{Seidel2019, Seidel2021}. Finally we detect or tentatively detect all of the species that \citet{Tabernero2021} were able to identify as single lines (Ca II, Mn I, Fe I, Mg I, Na I, Li I, and K I), thus validating our cross correlation methods. The SNRs of our detections are comparable to those from \citet{Tabernero2021} for the species that are dominated by singlet or doublet absorption (e.g., we measure SNR=7.8 while \citet{Tabernero2021} measures an average SNR=7.4 for Ca II), but for species that absorb with a forest of lines our cross correlation analysis is able to detect a higher SNR (e.g., SNR=5.7 for the single Fe I line at 4403 \AA\ in \citet{Tabernero2021} versus SNR=14.24 for our cross correlation analysis of Fe I).

Recent high-resolution studies in the NIR have added more evidence for molecular absorption in WASP-76b. \citet{Kesseli2020} was unable to detect FeH, but \citet{SanchezLopez2021} confirmed the presence of H$_2$O that was seen with HST \citep{Tsiaras2018, Edwards2020, Fu2020} and also found evidence for HCN and tentative evidence for NH$_3$. \citet{Landman2021} detected OH, which demonstrated that H$_2$O was dissociating in the upper atmosphere. Curiously, \citet{SanchezLopez2021} and \citet{Landman2021} found the peak $K_p$ for their signals to occur around 230$-$250 km s$^{-1}$, which is significantly more than the expected $K_p$ of 196.5. While the majority of our signals have peak $K_p$ values less than 196.5, as is expected for condensing species, Na I, Li I, and H I all have peak $K_p$ values $\geq200$. While the signal for H I is tentative, if the offset is shown to be real, this might demonstrate that H I, OH, H$_2$O, HCN, and NH$_3$ are colocated and exhibit similar behavior in the atmosphere of WASP-76b. More detailed work on H I absorption in WASP-76b as well as NIR follow-up of the intriguing signals seen in \citet{SanchezLopez2021} would help to determine the nature of this potential $K_p$ offset.

Finally, recent $Spitzer$ phase curve observations and accompanying GCM from \citet{May2021} show evidence for a large day-to-night contrast and hence favor a cold-interior model. The large day-to-night contrast is necessary to generate the high-speed winds and velocity shifts we measure at high spectral resolution \citep{Savel2021}. \citet{May2021} found little offset between the substellar point and the hottest point on the planet and therefore found that the morning and evening limbs should have similar temperatures. By incorporating magnetic drag into their GCMs, \citet{Beltz2021} were able to physically explain the minimal hotspot offset from the substellar point. However, our observations and GCM modeling of the high-spectral resolution observations (see Section \ref{s:modelcompare}) require a strong morning-evening temperature asymmetry. The significantly larger amplitudes that we find in the cross correlation during the second half of the transit for all of the transition metals point toward a hotter atmosphere on the eveningside of the planet. We hypothesize that the larger amplitudes during the first half of the transit for Na I and K I indicate that the morningside of the planet is cooler. Equilibrium chemistry predicts that at temperatures of 3000 K, most Na and K atoms are ionized, and so with cooler temperatures, the abundances of Na I and K I would increase and hence show stronger absorption signatures.
 
\subsection{Comparisons to GCM modeling results}
\label{s:modelcompare}

Two papers have recently performed in-depth modeling of the atmosphere of WASP-76b to explain the original asymmetric transit signal of Fe I from \citet{Ehrenreich2020} and \citet{Kesseli2021}. \citet{Wardenier2021} found that in order to reproduce the asymmetry and highly blueshifted signal from the eveningside of the planet, a large temperature difference between the cooler morningside and hotter eveningside of the planet or removing iron from the morningside of the planet were required. Both cases would result in a stronger signal that is more blueshifted during the second half of the transit. \citet{Savel2021} also found that a large temperature contrast between the evening and morning limbs, modeled with weak atmospheric drag, and modifications to the general GCM model were required to reproduce the iron signal. However, instead of condensation, they favored the formation of thick clouds (with a slight preference for Al$_2$O$_3$ over Fe clouds) on the cooler morningside of the planet, which effectively blocks much of the Fe I absorption on this side of the planet. 

As V I, Cr I, Mn I, Ni I, and Co I all show similar behavior, the potential explanations from modeling (clouds, condensation, or temperature asymmetries) all provide good explanations for these signals as well. Indeed, \citet{Savel2021} modeled how the radial velocities of the Fe I, Ti I, Mn I, Cr I and Mg I signals varied over the course of the observations and found nearly identical behavior. Alternatively, \citet{Savel2021} found that Ca II does not have this same behavior, and in their models, the peak radial velocity was less blueshifted throughout the observations. In their weak atmospheric drag case (the case favored by the Fe I results), the Ca II absorption was centered at $-2.5$ km s$^{-1}$ for the majority of the transit. We measure the peak radial velocity of Ca II for the second part of the transit to be $-2.52 \pm1.53$ km s$^{-1}$. During the first part of the transit, we also see absorption around $-2.5$ km s$^{-1}$, but in addition, we see stronger absorption closer to $+10$ km s$^{-1}$. \citet{Savel2021} found a slight uptick in the Ca II radial velocity at the beginning of the transit to radial velocities of about 0 km s$^{-1}$, but this is not nearly as redshifted as what we measure. Overall our results align very well with those proposed in \citet{Savel2021}, and any differences in our Ca II radial velocity measurements could be due to non-hydrostatic effects, as \citet{Savel2021} mention. 

\section{Conclusions} 
\label{s:conclusions}

We searched for absorption from $>30$ atoms and ions in the atmosphere of WASP-76b using the cross correlation technique on two archival ESPRESSO transits. We detected absorption from Li I, Na I, Mg I, Ca II, V I, Cr I, Mn I, Fe I, Ni I, and Sr II, and tentatively detect H I, K I, and Co I. The detected absorption from V I, Cr I, Ni I, Sr II and Co I are novel for this planet. 

For each of the detected or tentatively detected atoms and ions, we analyzed the beginning of the transit separately from the end of the transit to measure two different radial velocities and amplitudes of the planet absorption. We measure stronger absorption during the second half of the transit for all of the transition metals, and stronger absorption during the first half of the transit for Na I and K I. This observation points toward a temperature asymmetry between the morning and eveningside of the planet, as Na I and K I are expected to be less abundant (mostly ionized) on the the hotter eveningside of the planet, while the transition metals are more visible on this side of the planet due to the hotter atmosphere, which has a larger scale height.

We find evidence for a trend between the strength of the measured asymmetry (difference between the expected radial velocity at the beginning and end of the transit) and the condensation or ionization temperature. Na I and Li I show consistent radial velocities, while the transition metals show differences of $\sim7$ km s$^{-1}$, and the ions show the largest asymmetries. This could be evidence that rain-out or recombination in the cooler morningside of the planet is causing the asymmetries. Alternatively, the differing degrees of asymmetry could be caused by the different species absorbing at different locations in the atmosphere, as we find correlation between the FWHM of the signals and the strength of the asymmetry. The lower atmosphere could be dominated by a day-to-night wind, which would cause narrower absorption lines with varying Doppler shifts, while the upper atmosphere could be dominated by a vertical wind or outflow, which would cause less phase dependence but broader lines. 

Finally, we compare all of the detection SNRs to what we expected to be able to detect and find a few notable absences. Ti I and Ti II should both be observable, but we find neither, and \citet{Tabernero2021} also did not detect TiO. Furthermore, Al I has a strong doublet at 4000 \AA, which is also not detected. These non-detections could point toward condensation into Al$_2$O$_3$ and TiO$_2$ clouds, as these are both expected to condense at temperatures around 2000 K \citep{Gao2021}. More GCM and chemical modeling, especially disequilibrium chemical modeling, would help to interpret these non-detections, as well as the many new time resolved asymmetric Doppler shifts observed in the atmosphere of WASP-76b. 

\acknowledgements

A.K., I.S., N.C., and A.S. acknowledge funding from the European Research Council (ERC) under the European Union's Horizon 2020 research and innovation program under grant agreement No. 694513. P.M. acknowledges support from the European Research Council under the European Union's Horizon 2020 research and innovation program under grant agreement No. 832428-Origins.
The authors would like to thank the ESPRESSO team for sharing their data, without which we could not perform this work. We would also like to thank Arjun Savel for providing extra insight on recent GCM modeling work of WASP-76b. Finally we would like to thank Karan Molaverdikhani for computing many of the atom and ion opacities available in petitRADTRANS.

\bibliographystyle{aasjournal}
\bibliography{bib.bib}

\begin{thebibliography}{}
\expandafter\ifx\csname natexlab\endcsname\relax\def\natexlab#1{#1}\fi
\providecommand{\url}[1]{\href{#1}{#1}}
\providecommand{\dodoi}[1]{doi:~\href{http://doi.org/#1}{\nolinkurl{#1}}}
\providecommand{\doeprint}[1]{\href{http://ascl.net/#1}{\nolinkurl{http://ascl.net/#1}}}
\providecommand{\doarXiv}[1]{\href{https://arxiv.org/abs/#1}{\nolinkurl{https://arxiv.org/abs/#1}}}

\bibitem[{{Allart} {et~al.}(2017){Allart}, {Lovis}, {Pino}, {Wyttenbach},
  {Ehrenreich}, \& {Pepe}}]{Allart2017}
{Allart}, R., {Lovis}, C., {Pino}, L., {et~al.} 2017, \aap, 606, A144,
  \dodoi{10.1051/0004-6361/201730814}

\bibitem[{{Allart} {et~al.}(2020){Allart}, {Pino}, {Lovis}, {Sousa},
  {Casasayas-Barris}, {Zapatero Osorio}, {Cretignier}, {Palle}, {Pepe},
  {Cristiani}, {Rebolo}, {Santos}, {Borsa}, {Bourrier}, {Demangeon},
  {Ehrenreich}, {Lavie}, {Lendl}, {Lillo-Box}, {Micela}, {Oshagh}, {Sozzetti},
  {Tabernero}, {Adibekyan}, {Allende Prieto}, {Alibert}, {Amate}, {Benz},
  {Bouchy}, {Cabral}, {Dekker}, {D'Odorico}, {Di Marcantonio}, {Dumusque},
  {Figueira}, {Genova Santos}, {Gonz{\'a}lez Hern{\'a}ndez}, {Lo Curto},
  {Manescau}, {Martins}, {M{\'e}gevand}, {Mehner}, {Molaro}, {Nunes},
  {Poretti}, {Riva}, {Su{\'a}rez Mascare{\~n}o}, {Udry}, \&
  {Zerbi}}]{Allart2020}
{Allart}, R., {Pino}, L., {Lovis}, C., {et~al.} 2020, \aap, 644, A155,
  \dodoi{10.1051/0004-6361/202039234}

\bibitem[{{Asplund} {et~al.}(2009){Asplund}, {Grevesse}, {Sauval}, \&
  {Scott}}]{Asplund2009}
{Asplund}, M., {Grevesse}, N., {Sauval}, A.~J., \& {Scott}, P. 2009, \araa, 47,
  481, \dodoi{10.1146/annurev.astro.46.060407.145222}

\bibitem[{{Beltz} {et~al.}(2022){Beltz}, {Rauscher}, {Roman}, \&
  {Guilliat}}]{Beltz2021}
{Beltz}, H., {Rauscher}, E., {Roman}, M.~T., \& {Guilliat}, A. 2022, \aj, 163,
  35, \dodoi{10.3847/1538-3881/ac3746}

\bibitem[{{Ben-Yami} {et~al.}(2020){Ben-Yami}, {Madhusudhan}, {Cabot},
  {Constantinou}, {Piette}, {Gandhi}, \& {Welbanks}}]{Ben-Yami2020}
{Ben-Yami}, M., {Madhusudhan}, N., {Cabot}, S. H.~C., {et~al.} 2020, \apjl,
  897, L5, \dodoi{10.3847/2041-8213/ab94aa}

\bibitem[{{Borsa} {et~al.}(2021){Borsa}, {Allart}, {Casasayas-Barris},
  {Tabernero}, {Zapatero Osorio}, {Cristiani}, {Pepe}, {Rebolo}, {Santos},
  {Adibekyan}, {Bourrier}, {Demangeon}, {Ehrenreich}, {Pall{\'e}}, {Sousa},
  {Lillo-Box}, {Lovis}, {Micela}, {Oshagh}, {Poretti}, {Sozzetti}, {Allende
  Prieto}, {Alibert}, {Amate}, {Benz}, {Bouchy}, {Cabral}, {Dekker},
  {D'Odorico}, {Di Marcantonio}, {Figueira}, {Genova Santos}, {Gonz{\'a}lez
  Hern{\'a}ndez}, {Lo Curto}, {Manescau}, {Martins}, {M{\'e}gevand}, {Mehner},
  {Molaro}, {Nunes}, {Riva}, {Su{\'a}rez Mascare{\~n}o}, {Udry}, \&
  {Zerbi}}]{Borsa2021}
{Borsa}, F., {Allart}, R., {Casasayas-Barris}, N., {et~al.} 2021, \aap, 645,
  A24, \dodoi{10.1051/0004-6361/202039344}

\bibitem[{{Casasayas-Barris} {et~al.}(2019){Casasayas-Barris}, {Pall{\'e}},
  {Yan}, {Chen}, {Kohl}, {Stangret}, {Parviainen}, {Helling}, {Watanabe},
  {Czesla}, {Fukui}, {Monta{\~n}{\'e}s-Rodr{\'\i}guez}, {Nagel}, {Narita},
  {Nortmann}, {Nowak}, {Schmitt}, \& {Zapatero Osorio}}]{Casasayas2019}
{Casasayas-Barris}, N., {Pall{\'e}}, E., {Yan}, F., {et~al.} 2019, \aap, 628,
  A9, \dodoi{10.1051/0004-6361/201935623}

\bibitem[{{Casasayas-Barris} {et~al.}(2021{\natexlab{a}}){Casasayas-Barris},
  {Palle}, {Stangret}, {Bourrier}, {Tabernero}, {Yan}, {Borsa}, {Allart},
  {Zapatero Osorio}, {Lovis}, {Sousa}, {Chen}, {Oshagh}, {Santos}, {Pepe},
  {Rebolo}, {Molaro}, {Cristiani}, {Adibekyan}, {Alibert}, {Allende Prieto},
  {Bouchy}, {Demangeon}, {Di Marcantonio}, {D'Odorico}, {Ehrenreich},
  {Figueira}, {G{\'e}nova Santos}, {Gonz{\'a}lez Hern{\'a}ndez}, {Lavie},
  {Lillo-Box}, {Lo Curto}, {Martins}, {Mehner}, {Micela}, {Nunes}, {Poretti},
  {Sozzetti}, {Su{\'a}rez Mascare{\~n}o}, \& {Udry}}]{Casasayas2021}
{Casasayas-Barris}, N., {Palle}, E., {Stangret}, M., {et~al.}
  2021{\natexlab{a}}, \aap, 647, A26, \dodoi{10.1051/0004-6361/202039539}

\bibitem[{{Casasayas-Barris} {et~al.}(2021{\natexlab{b}}){Casasayas-Barris},
  {Orell-Miquel}, {Stangret}, {Nortmann}, {Yan}, {Oshagh}, {Palle},
  {Sanz-Forcada}, {L{\'o}pez-Puertas}, {Nagel}, {Luque}, {Morello}, {Snellen},
  {Zechmeister}, {Quirrenbach}, {Caballero}, {Ribas}, {Reiners}, {Amado},
  {Bergond}, {Czesla}, {Henning}, {Khalafinejad}, {Molaverdikhani}, {Montes},
  {Perger}, {S{\'a}nchez-L{\'o}pez}, \& {Sedaghati}}]{Casasayas2021b}
{Casasayas-Barris}, N., {Orell-Miquel}, J., {Stangret}, M., {et~al.}
  2021{\natexlab{b}}, \aap, 654, A163, \dodoi{10.1051/0004-6361/202141669}

\bibitem[{{Deibert} {et~al.}(2021){Deibert}, {de Mooij}, {Jayawardhana},
  {Turner}, {Ridden-Harper}, {Fossati}, {Hood}, {Fortney}, {Flagg},
  {MacDonald}, {Allart}, \& {Sing}}]{Deibert2021}
{Deibert}, E.~K., {de Mooij}, E. J.~W., {Jayawardhana}, R., {et~al.} 2021,
  \apjl, 919, L15, \dodoi{10.3847/2041-8213/ac2513}

\bibitem[{{Delrez} {et~al.}(2016){Delrez}, {Santerne}, {Almenara}, {Anderson},
  {Collier-Cameron}, {D{\'\i}az}, {Gillon}, {Hellier}, {Jehin}, {Lendl},
  {Maxted}, {Neveu-VanMalle}, {Pepe}, {Pollacco}, {Queloz}, {S{\'e}gransan},
  {Smalley}, {Smith}, {Triaud}, {Udry}, {Van Grootel}, \& {West}}]{Delrez2016}
{Delrez}, L., {Santerne}, A., {Almenara}, J.~M., {et~al.} 2016, \mnras, 458,
  4025, \dodoi{10.1093/mnras/stw522}

\bibitem[{{Dobbs-Dixon} \& {Agol}(2013)}]{Dobbs2013}
{Dobbs-Dixon}, I., \& {Agol}, E. 2013, \mnras, 435, 3159,
  \dodoi{10.1093/mnras/stt1509}

\bibitem[{{Edwards} {et~al.}(2020){Edwards}, {Changeat}, {Baeyens}, {Tsiaras},
  {Al-Refaie}, {Taylor}, {Yip}, {Bieger}, {Blain}, {Gressier}, {Guilluy},
  {Jaziri}, {Kiefer}, {Modirrousta-Galian}, {Morvan}, {Mugnai}, {Pluriel},
  {Poveda}, {Skaf}, {Whiteford}, {Wright}, {Zingales}, {Charnay}, {Drossart},
  {Leconte}, {Venot}, {Waldmann}, \& {Beaulieu}}]{Edwards2020}
{Edwards}, B., {Changeat}, Q., {Baeyens}, R., {et~al.} 2020, \aj, 160, 8,
  \dodoi{10.3847/1538-3881/ab9225}

\bibitem[{{Ehrenreich} {et~al.}(2020){Ehrenreich}, {Lovis}, {Allart}, {Zapatero
  Osorio}, {Pepe}, {Cristiani}, {Rebolo}, {Santos}, {Borsa}, {Demangeon},
  {Dumusque}, {Gonz{\'a}lez Hern{\'a}ndez}, {Casasayas-Barris},
  {S{\'e}gransan}, {Sousa}, {Abreu}, {Adibekyan}, {Affolter}, {Allende Prieto},
  {Alibert}, {Aliverti}, {Alves}, {Amate}, {Avila}, {Baldini}, {Bandy}, {Benz},
  {Bianco}, {Bolmont}, {Bouchy}, {Bourrier}, {Broeg}, {Cabral}, {Calderone},
  {Pall{\'e}}, {Cegla}, {Cirami}, {Coelho}, {Conconi}, {Coretti}, {Cumani},
  {Cupani}, {Dekker}, {Delabre}, {Deiries}, {D'Odorico}, {Di Marcantonio},
  {Figueira}, {Fragoso}, {Genolet}, {Genoni}, {G{\'e}nova Santos}, {Hara},
  {Hughes}, {Iwert}, {Kerber}, {Knudstrup}, {Land oni}, {Lavie}, {Lizon},
  {Lendl}, {Lo Curto}, {Maire}, {Manescau}, {Martins}, {M{\'e}gevand },
  {Mehner}, {Micela}, {Modigliani}, {Molaro}, {Monteiro}, {Monteiro},
  {Moschetti}, {M{\"u}ller}, {Nunes}, {Oggioni}, {Oliveira}, {Pariani},
  {Pasquini}, {Poretti}, {Rasilla}, {Redaelli}, {Riva}, {Santana Tschudi},
  {Santin}, {Santos}, {Segovia Milla}, {Seidel}, {Sosnowska}, {Sozzetti},
  {Span{\`o}}, {Su{\'a}rez Mascare{\~n}o}, {Tabernero}, {Tenegi}, {Udry},
  {Zanutta}, \& {Zerbi}}]{Ehrenreich2020}
{Ehrenreich}, D., {Lovis}, C., {Allart}, R., {et~al.} 2020, \nat, 580, 597,
  \dodoi{10.1038/s41586-020-2107-1}

\bibitem[{{Espinoza} \& {Jones}(2021)}]{Espinoza2021}
{Espinoza}, N., \& {Jones}, K. 2021, \aj, 162, 165,
  \dodoi{10.3847/1538-3881/ac134d}

\bibitem[{{Fu} {et~al.}(2021){Fu}, {Deming}, {Lothringer}, {Nikolov}, {Sing},
  {Kempton}, {Ih}, {Evans}, {Stevenson}, {Wakeford}, {Rodriguez}, {Eastman},
  {Stassun}, {Henry}, {L{\'o}pez-Morales}, {Lendl}, {Conti}, {Stockdale},
  {Collins}, {Kielkopf}, {Barstow}, {Sanz-Forcada}, {Ehrenreich}, {Bourrier},
  \& {dos Santos}}]{Fu2020}
{Fu}, G., {Deming}, D., {Lothringer}, J., {et~al.} 2021, \aj, 162, 108,
  \dodoi{10.3847/1538-3881/ac1200}

\bibitem[{{Gao} {et~al.}(2021){Gao}, {Wakeford}, {Moran}, \&
  {Parmentier}}]{Gao2021}
{Gao}, P., {Wakeford}, H.~R., {Moran}, S.~E., \& {Parmentier}, V. 2021, Journal
  of Geophysical Research (Planets), 126, e06655, \dodoi{10.1029/2020JE006655}

\bibitem[{{Gao} {et~al.}(2020){Gao}, {Thorngren}, {Lee}, {Fortney}, {Morley},
  {Wakeford}, {Powell}, {Stevenson}, \& {Zhang}}]{Gao2020}
{Gao}, P., {Thorngren}, D.~P., {Lee}, E. K.~H., {et~al.} 2020, Nature
  Astronomy, 4, 951, \dodoi{10.1038/s41550-020-1114-3}

\bibitem[{{Gaudi} {et~al.}(2017){Gaudi}, {Stassun}, {Collins}, {Beatty},
  {Zhou}, {Latham}, {Bieryla}, {Eastman}, {Siverd}, {Crepp}, {Gonzales},
  {Stevens}, {Buchhave}, {Pepper}, {Johnson}, {Colon}, {Jensen}, {Rodriguez},
  {Bozza}, {Novati}, {D'Ago}, {Dumont}, {Ellis}, {Gaillard}, {Jang-Condell},
  {Kasper}, {Fukui}, {Gregorio}, {Ito}, {Kielkopf}, {Manner}, {Matt}, {Narita},
  {Oberst}, {Reed}, {Scarpetta}, {Stephens}, {Yeigh}, {Zambelli}, {Fulton},
  {Howard}, {James}, {Penny}, {Bayliss}, {Curtis}, {Depoy}, {Esquerdo},
  {Gould}, {Joner}, {Kuhn}, {Labadie-Bartz}, {Lund}, {Marshall}, {McLeod},
  {Pogge}, {Relles}, {Stockdale}, {Tan}, {Trueblood}, \&
  {Trueblood}}]{Gaudi2017}
{Gaudi}, B.~S., {Stassun}, K.~G., {Collins}, K.~A., {et~al.} 2017, \nat, 546,
  514, \dodoi{10.1038/nature22392}

\bibitem[{{Grimm} \& {Heng}(2015)}]{Grimm2015}
{Grimm}, S.~L., \& {Heng}, K. 2015, \apj, 808, 182,
  \dodoi{10.1088/0004-637X/808/2/182}

\bibitem[{{Grimm} {et~al.}(2021){Grimm}, {Malik}, {Kitzmann},
  {Guzm{\'a}n-Mesa}, {Hoeijmakers}, {Fisher}, {Mendon{\c{c}}a}, {Yurchenko},
  {Tennyson}, {Alesina}, {Buchschacher}, {Burnier}, {Segransan}, {Kurucz}, \&
  {Heng}}]{Grimm2021}
{Grimm}, S.~L., {Malik}, M., {Kitzmann}, D., {et~al.} 2021, \apjs, 253, 30,
  \dodoi{10.3847/1538-4365/abd773}

\bibitem[{{Hoeijmakers} {et~al.}(2018){Hoeijmakers}, {Ehrenreich}, {Heng},
  {Kitzmann}, {Grimm}, {Allart}, {Deitrick}, {Wyttenbach}, {Oreshenko}, {Pino},
  {Rimmer}, {Molinari}, \& {Di Fabrizio}}]{Hoeijmakers2018}
{Hoeijmakers}, H.~J., {Ehrenreich}, D., {Heng}, K., {et~al.} 2018, \nat, 560,
  453, \dodoi{10.1038/s41586-018-0401-y}

\bibitem[{{Hoeijmakers} {et~al.}(2019){Hoeijmakers}, {Ehrenreich}, {Kitzmann},
  {Allart}, {Grimm}, {Seidel}, {Wyttenbach}, {Pino}, {Nielsen}, {Fisher},
  {Rimmer}, {Bourrier}, {Cegla}, {Lavie}, {Lovis}, {Patzer}, {Stock}, {Pepe},
  \& {Heng}}]{Hoeijmakers2019}
{Hoeijmakers}, H.~J., {Ehrenreich}, D., {Kitzmann}, D., {et~al.} 2019, \aap,
  627, A165, \dodoi{10.1051/0004-6361/201935089}

\bibitem[{{Hoeijmakers} {et~al.}(2020{\natexlab{a}}){Hoeijmakers}, {Seidel},
  {Pino}, {Kitzmann}, {Sindel}, {Ehrenreich}, {Oza}, {Bourrier}, {Allart},
  {Gebek}, {Lovis}, {Yurchenko}, {Astudillo-Defru}, {Bayliss}, {Cegla},
  {Lavie}, {Lendl}, {Melo}, {Murgas}, {Nascimbeni}, {Pepe}, {S{\'e}gransan},
  {Udry}, {Wyttenbach}, \& {Heng}}]{Hoeijmakers2020}
{Hoeijmakers}, H.~J., {Seidel}, J.~V., {Pino}, L., {et~al.} 2020{\natexlab{a}},
  \aap, 641, A123, \dodoi{10.1051/0004-6361/202038365}

\bibitem[{{Hoeijmakers} {et~al.}(2020{\natexlab{b}}){Hoeijmakers}, {Cabot},
  {Zhao}, {Buchhave}, {Tronsgaard}, {Davis}, {Kitzmann}, {Grimm}, {Cegla},
  {Bourrier}, {Ehrenreich}, {Heng}, {Lovis}, \& {Fischer}}]{Hoeijmakers2020M2}
{Hoeijmakers}, H.~J., {Cabot}, S. H.~C., {Zhao}, L., {et~al.}
  2020{\natexlab{b}}, \aap, 641, A120, \dodoi{10.1051/0004-6361/202037437}

\bibitem[{{Kataria} {et~al.}(2016){Kataria}, {Sing}, {Lewis}, {Visscher},
  {Showman}, {Fortney}, \& {Marley}}]{Kataria2016}
{Kataria}, T., {Sing}, D.~K., {Lewis}, N.~K., {et~al.} 2016, \apj, 821, 9,
  \dodoi{10.3847/0004-637X/821/1/9}

\bibitem[{{Kesseli} \& {Snellen}(2021)}]{Kesseli2021}
{Kesseli}, A.~Y., \& {Snellen}, I.~A.~G. 2021, \apjl, 908, L17,
  \dodoi{10.3847/2041-8213/abe047}

\bibitem[{{Kesseli} {et~al.}(2020){Kesseli}, {Snellen}, {Alonso-Floriano},
  {Molli{\`e}re}, \& {Serindag}}]{Kesseli2020}
{Kesseli}, A.~Y., {Snellen}, I.~A.~G., {Alonso-Floriano}, F.~J.,
  {Molli{\`e}re}, P., \& {Serindag}, D.~B. 2020, \aj, 160, 228,
  \dodoi{10.3847/1538-3881/abb59c}

\bibitem[{{Kramida} {et~al.}(2019){Kramida}, {Ralchenko}, {Reader}, \& {NIST
  ASD Team}}]{Kramida2019}
{Kramida}, A., {Ralchenko}, Y., {Reader}, J., \& {NIST ASD Team}. 2019, NIST
  Atomic Spectra Database (ver. 5.7.1), [Online]., National Institute of
  Standards and Technology, Gaithersburg, MD.,
  \dodoi{https://doi.org/10.18434/T4W30F}

\bibitem[{{Kurucz}(2018)}]{Kurucz2018}
{Kurucz}, R.~L. 2018, in Astronomical Society of the Pacific Conference Series,
  Vol. 515, Workshop on Astrophysical Opacities, 47

\bibitem[{{Landman} {et~al.}(2021){Landman}, {S{\'a}nchez-L{\'o}pez},
  {Molli{\`e}re}, {Kesseli}, {Louca}, \& {Snellen}}]{Landman2021}
{Landman}, R., {S{\'a}nchez-L{\'o}pez}, A., {Molli{\`e}re}, P., {et~al.} 2021,
  \aap, 656, A119, \dodoi{10.1051/0004-6361/202141696}

\bibitem[{{Lodders}(1999)}]{Lodders1999}
{Lodders}, K. 1999, \apj, 519, 793, \dodoi{10.1086/307387}

\bibitem[{{Lodders}(2002)}]{Lodders2002}
---. 2002, \apj, 577, 974, \dodoi{10.1086/342241}

\bibitem[{{Lothringer} {et~al.}(2018){Lothringer}, {Barman}, \&
  {Koskinen}}]{Lothringer2018}
{Lothringer}, J.~D., {Barman}, T., \& {Koskinen}, T. 2018, \apj, 866, 27,
  \dodoi{10.3847/1538-4357/aadd9e}

\bibitem[{{Lothringer} {et~al.}(2020){Lothringer}, {Fu}, {Sing}, \&
  {Barman}}]{Lothringer2020}
{Lothringer}, J.~D., {Fu}, G., {Sing}, D.~K., \& {Barman}, T.~S. 2020, \apjl,
  898, L14, \dodoi{10.3847/2041-8213/aba265}

\bibitem[{{May} {et~al.}(2021){May}, {Komacek}, {Stevenson}, {Kempton}, {Bean},
  {Malik}, {Ih}, {Mansfield}, {Savel}, {Deming}, {Desert}, {Feng}, {Fortney},
  {Kataria}, {Lewis}, {Morley}, {Rauscher}, \& {Showman}}]{May2021}
{May}, E.~M., {Komacek}, T.~D., {Stevenson}, K.~B., {et~al.} 2021, \aj, 162,
  158, \dodoi{10.3847/1538-3881/ac0e30}

\bibitem[{{Mayne} {et~al.}(2014){Mayne}, {Baraffe}, {Acreman}, {Smith},
  {Browning}, {Sk{\r{a}}lid Amundsen}, {Wood}, {Thuburn}, \&
  {Jackson}}]{Mayne2014}
{Mayne}, N.~J., {Baraffe}, I., {Acreman}, D.~M., {et~al.} 2014, \aap, 561, A1,
  \dodoi{10.1051/0004-6361/201322174}

\bibitem[{{Mendon{\c{c}}a} {et~al.}(2016){Mendon{\c{c}}a}, {Grimm},
  {Grosheintz}, \& {Heng}}]{Mendonca2016}
{Mendon{\c{c}}a}, J.~M., {Grimm}, S.~L., {Grosheintz}, L., \& {Heng}, K. 2016,
  \apj, 829, 115, \dodoi{10.3847/0004-637X/829/2/115}

\bibitem[{{Merritt} {et~al.}(2020){Merritt}, {Gibson}, {Nugroho}, {de Mooij},
  {Hooton}, {Matthews}, {McKemmish}, {Mikal-Evans}, {Nikolov}, {Sing}, {Spake},
  \& {Watson}}]{Merritt2020}
{Merritt}, S.~R., {Gibson}, N.~P., {Nugroho}, S.~K., {et~al.} 2020, \aap, 636,
  A117, \dodoi{10.1051/0004-6361/201937409}

\bibitem[{{Merritt} {et~al.}(2021){Merritt}, {Gibson}, {Nugroho}, {de Mooij},
  {Hooton}, {Lothringer}, {Matthews}, {Mikal-Evans}, {Nikolov}, {Sing}, \&
  {Watson}}]{Merritt2021}
---. 2021, \mnras, 506, 3853, \dodoi{10.1093/mnras/stab1878}

\bibitem[{{Molli{\`e}re} \& {Snellen}(2019)}]{Molliere2019isotope}
{Molli{\`e}re}, P., \& {Snellen}, I.~A.~G. 2019, \aap, 622, A139,
  \dodoi{10.1051/0004-6361/201834169}

\bibitem[{{Molli{\`e}re} {et~al.}(2017){Molli{\`e}re}, {van Boekel}, {Bouwman},
  {Henning}, {Lagage}, \& {Min}}]{Molliere2017}
{Molli{\`e}re}, P., {van Boekel}, R., {Bouwman}, J., {et~al.} 2017, \aap, 600,
  A10, \dodoi{10.1051/0004-6361/201629800}

\bibitem[{{Molli{\`e}re} {et~al.}(2019){Molli{\`e}re}, {Wardenier}, {van
  Boekel}, {Henning}, {Molaverdikhani}, \& {Snellen}}]{Molliere2019}
{Molli{\`e}re}, P., {Wardenier}, J.~P., {van Boekel}, R., {et~al.} 2019, \aap,
  627, A67, \dodoi{10.1051/0004-6361/201935470}

\bibitem[{{Nugroho} {et~al.}(2020){Nugroho}, {Gibson}, {de Mooij}, {Watson},
  {Kawahara}, \& {Merritt}}]{Nugroho2020}
{Nugroho}, S.~K., {Gibson}, N.~P., {de Mooij}, E. J.~W., {et~al.} 2020, \mnras,
  496, 504, \dodoi{10.1093/mnras/staa1459}

\bibitem[{{Parmentier} {et~al.}(2013){Parmentier}, {Showman}, \&
  {Lian}}]{Parmentier2013}
{Parmentier}, V., {Showman}, A.~P., \& {Lian}, Y. 2013, \aap, 558, A91,
  \dodoi{10.1051/0004-6361/201321132}

\bibitem[{{Parmentier} {et~al.}(2018){Parmentier}, {Line}, {Bean}, {Mansfield},
  {Kreidberg}, {Lupu}, {Visscher}, {D{\'e}sert}, {Fortney}, {Deleuil},
  {Arcangeli}, {Showman}, \& {Marley}}]{Parmentier2018}
{Parmentier}, V., {Line}, M.~R., {Bean}, J.~L., {et~al.} 2018, \aap, 617, A110,
  \dodoi{10.1051/0004-6361/201833059}

\bibitem[{{Pepe} {et~al.}(2010){Pepe}, {Cristiani}, {Rebolo Lopez}, {Santos},
  {Amorim}, {Avila}, {Benz}, {Bonifacio}, {Cabral}, {Carvas}, {Cirami},
  {Coelho}, {Comari}, {Coretti}, {De Caprio}, {Dekker}, {Delabre}, {Di
  Marcantonio}, {D'Odorico}, {Fleury}, {Garc{\'\i}a}, {Herreros Linares},
  {Hughes}, {Iwert}, {Lima}, {Lizon}, {Lo Curto}, {Lovis}, {Manescau},
  {Martins}, {M{\'e}gevand}, {Moitinho}, {Molaro}, {Monteiro}, {Monteiro},
  {Pasquini}, {Mordasini}, {Queloz}, {Rasilla}, {Rebord{\~a}o}, {Santana
  Tschudi}, {Santin}, {Sosnowska}, {Span{\`o}}, {Tenegi}, {Udry}, {Vanzella},
  {Viel}, {Zapatero Osorio}, \& {Zerbi}}]{Pepe2010}
{Pepe}, F.~A., {Cristiani}, S., {Rebolo Lopez}, R., {et~al.} 2010, in Society
  of Photo-Optical Instrumentation Engineers (SPIE) Conference Series, Vol.
  7735, Ground-based and Airborne Instrumentation for Astronomy III, ed. I.~S.
  {McLean}, S.~K. {Ramsay}, \& H.~{Takami}, 77350F, \dodoi{10.1117/12.857122}

\bibitem[{{Rathcke} {et~al.}(2021){Rathcke}, {MacDonald}, {Barstow}, {Goyal},
  {Lopez-Morales}, {Mendon{\c{c}}a}, {Sanz-Forcada}, {Henry}, {Sing}, {Alam},
  {Lewis}, {Chubb}, {Taylor}, {Nikolov}, \& {Buchhave}}]{Rathcke2021}
{Rathcke}, A.~D., {MacDonald}, R.~J., {Barstow}, J.~K., {et~al.} 2021, \aj,
  162, 138, \dodoi{10.3847/1538-3881/ac0e99}

\bibitem[{{Ryabchikova} {et~al.}(2015){Ryabchikova}, {Piskunov}, {Kurucz},
  {Stempels}, {Heiter}, {Pakhomov}, \& {Barklem}}]{vald3}
{Ryabchikova}, T., {Piskunov}, N., {Kurucz}, R.~L., {et~al.} 2015, \physscr,
  90, 054005, \dodoi{10.1088/0031-8949/90/5/054005}

\bibitem[{{Sanchez-Lopez} {et~al.}(2021){Sanchez-Lopez}, {Landman}, {Molliere},
  {Casasayas-Barris}, {Kesseli}, \& {Snellen}}]{SanchezLopez2021}
{Sanchez-Lopez}, A., {Landman}, R., {Molliere}, P., {et~al.} 2021, submitted to
  A\&A

\bibitem[{{S{\'a}nchez-L{\'o}pez} {et~al.}(2019){S{\'a}nchez-L{\'o}pez},
  {Alonso-Floriano}, {L{\'o}pez-Puertas}, {Snellen}, {Funke}, {Nagel}, {Bauer},
  {Amado}, {Caballero}, {Czesla}, {Nortmann}, {Pall{\'e}}, {Salz}, {Reiners},
  {Ribas}, {Quirrenbach}, {Anglada-Escud{\'e}}, {B{\'e}jar},
  {Casasayas-Barris}, {Galad{\'\i}-Enr{\'\i}quez}, {Guenther}, {Henning},
  {Kaminski}, {K{\"u}rster}, {Lamp{\'o}n}, {Lara}, {Montes}, {Morales},
  {Stangret}, {Tal-Or}, {Sanz-Forcada}, {Schmitt}, {Zapatero Osorio}, \&
  {Zechmeister}}]{SanchezLopez2019}
{S{\'a}nchez-L{\'o}pez}, A., {Alonso-Floriano}, F.~J., {L{\'o}pez-Puertas}, M.,
  {et~al.} 2019, \aap, 630, A53, \dodoi{10.1051/0004-6361/201936084}

\bibitem[{{Savel} {et~al.}(2021){Savel}, {Kempton}, {Malik}, {Komacek}, {Bean},
  {May}, {Stevenson}, {Mansfield}, \& {Rauscher}}]{Savel2021}
{Savel}, A.~B., {Kempton}, E. M.~R., {Malik}, M., {et~al.} 2021, arXiv
  e-prints, arXiv:2109.00163.
\newblock \doarXiv{2109.00163}

\bibitem[{{Seidel} {et~al.}(2019){Seidel}, {Ehrenreich}, {Wyttenbach},
  {Allart}, {Lendl}, {Pino}, {Bourrier}, {Cegla}, {Lovis}, {Barrado},
  {Bayliss}, {Astudillo-Defru}, {Deline}, {Fisher}, {Heng}, {Joseph}, {Lavie},
  {Melo}, {Pepe}, {S{\'e}gransan}, \& {Udry}}]{Seidel2019}
{Seidel}, J.~V., {Ehrenreich}, D., {Wyttenbach}, A., {et~al.} 2019, \aap, 623,
  A166, \dodoi{10.1051/0004-6361/201834776}

\bibitem[{{Seidel} {et~al.}(2021){Seidel}, {Ehrenreich}, {Allart},
  {Hoeijmakers}, {Lovis}, {Bourrier}, {Pino}, {Wyttenbach}, {Adibekyan},
  {Alibert}, {Borsa}, {Casasayas-Barris}, {Cristiani}, {Demangeon}, {Di
  Marcantonio}, {Figueira}, {Gonz{\'a}lez Hern{\'a}ndez}, {Lillo-Box},
  {Martins}, {Mehner}, {Molaro}, {Nunes}, {Palle}, {Pepe}, {Santos}, {Sousa},
  {Sozzetti}, {Tabernero}, \& {Zapatero Osorio}}]{Seidel2021}
{Seidel}, J.~V., {Ehrenreich}, D., {Allart}, R., {et~al.} 2021, \aap, 653, A73,
  \dodoi{10.1051/0004-6361/202140569}

\bibitem[{{Smette} {et~al.}(2015){Smette}, {Sana}, {Noll}, {Horst}, {Kausch},
  {Kimeswenger}, {Barden}, {Szyszka}, {Jones}, {Gallenne}, {Vinther},
  {Ballester}, \& {Taylor}}]{Smette2015}
{Smette}, A., {Sana}, H., {Noll}, S., {et~al.} 2015, \aap, 576, A77,
  \dodoi{10.1051/0004-6361/201423932}

\bibitem[{{Snellen} {et~al.}(2010){Snellen}, {de Kok}, {de Mooij}, \&
  {Albrecht}}]{Snellen2010}
{Snellen}, I. A.~G., {de Kok}, R.~J., {de Mooij}, E. J.~W., \& {Albrecht}, S.
  2010, \nat, 465, 1049, \dodoi{10.1038/nature09111}

\bibitem[{{Spiegel} {et~al.}(2009){Spiegel}, {Silverio}, \&
  {Burrows}}]{Spiegel2009}
{Spiegel}, D.~S., {Silverio}, K., \& {Burrows}, A. 2009, \apj, 699, 1487,
  \dodoi{10.1088/0004-637X/699/2/1487}

\bibitem[{{Stangret} {et~al.}(2020){Stangret}, {Casasayas-Barris}, {Pall{\'e}},
  {Yan}, {S{\'a}nchez-L{\'o}pez}, \& {L{\'o}pez-Puertas}}]{Stangret2020}
{Stangret}, M., {Casasayas-Barris}, N., {Pall{\'e}}, E., {et~al.} 2020, \aap,
  638, A26, \dodoi{10.1051/0004-6361/202037541}

\bibitem[{{Tabernero} {et~al.}(2021){Tabernero}, {Zapatero Osorio}, {Allart},
  {Borsa}, {Casasayas-Barris}, {Demangeon}, {Ehrenreich}, {Lillo-Box}, {Lovis},
  {Pall{\'e}}, {Sousa}, {Rebolo}, {Santos}, {Pepe}, {Cristiani}, {Adibekyan},
  {Allende Prieto}, {Alibert}, {Barros}, {Bouchy}, {Bourrier}, {D'Odorico},
  {Dumusque}, {Faria}, {Figueira}, {G{\'e}nova Santos}, {Gonz{\'a}lez
  Hern{\'a}ndez}, {Hojjatpanah}, {Lo Curto}, {Lavie}, {Martins}, {Martins},
  {Mehner}, {Micela}, {Molaro}, {Nunes}, {Poretti}, {Seidel}, {Sozzetti},
  {Su{\'a}rez Mascare{\~n}o}, {Udry}, {Aliverti}, {Affolter}, {Alves}, {Amate},
  {Avila}, {Bandy}, {Benz}, {Bianco}, {Broeg}, {Cabral}, {Conconi}, {Coelho},
  {Cumani}, {Deiries}, {Dekker}, {Delabre}, {Fragoso}, {Genoni}, {Genolet},
  {Hughes}, {Knudstrup}, {Kerber}, {Landoni}, {Lizon}, {Maire}, {Manescau}, {Di
  Marcantonio}, {M{\'e}gevand}, {Monteiro}, {Monteiro}, {Moschetti}, {Mueller},
  {Modigliani}, {Oggioni}, {Oliveira}, {Pariani}, {Pasquini}, {Rasilla},
  {Redaelli}, {Riva}, {Santana-Tschudi}, {Santin}, {Santos}, {Segovia},
  {Sosnowska}, {Span{\`o}}, {Tenegi}, {Iwert}, {Zanutta}, \&
  {Zerbi}}]{Tabernero2021}
{Tabernero}, H.~M., {Zapatero Osorio}, M.~R., {Allart}, R., {et~al.} 2021,
  \aap, 646, A158, \dodoi{10.1051/0004-6361/202039511}

\bibitem[{{Talens} {et~al.}(2018){Talens}, {Justesen}, {Albrecht}, {McCormac},
  {Van Eylen}, {Otten}, {Murgas}, {Palle}, {Pollacco}, {Stuik}, {Spronck},
  {Lesage}, {Grundahl}, {Fredslund Andersen}, {Antoci}, \&
  {Snellen}}]{Talens2018}
{Talens}, G.~J.~J., {Justesen}, A.~B., {Albrecht}, S., {et~al.} 2018, \aap,
  612, A57, \dodoi{10.1051/0004-6361/201731512}

\bibitem[{{Th{\'e}venin} {et~al.}(2017){Th{\'e}venin}, {Oreshina}, {Baturin},
  {Gorshkov}, {Morel}, \& {Provost}}]{Thevenin2017}
{Th{\'e}venin}, F., {Oreshina}, A.~V., {Baturin}, V.~A., {et~al.} 2017, \aap,
  598, A64, \dodoi{10.1051/0004-6361/201629385}

\bibitem[{{Tsiaras} {et~al.}(2018){Tsiaras}, {Waldmann}, {Zingales},
  {Rocchetto}, {Morello}, {Damiano}, {Karpouzas}, {Tinetti}, {McKemmish},
  {Tennyson}, \& {Yurchenko}}]{Tsiaras2018}
{Tsiaras}, A., {Waldmann}, I.~P., {Zingales}, T., {et~al.} 2018, \aj, 155, 156,
  \dodoi{10.3847/1538-3881/aaaf75}

\bibitem[{{von Essen} {et~al.}(2020){von Essen}, {Mallonn}, {Hermansen},
  {Nixon}, {Madhusudhan}, {Kjeldsen}, \&
  {Tautvai{\v{s}}ien{\.{e}}}}]{vonEssen2020}
{von Essen}, C., {Mallonn}, M., {Hermansen}, S., {et~al.} 2020, \aap, 637, A76,
  \dodoi{10.1051/0004-6361/201937169}

\bibitem[{{Wardenier} {et~al.}(2021){Wardenier}, {Parmentier}, {Lee}, {Line},
  \& {Gharib-Nezhad}}]{Wardenier2021}
{Wardenier}, J.~P., {Parmentier}, V., {Lee}, E. K.~H., {Line}, M.~R., \&
  {Gharib-Nezhad}, E. 2021, \mnras, 506, 1258, \dodoi{10.1093/mnras/stab1797}

\bibitem[{{Welbanks} \& {Madhusudhan}(2019)}]{Welbanks2019}
{Welbanks}, L., \& {Madhusudhan}, N. 2019, \aj, 157, 206,
  \dodoi{10.3847/1538-3881/ab14de}

\bibitem[{{West} {et~al.}(2016){West}, {Hellier}, {Almenara}, {Anderson},
  {Barros}, {Bouchy}, {Brown}, {Collier Cameron}, {Deleuil}, {Delrez}, {Doyle},
  {Faedi}, {Fumel}, {Gillon}, {G{\'o}mez Maqueo Chew}, {H{\'e}brard}, {Jehin},
  {Lendl}, {Maxted}, {Pepe}, {Pollacco}, {Queloz}, {S{\'e}gransan}, {Smalley},
  {Smith}, {Southworth}, {Triaud}, \& {Udry}}]{West2016}
{West}, R.~G., {Hellier}, C., {Almenara}, J.~M., {et~al.} 2016, \aap, 585,
  A126, \dodoi{10.1051/0004-6361/201527276}

\bibitem[{{Yan} \& {Henning}(2018)}]{Yan2018}
{Yan}, F., \& {Henning}, T. 2018, Nature Astronomy, 2, 714,
  \dodoi{10.1038/s41550-018-0503-3}

\end{thebibliography}

\end{document}